\pdfoutput=1
\documentclass[PAPER, atlasdraft=false, texlive=2025, cernpreprint, UKenglish, texmf, orcidlogo]{atlasdoc}
\usepackage[subfigure=true,block=none]{atlaspackage}
\usepackage{multirow}
\usepackage{atlasbiblatex}

\usepackage{atlasphysics}
\usepackage{rotating}
\graphicspath{{logos/}{figures/}}

\usepackage{ANA-STDM-2018-27-PAPER-defs}

\AtlasTitle{Electroweak diboson production in
association with a high-mass dijet system in
semileptonic final states from $pp$ collisions at
$\sqrt{s} = 13$ TeV with the ATLAS detector}

\AtlasVersion{2.5}

\AtlasAbstract{%
This paper reports the observation of electroweak diboson ($WW/WZ/ZZ$) production in association with a high-mass dijet system, in which final states with one boson decaying leptonically and the other boson decaying hadronically are studied.
The hadronically decaying $W/Z$ boson is reconstructed as either two small-radius jets or one large-radius jet with jet substructure requirements.
The data analyzed correspond to an integrated luminosity of
$140 \ \text{fb}^{-1}$ of proton--proton collisions at a center-of-mass energy of $\sqrt{s}=13$ TeV collected with the ATLAS detector during the 2015-2018 data taking at the Large Hadron Collider.
The electroweak production of $WW/WZ/ZZ$ in association with two jets is observed in a phase space dominated by vector-boson scattering with a significance of $7.4\sigma$ (expected $6.1\sigma$)
and the signal strength is determined to be $1.28^{+0.23}_{-0.21}$. The corresponding production cross section in a fiducial phase space is measured in addition.
The signal strengths of both electroweak and QCD associated diboson productions are furthermore measured in a two-dimensional fit, the result of which agrees with the Standard Model prediction.
The data are interpreted in the context of a dimension-8 effective field theory to probe anomalous quartic gauge couplings resulting in the first set of exclusion limits on the Wilson coefficients in the semileptonic channel reported by the ATLAS Collaboration.
The observed limits for the S02, T0 and M0 operators are
$(-3.96 < f_{S02} / \Lambda^4 < 3.96)$~TeV$^{-4}$,
$(-0.25 < f_{T0} / \Lambda^4 < 0.22)$~TeV$^{-4}$,
$(-1.26 < f_{M0} / \Lambda^4 < 1.25)$~TeV$^{-4}$.
}

\AtlasRefCode{STDM-2018-27}

\PreprintIdNumber{CERN-EP-2025-050}

\AtlasJournal{EPJC}
\AtlasJournalRef{\EPJC 86 (2026) 433}
\AtlasDOI{10.1140/epjc/s10052-025-15217-3}

\hypersetup{pdftitle={ATLAS document},pdfauthor={The ATLAS Collaboration}}

\begin{document}

\maketitle

\tableofcontents

\section{Introduction}

%

%

Vector Boson Scattering (VBS) involves both the self-couplings of the gauge bosons and their coupling to the Higgs boson. It is therefore crucial for probing the non-Abelian gauge structure of the electroweak (EWK) sector of the Standard Model (SM). In the absence of the SM Higgs boson, the amplitudes for VBS would increase as a function of partonic center-of-mass energy and ultimately violate unitarity~\cite{Lee:1977yc,Lee:1977eg}. The discovery of a Higgs boson in 2012 at the Large Hadron Collider (LHC)~\cite{HIGG-2012-27,CMS-HIG-12-028}, with measured properties~\cite{HIGG-2014-06,HIGG-2013-17,CMS-HIG-14-009,CMS-HIG-14-018} consistent with those of the SM Higgs boson, represents a major milestone in understanding the EWK symmetry breaking.
The investigation of the VBS component of EWK diboson production offers a crucial test of the SM and of the Higgs mechanism, and helps to validate the current understanding of the EWK symmetry breaking.
The interest has now shifted to probing the EWK diboson scattering at higher energies.
In this regard, the semileptonic channel offers a unique possibility to access EWK diboson events at a higher boson transverse momentum (\pt) than in other final states.

Theories of new phenomena beyond the SM (BSM) that foresee anomalous quartic gauge couplings (aQGC)
~\cite{Eboli:2003nq,eboli2006p}
or include the presence of additional resonances~\cite{Chang:2013aya,Espriu:2012ih}
predict enhancements of VBS at high \pt
of the vector bosons and at high invariant mass of the diboson system.
The sensitivity of such BSM scenarios is closely tied to the energy growth
of the longitudinal vector boson scattering amplitudes,
which is governed over many orders of magnitude by the Goldstone Boson Equivalence Theorem.
This energy-dependent behavior amplifies potential deviations from the Standard Model,
making VBS a particularly powerful probe of electroweak symmetry breaking and new physics.
The aQGC can be studied in EWK diboson production by comparing measured cross sections to SM predictions.
The effects of these anomalous couplings can be parameterized in the SM effective field theory (EFT) framework.
In addition to VBS diboson signatures, aQGC can also be constrained with triboson final states \cite{STDM-2021-09, CMS-SMP-15-008}.
To model possible BSM contributions to the \VBS process in an EFT framework, the Eboli model~\cite{eboli2006p} is used.
This model introduces 21 new dimension-8 operators which satisfy the SM $SU(2) \times U(1)_Y$ symmetry.
Among these operators, 19 affect the semileptonic final state.

\begin{figure}[!t]
\centering
\subfigure[]{\includegraphics[width=0.325\textwidth]{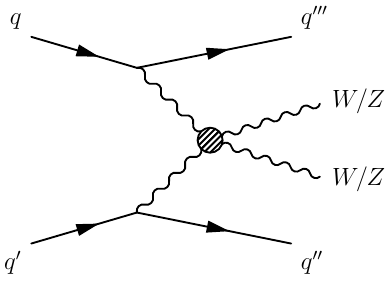}}
\subfigure[]{\includegraphics[width=0.325\textwidth]{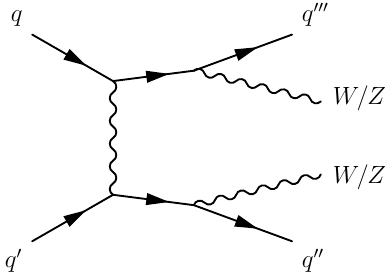}}
\subfigure[]{\includegraphics[width=0.325\textwidth]{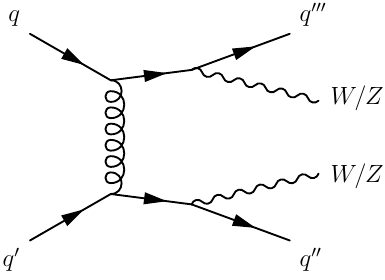}}
\caption{\label{fig:diag} Representative Feynman diagrams for (a) EWK $VVjj$\ production via VBS, (b) EWK $VVjj$\ production via a non-VBS contribution, and (c) QCD $VVjj$\ production.
}
\end{figure}

The experimental signature of VBS is characterized by a pair of SM vector bosons ($V$ = $W$ or $Z$) and two forward jets with a large rapidity separation and a large invariant mass. The SM predicts several processes with the same final state of two bosons and two jets. At tree level, the production of $VVjj$ has both an EWK contribution involving only EWK-interaction vertices, and a strong contribution (QCD-induced) involving two strong-interaction vertices. The EWK production can be further characterized into two components. The first component is EWK VBS production with actual scattering of the two EWK bosons. The scattering occurs via quartic gauge vertices, or triple gauge vertices involving the $s$- or $t$-channel exchange of a Higgs boson or a $W/Z$ boson. The second component is an EWK non-VBS production that also has only EWK vertices, but without the scattering of the two bosons.
The EWK VBS component cannot be separated from the other components in a gauge invariant way~\cite{Barger:1990py}. Despite the analysis being designed to enhance the VBS component, all the components introduced are included in the signal generation and all of them contribute to the total cross section. Representative Feynman diagrams at tree level are shown in Figure~\ref{fig:diag}.

The amount of diboson events in the semileptonic decay channel is higher than the one expected in fully leptonic final states;
indeed, the branching fraction of a $V$ decaying into quarks is more than a factor two larger than the one of the $V$ decaying into charged leptons or neutrinos.
On the other side, this decay channel is characterised by a larger contamination from other SM processes.
Nevertheless, this final state offers a larger set of data in the high momentum phase space that can be analyzed to probe further the EWK interactions foreseen by the SM (by measuring the \VBS process) and to look further for deviations from the SM predictions (search for aQGC).

The effect of new physics introduced by aQGC can be estimated using an EFT
~\cite{Degrande_2013} and parameterized linearly by an effective Lagrangian as:
\begin{equation*}
\mathcal{L}=\mathcal{L}_{sm}+\sum_{i}\frac{c_i}{\Lambda^{2}}\mathcal{L}_i+\sum_{n}\frac{f_n}{\Lambda^{4}}\mathcal{L}_n,
\end{equation*}
where $\mathcal{L}_{sm}$ represents the SM term
and $\mathcal{L}_{i}$ and $\mathcal{L}_{n}$ represent possible dimension-6 and dimension-8 operators;
$c_i$ and $f_n$ are their corresponding Wilson coefficients, and $\Lambda$ represents the energy scale at which the new degrees of freedom are integrated out.
Both dimension-6 and dimension-8 operators contribute to the \VBS process,
but due to the tight constraints on dimension-6 operators from inclusive diboson and Higgs measurements~\cite{ATL-PHYS-PUB-2022-037}, only dimension-8 operators are discussed in this article. %

Both the ATLAS and CMS Collaborations have searched for experimental evidence of EWK diboson production in a VBS-enhanced phase space.
So far, evidence or observation of EWK $VVjj$\ production has been reported in
the same-sign $W^\pm W^\pm jj$ ~\cite{STDM-2017-06, CMS-SMP-17-004},
opposite-sign $W^\pm W^\mp jj$ ~\cite{STDM-2022-06, CMS-SMP-21-001},
$WZjj$ ~\cite{STDM-2018-35, CMS-SMP-18-001},
$ZZ jj$ ~\cite{STDM-2017-19, CMS-SMP-20-001},
$W \gamma jj$ ~\cite{STDM-2018-31, CMS-SMP-21-011},
$Z (\rightarrow \nu\nu) \gamma jj$ ~\cite{STDM-2018-59},
and
$Z (\rightarrow ll) \gamma jj$ ~\cite{STDM-2018-36, CMS-SMP-20-016}
channels
with the weak bosons decaying in the fully leptonic final states
using data collected at a center-of-mass energy of $\sqrt{s} = 13$~\TeV.
The ATLAS Collaboration reported observation of EWK $VVjj$\ production in all these channels.
This paper presents a measurement
of EWK diboson production with two jets in a semileptonic final state.
The CMS Collaboration reported an evidence in the semileptonic final state $W(\rightarrow l \nu)V \pm jj$ \cite{CMS-SMP-20-013}.

Several of the diboson channels referenced above are sensitive and constrain the production of aQGC;
the same-sign $W^\pm W^\pm jj$, $WZjj$,
$W \gamma jj$ and $Z (\rightarrow \nu\nu) \gamma jj$ channels
provide constraints on EFT operators.
The CMS Collaboration performed a search for aQGC in semileptonic final states using a partial dataset of $35.9$~fb$^{-1}$, reporting EFT limits \cite{CMS-SMP-18-006}.
The ATLAS Collaboration had never reported EFT results in this final state, so this analysis allows EFT operators to be constrained at higher momentum than previous ATLAS results.

The ATLAS Collaboration reported a search for the EWK diboson production in the semileptonic final states~\cite{STDM-2017-20}, but using only a partial dataset of $35.5$ fb$^{-1}$.
This paper represents an improvement with respect to the cited  result insofar it benefits from
a larger dataset
and improved reconstruction and calibration of the objects used, and it extends the scope of the analysis strategy and data interpretation.

Three \VVjj semileptonic decay channels are explored:
a $Z$ boson decaying into a pair of neutrinos, $Z\to\nu\nu$;\footnote{To simplify the notation, antiparticles are not explicitly labeled in this paper.}
a $W$ boson decaying into a charged lepton~(an electron or muon, denoted by $\ell$) and a neutrino, $W\to\ell\nu$;
and a $Z$ boson decaying into a pair of light charged leptons~(electrons or muons, each denoted by $\ell$), $Z\to\ell\ell$. In all cases, the other vector boson $V$ is required to decay into a pair of quarks, \Vtoqq,
leading to three reconstructed diboson final states: \ZVvvqq, \WVlvqq\ and \ZVllqq.
The reconstructed events are split according to the number of charged leptons in three categories: 0-, 1- and 2-lepton channels.
A fiducial region of the measurement is defined to account for the geometrical, reconstruction and analysis acceptance of leptons and jets.

The \Vtoqq\ decay is considered over a large \pt range of the vector boson; as \pt increases, the hadronic activity from the decay is bound in an increasingly narrower solid angle with a large overlap of the two jets.
In this case, hadrons from the two quarks overlap in the detector and are more efficiently reconstructed as a single large-$R$ jet.
Therefore, two different reconstruction techniques for the \Vtoqq\ decay are considered: resolved and merged. The resolved reconstruction attempts to identify two separate small-radius jets~(small-$R$ jets denoted by $j$) of hadrons from the \Vtoqq\ decay, and the merged reconstruction uses a large-radius jet~(large-$R$ jet denoted by $J$) in conjunction with jet substructure techniques to identify the decay.
Although the introduction of a merged category appears to be more challenging, it is a crucial category for semileptonic final states that allows the exploration of a kinematic regime defined by a boson \pt > 200~GeV that would otherwise remain inaccessible with the resolved regime alone.
In terms of sensitivity, the merged regime significantly contributes to the \VBS measurement while it is the dominant regime for the EFT interpretation.
In total, six final states are included in this study: 0-, 1- and 2-lepton final states, each using resolved or merged \Vtoqq\ reconstruction techniques;
the merged final state is further divided in two regions as described in Section \ref{sec:eventselection}.

To probe the EWK production mechanism,
the diboson semileptonic events are selected by requiring two additional energetic jets in the forward region of the detector.
Dedicated kinematic requirements allow a further enhancement of the VBS component in the selected events.

To extract the signal and to measure the cross section for the EWK production of \VVjj, a machine learning (ML) discriminant, which exploits the kinematics and the track multiplicity of the reconstructed jets, is used to separate the EWK-induced \VVjj\ production from other processes in the SM.
Candidate signal events are used to measure the cross section of EWK \VVjj\ production in a region of kinematic phase space close to the detector and analysis acceptance.
Both the EWK and QCD associated diboson production are measured in a two dimensional fit.
Fiducial cross sections are measured in the 0-, 1- and 2-lepton channels.
The diboson system invariant mass is combined with the ML discriminant to derive limits on the Wilson coefficients of the dimension-8 EFT Lagrangian.

%


\section{ATLAS detector}

%
%
%
\newcommand{\AtlasCoordFootnote}{%
ATLAS uses a right-handed coordinate system with its origin at the nominal interaction point (IP)
in the centre of the detector and the \(z\)-axis along the beam pipe.
The \(x\)-axis points from the IP to the centre of the LHC ring,
and the \(y\)-axis points upwards.
Cylindrical coordinates \((r,\phi)\) are used in the transverse plane,
\(\phi\) being the azimuthal angle around the \(z\)-axis.
The pseudorapidity is defined in terms of the polar angle \(\theta\) as \(\eta = -\ln \tan(\theta/2)\).
Angular distance is measured in units of \(\Delta R \equiv \sqrt{(\Delta y)^{2} + (\Delta\phi)^{2}}\).}

The ATLAS experiment~\cite{PERF-2007-01} at the LHC is a multipurpose particle detector
with a forward--backward symmetric cylindrical geometry and a near \(4\pi\) coverage in
solid angle.\footnote{\AtlasCoordFootnote}
It consists of an inner tracking detector surrounded by a thin superconducting solenoid
providing a \qty{2}{\tesla} axial magnetic field, electromagnetic and hadronic calorimeters, and a muon spectrometer.
The inner tracking detector (ID) covers the pseudorapidity range \(|\eta| < 2.5\).
It consists of silicon pixel, silicon microstrip, and transition radiation tracking detectors.
Lead/liquid-argon (LAr) sampling calorimeters provide electromagnetic (EM) energy measurements
with high granularity within the region \(|\eta|< 3.2\).
A steel/scintillator-tile hadronic calorimeter covers the central pseudorapidity range (\(|\eta| < 1.7\)).
The endcap and forward regions are instrumented with LAr calorimeters
for EM and hadronic energy measurements up to \(|\eta| = 4.9\).
The muon spectrometer (MS) surrounds the calorimeters and is based on
three large superconducting air-core toroidal magnets with eight coils each.
The field integral of the toroids ranges between \num{2.0} and \qty{6.0}{\tesla\metre}
across most of the detector.
The muon spectrometer includes a system of precision tracking chambers up to \(|\eta| = 2.7\) and fast detectors for triggering up to \(|\eta| = 2.4\).
The luminosity is measured mainly by the LUCID--2 detector which is located close to the beampipe.
A two-level trigger system is used to select events.
The first-level trigger is implemented in hardware and uses a subset of the detector information
to accept events at a rate close to \qty{100}{\kHz}.
This is followed by a software-based trigger that
reduces the accepted rate of complete events to \qty{1.25}{\kHz} on average,
depending on the data-taking conditions.
A software suite~\cite{SOFT-2022-02} is used in data simulation, in the reconstruction
and analysis of real and simulated data, in detector operations, and in the trigger and data acquisition
systems of the experiment.


\section{Data and simulated event samples}

%
%

\subsection{Data}

This analysis uses the 2015-2018 proton--proton (\pp) collision data at $\sqrt{s}=13$~\tev \
recorded with the ATLAS detector at the LHC for a total integrated luminosity of 140~\ifb \cite{DAPR-2021-01}.
Events used for the analysis were recorded with all detector systems operating normally and passing data-quality requirements \cite{ATLAS-CONF-2015-029}.

To optimize the trigger efficiency for the signal process, a combination of unprescaled single-lepton with tight isolation requirements and missing transverse momentum (\met) triggers were used \cite{TRIG-2016-01, TRIG-2018-05, TRIG-2018-01, TRIG-2019-01}.
During data-taking, the thresholds of the single-lepton triggers with tight isolation requirements were increased in stages alongside the increase in instantaneous luminosity.
The lepton triggers with tight isolation were also complemented by triggers with looser isolation but higher \et or \pt thresholds.
In the 0-lepton channel, \met triggers were used with the lowest threshold varying from 70 to 110~\gev.
The single-muon (single-electron) triggers were used in the 2-lepton channel with lowest thresholds of 20 (24) to 26 (26)~\gev.
For the 1-lepton channel, the same single-lepton triggers were also used, but the single-muon triggers were complemented with \met triggers.
Since at the trigger level, the \met calculation does not account for the presence of muons, the inclusion of \met triggers reduces the efficiency loss from the difference in trigger and reconstructed muon tracking.
The trigger efficiency for signal events in the 0 and 2-lepton channels is above 90\%, and above 80\% in the 1-lepton channel.

\subsection{Signal and background simulation}
\label{subsec:mc}

EWK signal samples,
$W(\rightarrow l \nu)V$, $Z(\rightarrow \nu\nu)V$, $Z(\rightarrow ll)V$,
with one boson decaying leptonically and the other hadronically,
were generated using \MGNLO 2.6.6~\cite{Alwall:2014hca} at the order $O(\alpha^6_{\mathrm{EWK}})$, where $\alpha_{\mathrm{EWK}}$ is the EWK coupling constant.
Both the VBS and non-VBS amplitudes of the \VVjj process were included.
The latter contains also diagrams with a $W$ boson and a top quark ($Wtb$) resulting into a \VVjj final state.
This contribution is suppressed by an event selection requirement, but it cannot be completely removed to preserve the gauge invariance. Therefore, the signal process is represented by the EWK production
and its vector-boson-scattering (\VBS) contribution will be enhanced by the analysis selection.
The \NNPDF[3.0nlo]~\cite{Ball:2014uwa} parton distribution functions (PDF) set was used and the parton shower and hadronization were modelled with Pythia 8.186 ~\cite{Sjostrand:2007gs} using the A14 set of tuned parameters (tune) for the underlying event~\cite{ATL-PHYS-PUB-2014-021}.

Potential modifications to the SM processes introduced by the Eboli model operators are implemented in the BSM signal samples.
For each operator two terms were considered, the one including the Lagrangian term of the BSM effect only (pure-BSM) and the one including the SM-BSM interference. Both terms were modelled using \MGNLO v2.7.2~\cite{Alwall:2014hca} at leading order (LO) with the boson decays simulated using \MADSPIN~\cite{Artoisenet:2012st}.
The \NNPDF[3.0lo]~\cite{Ball:2014uwa} PDF set was used with \PYTHIA[8.244]~\cite{Sjostrand:2014zea} for parton shower and hadronization.
The total \VBS process for a certain operator and coupling value was then modelled by summing the SM \VBS process with the pure-BSM and the SM-BSM interference terms. %

The background processes include the production of \Wboson and \Zboson bosons in association with jets (\Vjets), top-quark pair production (\ttbar) and single-top quark production (\singletop).
The QCD-induced diboson production at order $O(\alpha_{EWK}^4\alpha_{QCD}^2)$ was also treated as a background process (QCD VV). The QCD and EWK diboson processes were generated separately and therefore the effect of the interference between the two was not included in the simulated samples;
nevertheless this effect was estimated with a simplified procedure based on samples generated at the truth level without the full detector simulation. The interference contribution was found to be 5\% to 10\% of the EWK process for the $Z(\rightarrow \nu\nu/ll)V$ samples and 10\% to 20\% for the $W(\rightarrow l \nu)V$ samples;
the contribution is included in the analysis as a systematic effect.

The dominant \Vjets background process was simulated with the
\SHERPA[2.2.1]~\cite{Bothmann:2019yzt} generator using next-to-leading-order (NLO) matrix elements (ME) for up to two partons,
and LO matrix elements for up to four partons calculated with the \COMIX~\cite{Gleisberg:2008fv} and \OPENLOOPS~\cite{Buccioni:2019sur,Cascioli:2011va,Denner:2016kdg} libraries.
The \VVQCD processes with one of the bosons decaying hadronically and the other one leptonically were simulated
using \SHERPA 2.2.1 with up to one additional parton at NLO and up to three additional partons at LO using the \COMIX and \OPENLOOPS programs.
For both \Vjets and \VVQCD processes, the \NNPDF[3.0nnlo] set of PDFs
was used, and each sample matched with the \SHERPA parton shower~\cite{Schumann:2007mg} using the MEPS@NLO prescription~\cite{Hoeche:2011fd,Hoeche:2012yf,Catani:2001cc,Hoeche:2009rj} and the set of tuned parameters developed by the \SHERPA authors.

The \ttbar and single-top quark processes were modelled using the \POWHEGBOX[v2]~\cite{Re:2010bp,Frixione:2007nw,Nason:2004rx,Frixione:2007vw,Alioli:2010xd} generator at NLO with the \NNPDF[3.0nlo] PDF set.
The \hdamp parameter, which regulates the high-\pT radiation against which the \ttbar system recoils, was set to 1.5\,$m_{t}$\cite{ATL-PHYS-PUB-2016-020}.
The events were interfaced to \PYTHIA[8.230]
to model the parton shower, hadronization, and underlying event, with parameters set according to the A14 tune
and using the \NNPDF[2.3lo] set of PDFs~\cite{Ball:2012cx}.
For the \singletop channel, the top-quark decay was simulated using \MADSPIN, while the remaining samples with top quarks have the top quark spin correlations preserved.
The decays of bottom and charm hadrons were performed by \EVTGEN[1.6.0]~\cite{Lange:2001uf}.

All simulated processes were normalized using the best currently available theoretical predictions for their cross sections.
The \Vjets cross sections were calculated at next-to-next-to-leading-order (NNLO) in QCD ~\cite{Anastasiou:2003ds} and the \VVQCD at NLO including LO contributions with two additional partons~\cite{Gleisberg:2008ta,Hoeche:2009rj}.
The \ttbar production cross section was calculated at NNLO in QCD, including resummation of next-to-next-to-leading logarithmic
(NNLL) soft-gluon terms~\cite{Czakon:2013goa, Czakon:2011xx}.
The \singletop production cross sections were calculated to NLO in QCD~\cite{kidonakis2010next}, including the soft-gluon resummation at NNLL~\cite{Kidonakis:2010ux} for the $Wt$ process.
The normalization of the \Vjets and \ttbar processes in the relevant phase space were further improved by using data driven estimations in control regions as detailed in Section~\ref{sec:eventselection}.

Simulated events were processed with a detailed detector simulation \cite{SOFT-2010-01} based on Geant4~\cite{Agostinelli:2002hh}.
The generation of the simulated event samples includes the effect of multiple $pp$ interactions per bunch crossing, as well as the effect on the detector response due to interactions from bunch crossings before or after the one containing the hard interaction (pile-up).
The MC samples were reweighted to match the pile-up conditions in the data.
%


\section{Object reconstruction}

\label{sec:objectreconstruction}
Collision vertices are formed from tracks with $p_T > 500$~\MeV. The vertex candidate with the highest $\Sigma p_T^2$ of its associated tracks is selected as the primary vertex. All selected events
are required to contain a primary vertex with at least two associated tracks.

Electrons are identified as isolated energy clusters in the electromagnetic calorimeter matched to ID tracks, and are required to have $E_T > 7$~\GeV\ and $|\eta| < 2.47$, excluding the electromagnetic calorimeter  barrel–endcap transition region $1.37 < |\eta| < 1.52$. A likelihood-based requirement~\cite{EGAM-2021-01}
is imposed to reduce the background from non-prompt electrons or hadrons misidentified as electrons. Electrons are classified as either ‘loose’, ‘medium’ or ‘tight’ according to the likelihood-based identification criteria described in Ref. ~\cite{EGAM-2018-01}.

Muons are reconstructed by a combined fit to the ID and MS tracks and are required to have $\pt>7$~\GeV\ and $|\eta|<2.5$. Muons must pass identification requirements based on the number of hits in the ID and MS subsystems, and on the significance of the difference $|q/p_{\mathrm{MS}} - q/p_{\mathrm{ID}}|$~\cite{MUON-2018-03}
, where $q$ is the charge and $p_{\mathrm{MS}}\ (p_{\mathrm{ID}})$ is the momentum of the muon measured in the MS (ID). Similarly to electrons, muons are classified as either `loose', `medium' or `tight', following the criteria in Ref. ~\cite{PERF-2015-10}.

In this analysis, only ‘loose’ or ‘tight’ electrons and muons are selected.
All electrons and muons are required to be isolated
by using selections based on the sum
in small cones around the lepton direction
of the $\pt$ of tracks,
of calorimeter $E_T$, or both,
to further reduce backgrounds from non-isolated sources.
Leptons of loose quality with \pt < 100 GeV are required to pass a 'FixedCutLoose' isolation requirement (\cite{EGAM-2021-01, MUON-2018-03}) and no isolation requirement is applied for \pt > 100 GeV so as not to remove the leptons from $Z\to\ell\ell$ decays which are near one another; the same strategy was used in Ref. \cite{HDBS-2018-10}.
Furthermore, leptons are required to have associated tracks satisfying $|d_0/\sigma_{d_0}|<5~(3)$ and $|z_0\times\sin\theta|<0.5$ mm for electrons~(muons), where $d_0$ is the transverse impact parameter relative to the beam line, $\sigma_{d_0}$ is its uncertainty, and $z_0$ is the distance between the longitudinal position of the track along the beam line at the point where $d_0$ is measured and the longitudinal position of the primary vertex.

Two types of jets are employed in the analysis.
Both of them are reconstructed
using the anti-$k_t$ algorithm~\cite{Cacciari:2008gp,Fastjet}
but different values of the radius parameter $R$.
Small-$R$ jets are reconstructed by clustering particle-flow objects \cite{PERF-2015-09} with a radius parameter of $R=0.4$ and are required to have $\pT>20$~\GeV\ for $|\eta|<2.5$ and $\pT>30$~\GeV\ for $2.5<|\eta|<4.5$.
A jet vertex tagger (JVT)~\cite{PERF-2014-03} is applied to jets with $20$~\GeV\ $<\pT<60$~\GeV\ and $|\eta|<2.4$ to suppress jets from pile-up interactions.
This tagger uses information on tracks associated with the primary vertex and pile-up vertices.
The forward-JVT (fJVT)~\cite{PERF-2016-06} with the loose working point is also applied
to jets with  $20$~\GeV\ $<\pT<120$~\GeV\ and $2.5<|\eta|<4.5$.
Energy- and $\eta$-dependent correction factors derived from MC simulations are applied to correct jets to the particle level~\cite{PERF-2016-04}.
Several in situ techniques are employed to correct for differences between data and simulation and to measure the resolution of jets \cite{JETM-2018-05}.
Small-$R$ jets containing $b$-hadrons are identified using a multivariate algorithm ($b$-tagging) based on a deep feed-forward neural network (DL1 $b$-tagging)~\cite{FTAG-2018-01}.
The chosen $b$-tagging algorithm has an efficiency of 70\% for $b$-quark jets in simulated $t\bar{t}$ events, with a light-flavor jet rejection factor of about 390 and a $c$-jet rejection factor of about 9.4~\cite{FTAG-2018-01}.

Large-$R$ jets are reconstructed with a radius parameter of $R=1.0$ and they are built from topological clusters, which are calibrated to the hadronic scale using the local hadronic
cell weighting scheme \cite{PERF-2014-07}.
To mitigate the effects of pile-up and soft radiation, the large-$R$ jets are groomed following the trimming algorithm ~\cite{jet_trimming}.
Trimming takes the original constituents of the jet and reclusters them using the $k_t$ algorithm \cite{kt_algorithm} with a smaller radius parameter, $R_{\mathrm{subjet}}$, to produce a collection of subjets. These subjets are discarded if they carry less than a specific fraction ($f_{\mathrm{cut}}$) of the original jet \pT.
The trimming parameters were optimized for $W/Z$ boson tagging and are $R_{\mathrm{subjet}} =0.2 $ and $f_{\mathrm{cut}}=5\%$.
The large-$R$ jet four-momenta are recomputed from the remaining subjets, and the jet energies are calibrated to particle level using correction factors derived from MC simulations~\cite{PERF-2012-02}.
The jet energy scale and resolution are measured using SM events
relying on other and well calibrated objects as reference;
the jet mass response is measured using the masses of high-$p_T$ $W$ bosons and top quarks, and by comparing the jet mass measured using the energy deposited in the calorimeter with that using the momenta of charged-particle tracks \cite{JETM-2018-02}.
Large-$R$ jets are required to have $\pt > 200$~\GeV, $|\eta|<2.0$ and a mass $m_J > 50$~\GeV.
Moreover, the jet substructure variable, $D_{2}$, reconstructed from energy correlation functions based on energies and pair-wise angles of the sub-constituents \cite{Larkoski:2014gra,Larkoski:2015kga} is used.
The $D_{2}$ variable is sensitive to the 2-prong sub-structure expected from the $W/Z$ bosons decay.
Furthermore, the multiplicity of tracks associated to the large-$R$ jet before the grooming algorithm and passing standard reconstruction criteria with \pt $> 500$ MeV (\Ntrk) is used to further improve the background rejection.

\begin{figure}[t]
\centering
\subfigure[]{\includegraphics[width=.45\linewidth]{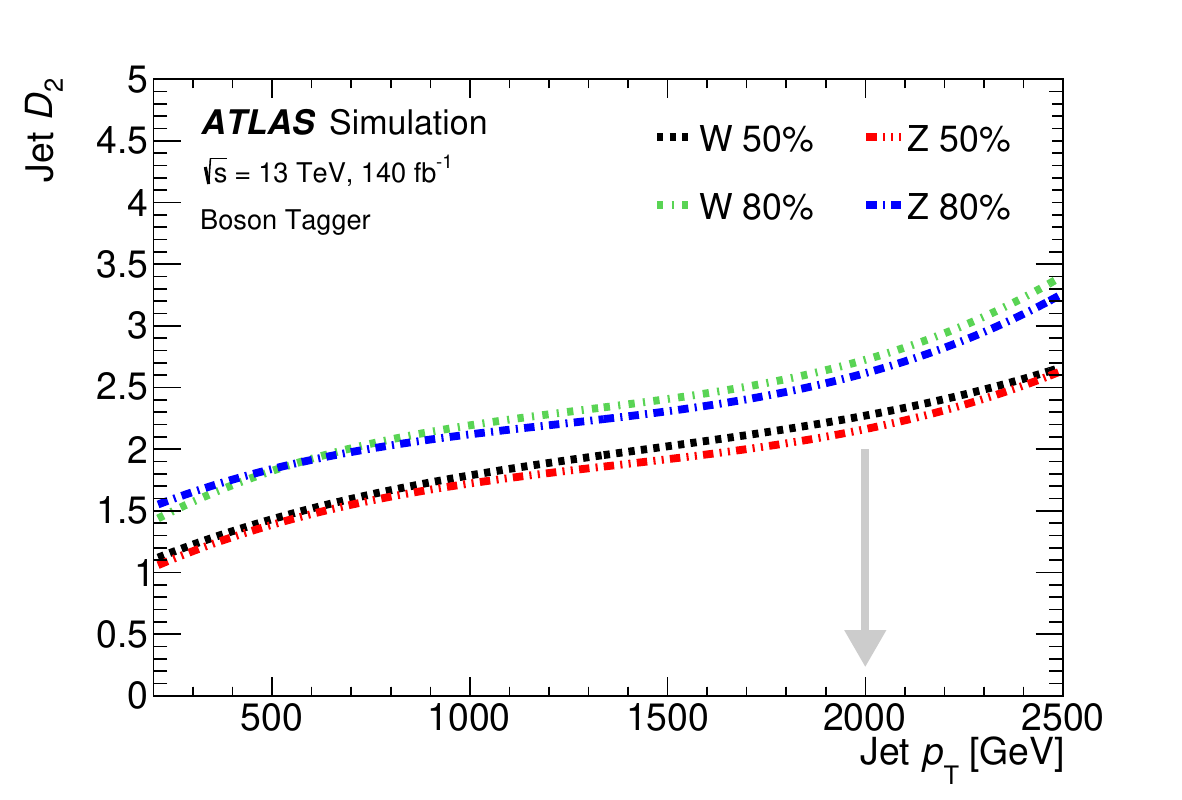}}
\subfigure[]{\includegraphics[width=.45\linewidth]{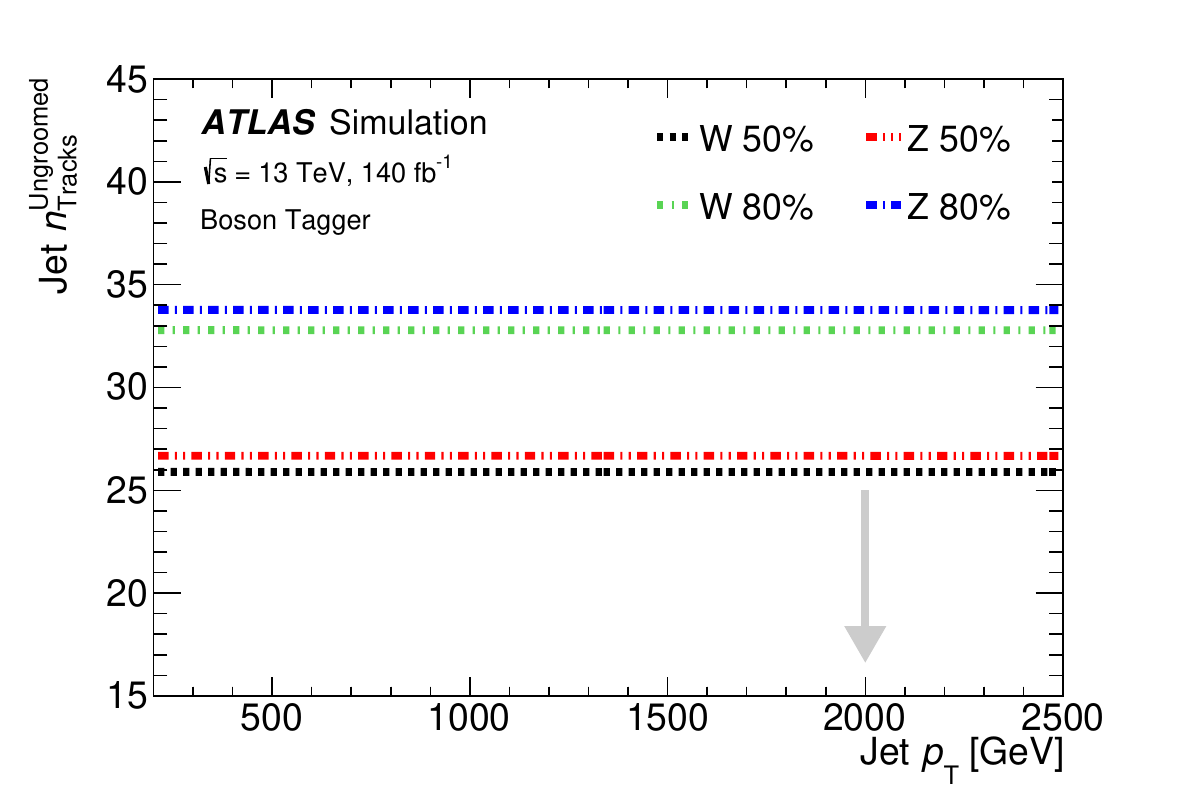}}
\subfigure[]{\includegraphics[width=.45\linewidth]{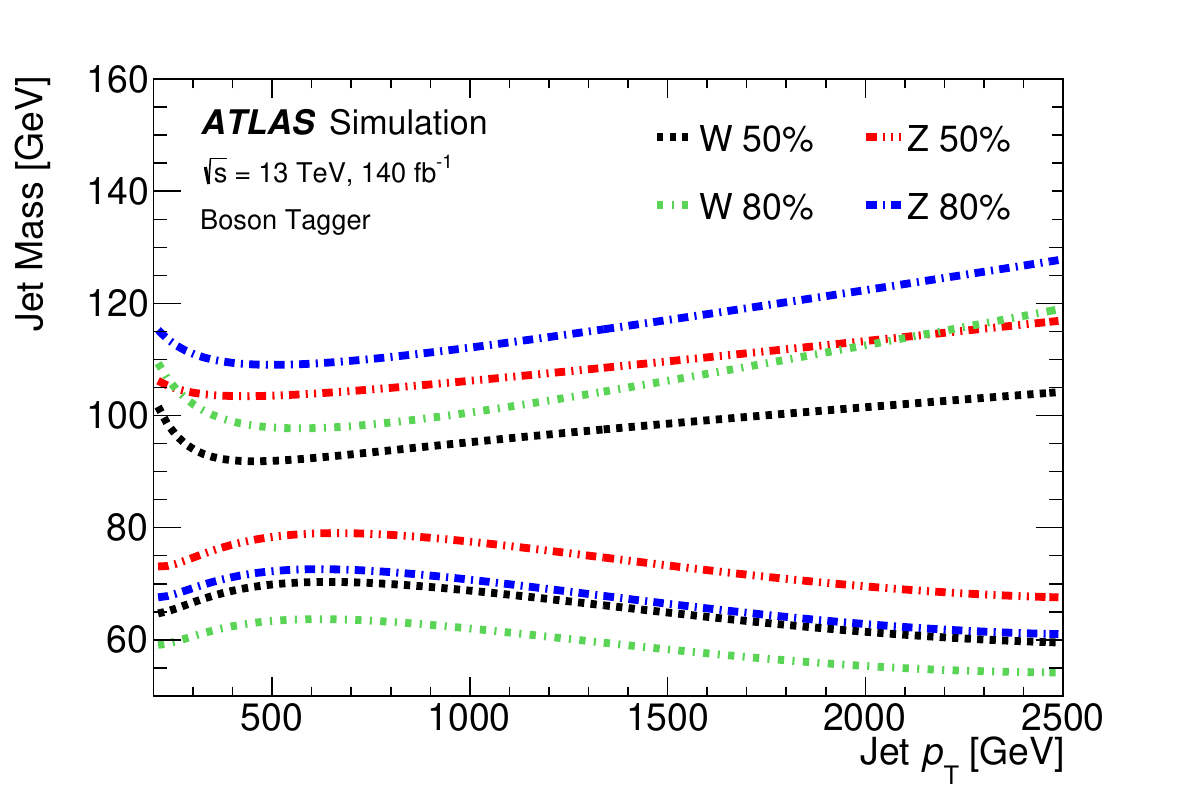}}
\caption{The requirement thresholds applied by the boson tagging algorithm on a) \Dtwo, b) \Ntrk, and c) jet-mass for different boson hypothesis and working points corresponding to 50\% and 80\% signal efficiency as a function of jet \pt. All three requirements are applied to derive the efficiency of the working point. The grey arrow shows the direction of the requirement applied for the \Dtwo and \Ntrk variables, while for the mass, jets within the window are selected.}
\label{fig:TaggerCuts}
\end{figure}

Boson tagging is applied to large-$R$ jets to select those consistent with $V\to qq$ decays. The tagger is based on three variables, namely the sub-structure variables, \Dtwo and \Ntrk, and the jet mass.
Large-$R$ jets from a $V\to qq$ decay are expected to have values for the \Dtwo and the \Ntrk variables smaller than the ones of jets initiated by quarks or gluons, while the mass is expected to peak around the $W$ or $Z$ boson mass.
The algorithm operating points are based on simultaneous requirements on the three variables;
they are designed to provide a constant efficiency independent of the large-$R$ jet $\pT$ for the signals studied and they are labelled according to their efficiency.
The requirement thresholds applied to the three variables therefore depend on the \pt of the jet and are shown in Figure~\ref{fig:TaggerCuts}.
A similar tagging scheme was used in previous diboson searches \cite{HDBS-2018-31, HDBS-2018-10}.
In this analysis, two working points, one with 50\% efficiency and the other one with 80\% efficiency, are used to define two analysis regions with different purities.

The missing transverse energy, $E_{\text{T}}^{\text{miss}}$
is reconstructed by taking the negative vectorial sum of all the reconstructed and calibrated electrons, muons, and small-$R$ jets.
An additional term is used to account
for soft radiation that leaves tracks in the inner detector that are not used in any reconstructed physics object~\cite{JETM-2020-03}. Large-$R$ jets are not included in this definition to avoid double counting with small-$R$ jets.
The track-based missing transverse momentum, $p_{T}^{\text{miss}}$, is the vectorial sum of the transverse momenta reconstructed from particle tracks.
Both \met and $p_{T}^{\text{miss}}$ are utilized in the event selection.

An overlap-removal procedure is applied to the selected leptons and jets
to remove ambiguities in the identification of the objects mentioned above
and to assign a unique hypothesis to each object.
If an electron and a small-$R$ jet are separated by
$\Delta R<$ 0.2, the jet is removed. The electron is removed if the separation satisfies 0.2 $<\Delta R<$ 0.4.
If a muon and a small-$R$ jet are separated by
$\Delta R<$ 0.2 and if the jet has less than three tracks or the energy and momentum differences between the muon and the jet are small, the jet is removed.
Otherwise the muon is removed if the separation satisfies $\Delta R<$ 0.4.
To prevent double-counting of energy from an electron inside a large-$R$ jet, the large-$R$ jet is removed if an electron and a large-$R$ jet are separated by $\Delta R < 1.0$. Small-$R$ jets and large-$R$ jets are reconstructed independently from the same energy depositions for a given event.
No overlap removal is applied between large-$R$ jets and small-$R$ jets.


\section{Event selection and background estimation}

\label{sec:eventselection}

According to the number of selected electrons and muons, each event is classified as belonging to the 0-, 1- or 2-lepton channel. After requiring a candidate boson decaying into leptons (\Vlep), the events in all three channels are required to contain two forward small-$R$ jets (called tagging jets) with a high invariant mass.
The V candidate decaying into hadrons (\Vhad) is reconstructed as either two small-$R$ jets in the resolved selection, or one large-$R$ jet in the merged selection.
The event selection is designed to guarantee the statistical independence of the channels and to maximize the combined sensitivity of the analysis. The selection results in nine non-overlapping distinct signal regions (SR):
for each of the three lepton channels, there are three categories based on the choice and selections on the \Vhad candidate, one for the resolved and two (one for each boson tagger working point) for merged \Vhad candidate.

The event selection for the entire analysis is summarized in Table \ref{tbl:evselTable}. More details are given in the following Sections.

\begin{table}[phtb]
\caption{Summary of the event selection in the 0-, 1- and 2-lepton channels.}
\small
\centering
\resizebox{\textwidth}{!}{\begin{tabular}{l  c | c | c }
\toprule
Selection & \zlep & \olep & \tlep \\
\midrule
\multirow{2}{*}{Trigger} & \met\ triggers & Single-electron triggers & Single-lepton triggers \\
&  & Single-muon or \met\ triggers & \\
\midrule
\multirow{6}{*}{$V \to \nu\nu/\ell\nu/\ell\ell$} & 0 `Loose' leptons $\pT>7$~\GeV & 1 `Tight' lepton $\pT>27$~\GeV & 2 `Loose' leptons $\pT>27$~\GeV\\
& $\met >$ 200~\textrm{\GeV} & 0 other `Loose' leptons with $\pT>7$~\GeV & \\
& $\mpt > 50$~\GeV & $\met >$ 80~\textrm{\GeV} & $83<m_{ee}<$ 99~\textrm{\GeV} \\
& $\dPhiEtmissPtmiss<\pi/2$  &  & $-0.0117\times \pT^{\mu\mu} + 85.63~\textrm{\GeV} < m_{\mu\mu}$ \\
& $\min[\dPhiEtmissj]>\pi/6$ &  & $0.0185\times \pT^{\mu\mu}+94~\textrm{\GeV} > m_{\mu\mu}$ \\
& $\dPhiEtmissVhad>\pi/9$    &  &  \\
\midrule
\multirow{2}{*}{VBS topology tagging jets} &  \multicolumn{3}{c}{ Largest \mjj pair with $\eta_{\mathrm{tag}\ j_1} \cdot \eta_{\mathrm{tag}\ j_2} <0$ }\\
&  \multicolumn{3}{c}{ $\mjjtag>400$~\GeV, $\pt^{tag\ j_{1,2}} >30$~\GeV }\\
\midrule
\multirow{3}{*}{$V \to J$} & \multicolumn{3}{c}{ Leading \pt large-$R$ jet}\\
& \multicolumn{3}{c}{$\pt> 200$ \textrm{\GeV}, $|\eta|<2$  }\\
& \multicolumn{3}{c}{ $V$ boson tagger requirements on $m_J$, \Dtwo, and \Ntrk}\\
\midrule
\multirow{4}{*}{$V \to jj$}  & \multicolumn{3}{c}{Two leading \pt  small-$R$ jets}\\
& \multicolumn{3}{c}{$\pt > 20$~\GeV\ if $|\eta| < 2.5$, and $\pt > 30$~\GeV\ if $2.5 < |\eta| < 4.5$}\\
& \multicolumn{3}{c}{$\pt^{j_{1}} >40$~\GeV}\\
& \multicolumn{3}{c}{$64 < \mjj <$ 106~\GeV}\\
\midrule
Top veto  & \multicolumn{3}{c}{$\mjjj >$ 220~\GeV (resolved only)} \\
\midrule
Additional b-jet veto & No & Yes & No \\
\bottomrule
\end{tabular}}

\label{tbl:evselTable}
\end{table}

\subsection{Event selection}

Signal events in the \zlep channel are typically characterized by a hadronically decaying $V$ boson recoiling against a large amount of \met stemming from either a \Ztovv\ decay or a \Wtolv\ decay with the lepton outside of the acceptance of the detector. An initial selection is made by requiring \met > 200~\GeV, and rejecting events with electrons or muons passing the loose quality requirements. Further selections are applied to reduce the amount of multijet background. This background originates primarily from the presence of misidentified jets and non-collision phenomena. It is suppressed
by requiring a \mpt~of at least 50~\GeV and an \met of at least 200~\GeV.
Further angular selection criteria are applied,
\dPhiEtmissPtmiss $< \pi/2$,
$\min[\dPhiEtmissj] > \pi/6$,
\dPhiEtmissVhad $> \pi/9$.
The multijet background is found to be negligible after this set of selections.

The \olep channel targets events with a leptonically decaying $W$ boson.
The \Wtolv\ candidates are selected by requiring one isolated lepton~(electron or muon) satisfying the tight criteria with $\pT> 27$~\GeV. Events are required to have \met$>80$~\GeV, and must not have any additional loose lepton. In order to reconstruct the invariant mass of the $WV$ system, the neutrino momentum four-vector is reconstructed by imposing a $W$ boson mass constraint on the lepton-neutrino system. The neutrino \pT components are set equal to the \met, and the unknown $z$-component of the momentum ($p_z$) is obtained from the resulting quadratic equation. The $p_z$ is chosen either as the smaller, in absolute value, of the two real solutions or, if the solution is complex, as its real part.

In the \tlep channel, the \Ztoll\ candidates are identified by requiring two isolated same-flavor leptons satisfying the loose criteria.
Both the leading- and subleading-\pt leptons must satisfy $\pT > 27$~\GeV.
The trigger efficiency is stable as a function of the lepton \pt when the lepton matching the trigger decision has a \pt above that threshold.
The same \pt requirement is applied on both leptons to ensure an optimal signal-to-background ratio.
Opposite charges are required for the muon pairs but not for the electron pairs, since electrons are more susceptible to charge mis-identification due to the conversion of photons from bremsstrahlung, especially at high \pT. The dilepton invariant mass is required to be consistent with that of the $Z$ boson: $83< m_{ee}<99$~\GeV\ in the case of electrons and $(-0.0117\times \pT^{\mu\mu} + 85.63~\GeV) < m_{\mu\mu}< (0.0185\times \pT^{\mu\mu}+94~\GeV)$ in the case of muons. The $\pT$-dependent requirement on $m_{\mu\mu}$ recovers the selection efficiency at high $\pT^{\mu\mu}$, which would otherwise fall due to the degraded dimuon invariant mass resolution~\cite{EXOT-2016-29}.

Tagging jets are selected from the entire small-$R$ jets collection.
Candidate jets are required to pass the basic kinematic selections introduced
in Section \ref{sec:objectreconstruction}.
The candidate jets are then
required to be in opposite hemispheres, $\etajo \cdot \etajt<0$,
and the candidate jet pair is selected as the one with the highest invariant mass. Once the candidate pair is identified, the invariant mass of the two tagging jets is further required to satisfy $\mjjtag > 400$~\GeV\ and both tagging jets are required to have $\pT >30$~\GeV.

Once the tagging jets are identified, the merged selection is applied as the first step in identifying a \Vhad candidate. If an event is not selected as merged, then the resolved selection is used.
The merged regime contributes less than the resolved one to the overall signal efficiency of the EWK process, roughly $30 \%$ of signal events have a boson with a \pT greater than $200$~\GeV,
but the lower background contamination results in a contribution to the sensitivity to the EWK process that is competitive with the one of the resolved regime.  The merged regime sensitivity is instead the leading one for the aQGC samples.
The \Vhad\ candidates are selected in each of the three lepton channels.
According to the merged regime selection,
events are required to have at least one large-$R$ jet;
large-$R$ jets overlapping with one of the selected tagging jets are discarded,
if there is more than one, the one with the leading \pt is selected as signal jet.
In the \tlep channel, events with the signal large-$R$ jet overlapping with one of the two selected leptons,
within $\Delta R<0.4$, are rejected as $Z \rightarrow ee$ candidates can potentially be reconstructed as large-$R$ jets.

Two SRs are defined in the merged regime by requirements on the boson tagging, as discussed in Section~\ref{sec:objectreconstruction}. The high-purity~(HP) SR comprises events which pass the 50\% boson tagger working point requirement.
An orthogonal low-purity~(LP) SR is defined with events which fail the high-purity requirement but still pass the 80\% boson tagger working point.
Both regions are defined without distinguishing between the $W$ and $Z$ hypotheses,
requiring that the large-$R$ jet satisfies either $W$ or $Z$ boson tagging.
In the resolved selection,
the tagging jets are excluded from the pool of jets used to identify the signal jets candidates;
the remaining two leading \pt small-$R$ jets are labelled as signal jets.
Events are required to have the dijet invariant mass falling in the $m_{W /Z}$ window:
$64 < m_{jj} < 106$~\GeV.
Out of the jets forming the selected \Vhad\ candidate, the leading-\pt one must have $\pT >40$~\GeV.

Finally, in the \olep channel,
the event is rejected if it contains any $b$-tagged small-$R$ jet
that is not a tagging jet
or a jet forming a \Vhad\ candidate (resolved)
or a jet that overlaps the \Vhad\ candidate jet (merged),
in order to reduce the contribution from top-quark processes.

The tagging jets selection is designed to enhance the VBS contribution over the other EWK contributions. Despite this selection, a significant contribution from the top-quark associated production, i.e. from diagrams with the $Wtb$ vertex, is found in the resolved regime. Therefore, a further selection is applied to improve the purity of the VBS contribution.
A triple-jet system is reconstructed selecting the jets coming from the \Vhad and an additional jet
(without any requirement on the $b$-tagging) such that the three-jet mass is the closest to the top-quark mass.
Events with a three-jet mass below $220$~\GeV\ are rejected,
allowing a reduction of the top-quark associated production
to 10\% in $WZjj$ and to 30\% in $WWjj$ samples in the resolved regime;
the corresponding efficiency for the VBS events if of around 65\%.
The contribution of the top-quark associated production is already suppressed in the boosted regime after the other selection requirements;
a further requirement based on the reconstructed top mass similar to the one used in the resolved regime does not reduce it further and is therefore not applied.

The regions as defined above are not completely orthogonal, and so to simplify the statistical procedure, a region prioritization scheme is used to maximize the signal sensitivity while removing overlapping events.
Merged regions are prioritized over resolved regions and all events are sorted into only one possible analysis region.
The flow is summarized schematically in Figure \ref{fig:eventcatflow}.

\begin{figure}[t]
\centering
\includegraphics[width=0.8\textwidth]{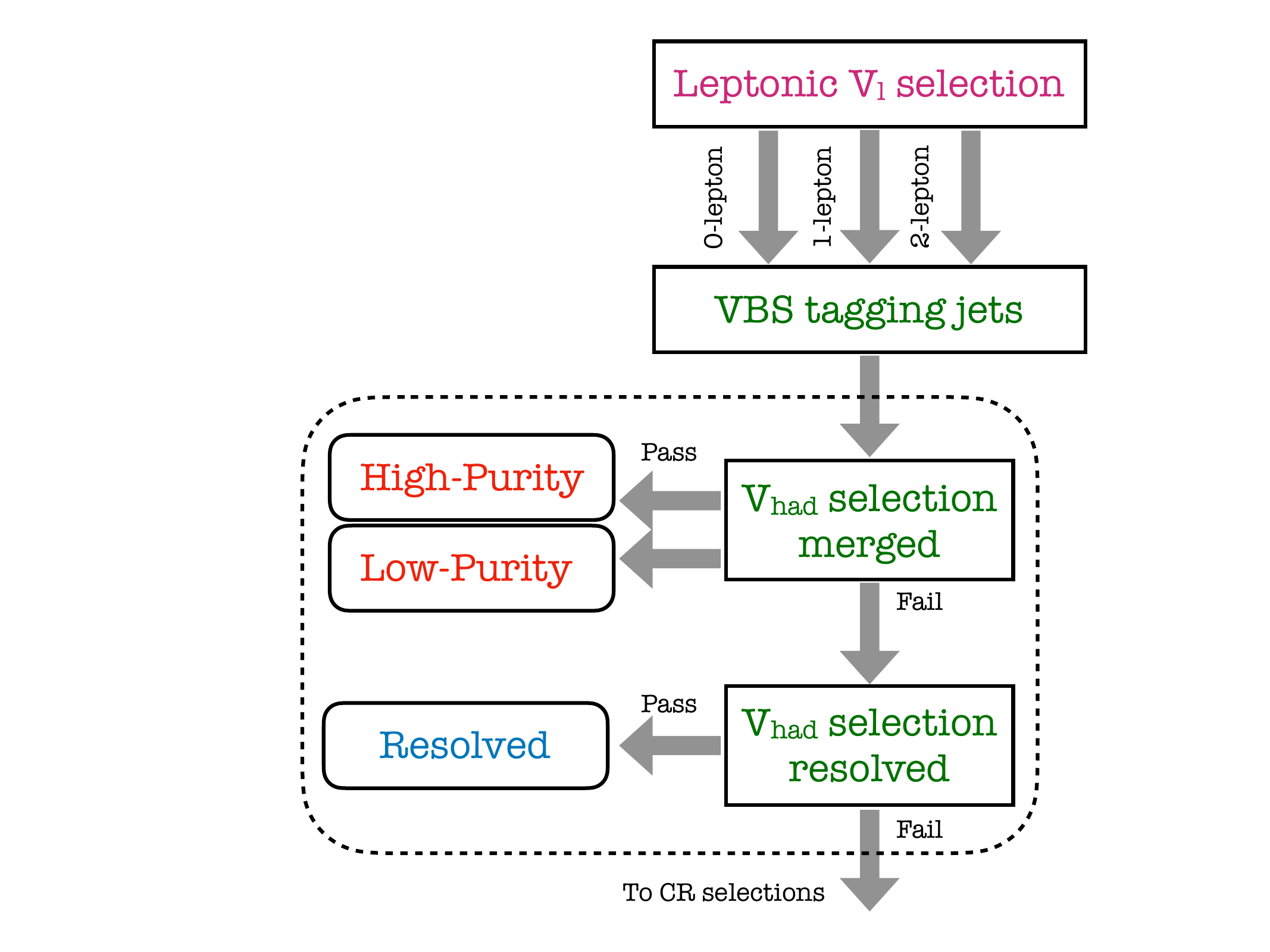}
\caption{\label{fig:eventcatflow} Schematic of the analysis categorisation flow.
}
\end{figure}

\subsection{Background estimation}

The dominant backgrounds in the \olep channel are \Wjets and $t\bar t$ production; in the \tlep channel it is \Zjets production; while in the \zlep channel, they all contribute significantly. Single-top and QCD-induced diboson production are small backgrounds for all channels. The background contributions are estimated using a combination of MC and data-driven techniques; indeed, the shapes of kinematic variable distributions are taken from MC simulations in all cases and the normalisation of the main backgrounds are constrained using dedicated control regions (CRs).

A \Vjets CR is defined for each of the three SRs in all the three lepton channels; specifically, these CRs are named ZCR in the \tlep channel, WCR in the \olep channel and VCR in the \zlep channel since the first two channels are mainly populated by \Zjets and \Wjets processes, respectively, while the \zlep channel is populated by a mixture of both processes.
These CRs are defined using exactly the same requirements as in the SRs but reverting the mass requirement, $m_J$ or $m_{jj}$, of the \Vhad candidate.
The ZCRs are dominated by the $Z$+jets contribution, with a purity higher than $90$\% in all regions. They are therefore used to constrain its contribution in SRs through simultaneous fits, as discussed in Section~\ref{sec:results}. The WCRs are enriched in the \Wjets process, with purities of 47\% and 64\% in the merged and resolved categories of the \olep channel, respectively; the remaining events are primarily from \ttbar production.

The three \ttbar CRs, TCRs, are defined to contain events satisfying the SRs selection of the \olep channel and requiring a $b$-jet instead of vetoing it.
These CRs are dominated by \ttbar production, with a purity of 90\% and 75\% for the merged and resolved categories respectively, with remaining background contributions from single-top, $V$+jets or diboson production, for both the merged and the resolved event topologies.

The \mjjtag\ distributions
are not well modelled by the {\textsc{Sherpa}} simulated \Wjets (\Zjets) events, as reported in Ref.~\cite{STDM-2017-20}.
A similar effect in the {\textsc{Sherpa}} modelling was also reported for diboson events~\cite{STDM-2018-32}.
The modelling is studied in the WCRs (ZCRs);
an example of the effect in the merged and resolved ZCRs
is shown in Figure~\ref{fig:mjjreweighting}.

\begin{figure}[!t]
\centering
\subfigure[]{\includegraphics[width=0.45\textwidth]{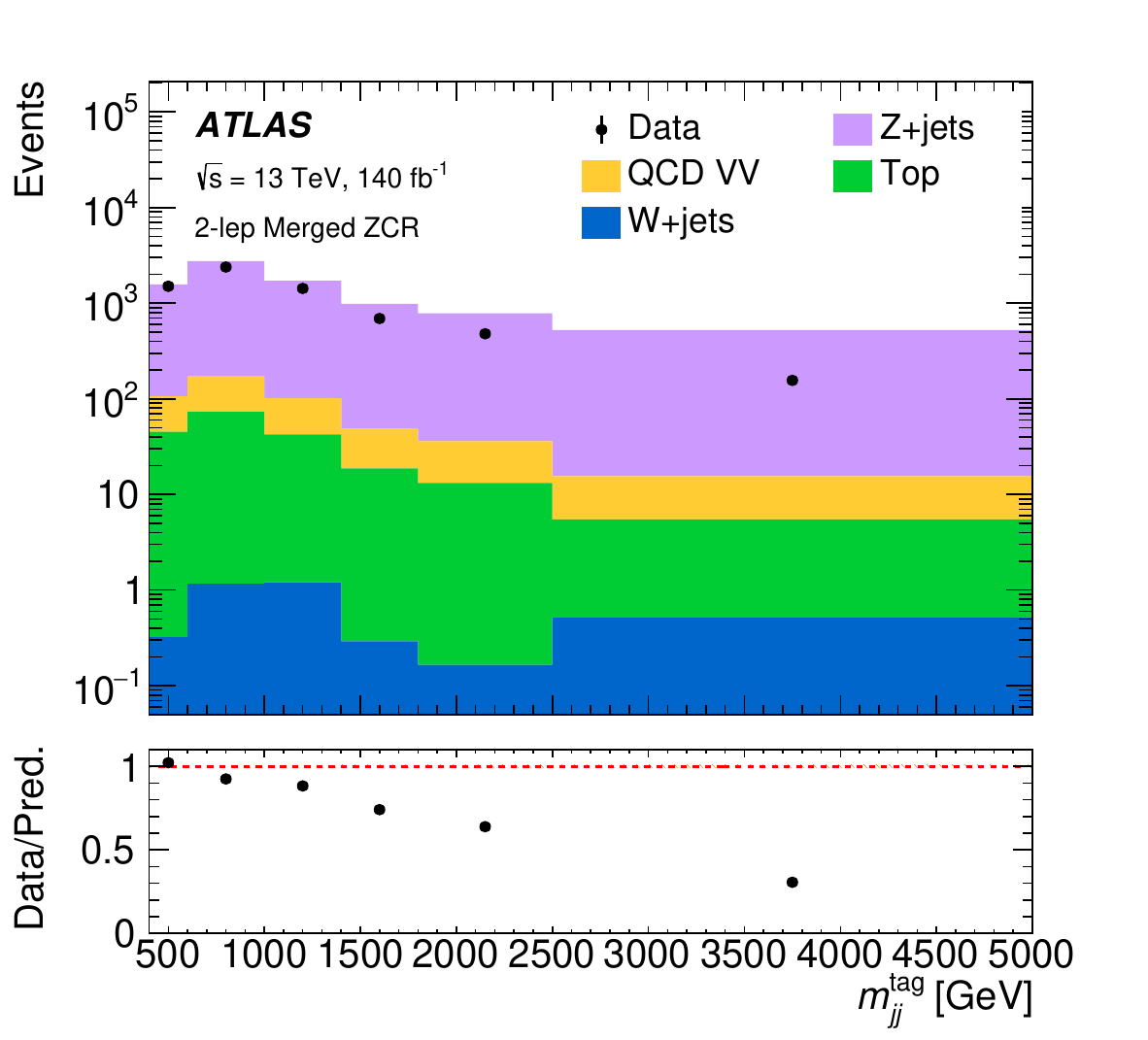}}
\subfigure[]{\includegraphics[width=0.45\textwidth]{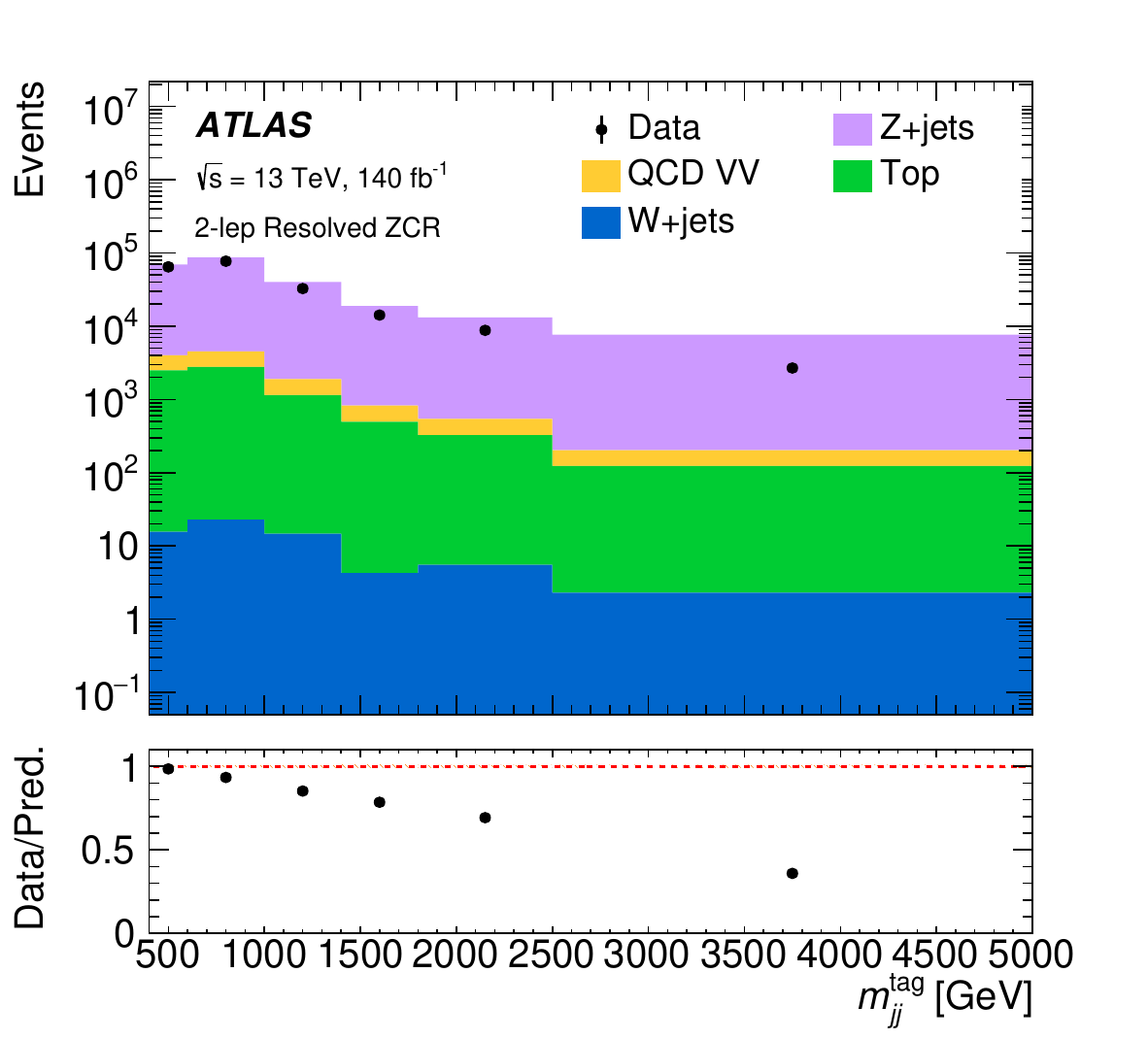}}
\caption{\label{fig:mjjreweighting} The measured (dots) distribution as a function of \mjjtag in the \tlep ZCR in the merged (a) and resolved (b) regions.
The last bin contains overflow events.
The different simulated processes are shown as stacked histograms.
The predictions are normalised to the generator cross section including the pile-up reweighting and other trigger and leptons related scale factors. The bottom panel shows the ratio of data to prediction, without systematic uncertainties.
}
\end{figure}

A data-driven procedure, similar to that described in \cite{STDM-2017-20},
is applied to the simulated \Wjets and \Zjets events to correct for this shape mis-modelling.
Reweighting factors are derived from WCRs and ZCRs as functions of \mjjtag,
and applied in all SRs and CRs (for the 0-, 1-, and 2-lepton regions)
to the MC simulation of \Wjets and \Zjets events.
The non-\Wjets(\Zjets) contributions, as predicted by MC, are subtracted from data
and the resulting distribution is normalized to unity;
the resulting shape is compared with the one from the MC simulation of \Wjets(\Zjets) processes
and the reweighting factors are estimated by performing a linear fit as a function of \mjjtag.
Reweighting factors are applied to events provided that the computed values are positive;
if a computed value is negative, a factor of zero is assigned instead.
Given the observed mismodeling, a linear model was found to be sufficient to describe and correct it;
more complex models, although capable of incorporating additional event information, did not lead to significant improvements.

The procedure is done separately for the merged and resolved analyses.
For the \Wjets process, the reweighting factor derived in the WCRs ranges from 1.04~(1.16) at $\mjjtag = 400$~\GeV\ to 0.58~(0.51) at $\mjjtag = 3000$~\GeV\ in the resolved~(merged) analysis.
For the \Zjets process, the reweighting factor derived in the ZCRs ranges from 1.10~(1.20) at $\mjjtag = 400$~\GeV\ in the resolved~(merged) analysis to 0.56~(0.55) at $\mjjtag = 3000$~\GeV.
A demonstration of the procedure for the \Zjets is shown in Figure \ref{fig:mjjreweighting_fit}.

\begin{figure}[!t]
\centering
\subfigure[]{\includegraphics[width=0.45\textwidth]{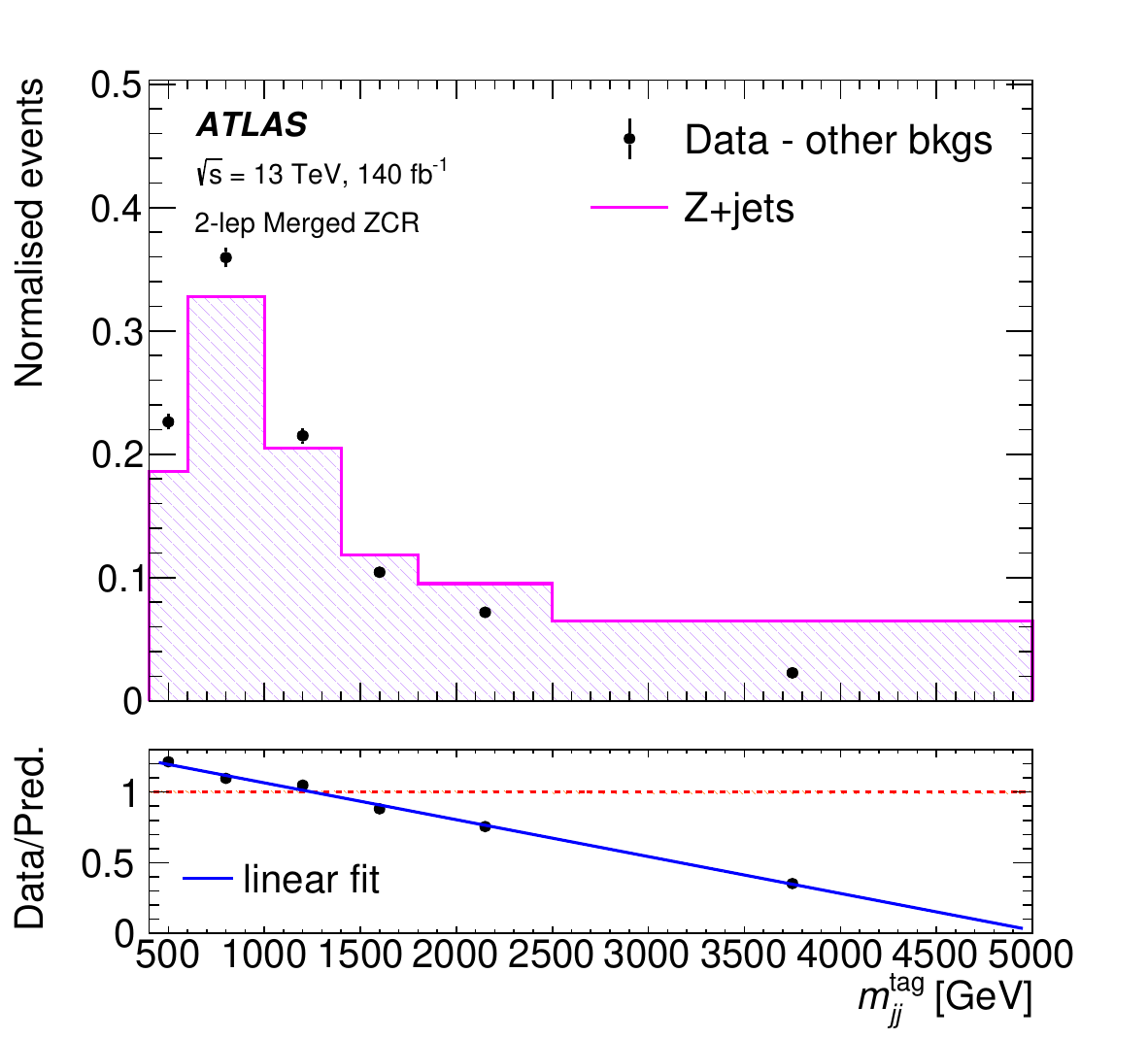}}
\subfigure[]{\includegraphics[width=0.45\textwidth]{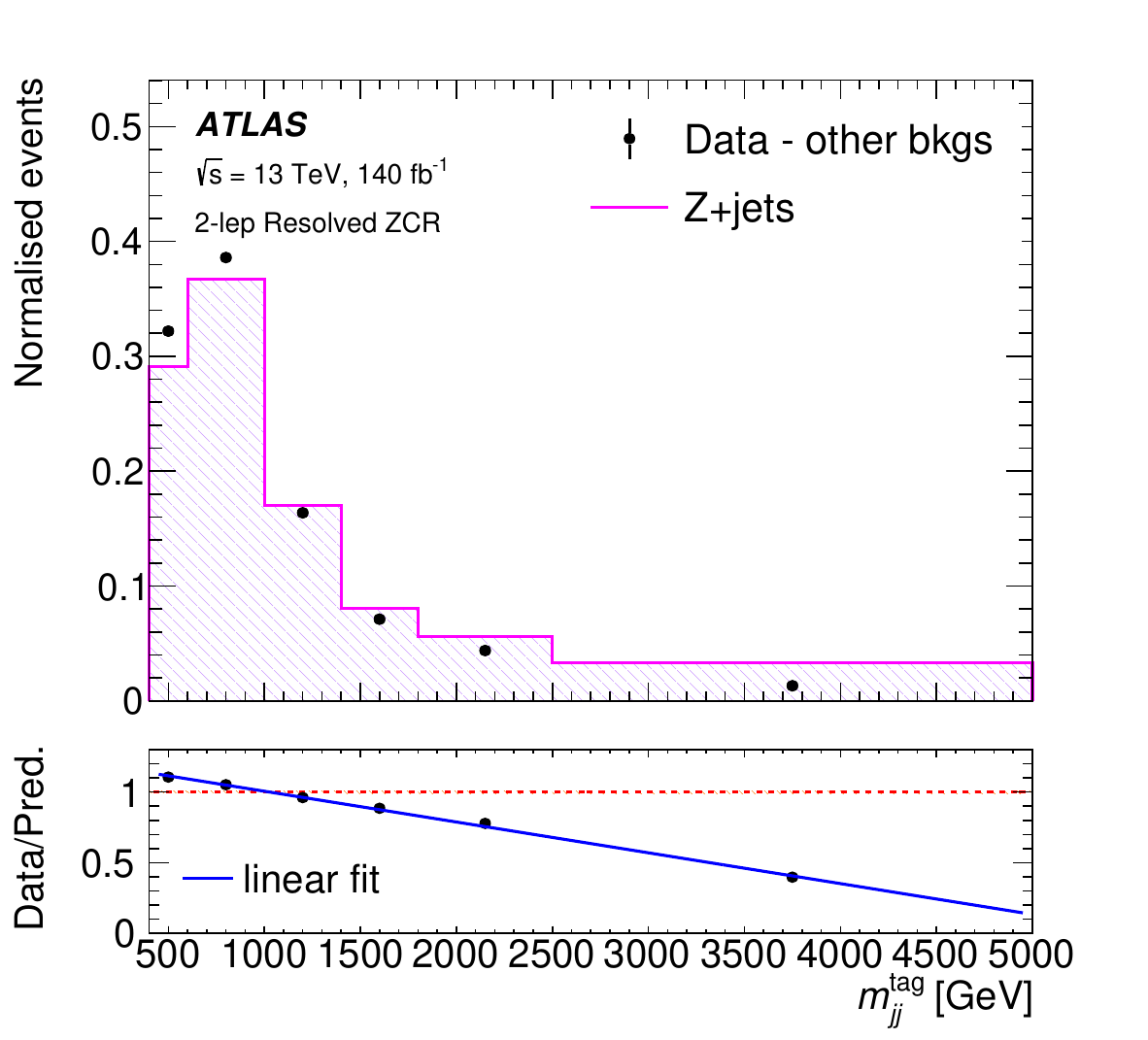}}
\caption{\label{fig:mjjreweighting_fit} The measured distribution after background subtraction and linear fit as a function of \mjjtag in the (a) merged and (b) resolved \tlep ZCR.
The simulated \Zjets process is shown as a histogram while the dots show the data after the subtraction of the other background processes; the distributions are normalised to unity. The bottom panel shows the ratio of the data to the prediction, together with the linear fit (solid line).
}
\end{figure}

Dedicated reweighting factors are needed for the MC simulation of \Wjets and \Zjets events in the \zlep\ channel because the phase space is different between the 0-lepton selection and the 1- and 2-lepton selections so that the reweightings described above are not applicable.
A similar linear fit procedure is applied to the sum of the \Wjets and \Zjets events in VCR as function of \mjjtag;
the reweighting factor derived in the VCR ranges from 1.03 (0.96) at $\mjjtag = 400$~\GeV\ to 0.74 (0.62)
at $\mjjtag = 3000$~\GeV\ in the resolved~(merged) analysis.

Good agreement between the prediction from MC simulation
and the data in the VCR is achieved after the reweighting procedure is applied.
The systematic uncertainty associated with this reweighting procedure
is described in Section \ref{sec:systematics}.
Conservative 100\% uncertainties in the fitted parameters are considered, which will be constrained by the fit as discussed in Section \ref{sec:systematics}.


\section{Machine learning approach for final discriminant}

%
%
The target signal in this analysis is the EWK \VVjj production and specifically its VBS contribution.
The VBS signal topology is characterized by two extra hard jets that are produced in the two opposite forward regions of the detector. Both of them usually have high $|\eta|$ and energy values and the dijet pair is characterized by a high invariant mass and a large angular separation. Furthermore, the invariant mass of the reconstructed diboson system is also higher for VBS signal events.
These represent some key discriminating variables.

To improve the separation from the background processes, a ML approach based on a recurrent neural network (RNN) architecture \cite{Sherstinsky_2020, HDBS-2018-10} is developed for this analysis.
The architecture of the network is based on a RNN in the resolved regime SRs.
In the merged regime, the architecture of the network is extended to also include a branch based on a deep neural network (DNN), with a few hidden flat layers, taking into account the large-R jet information.

\begin{figure}[h!]
\centering
\subfigure[]{\includegraphics[width=0.4\textwidth]{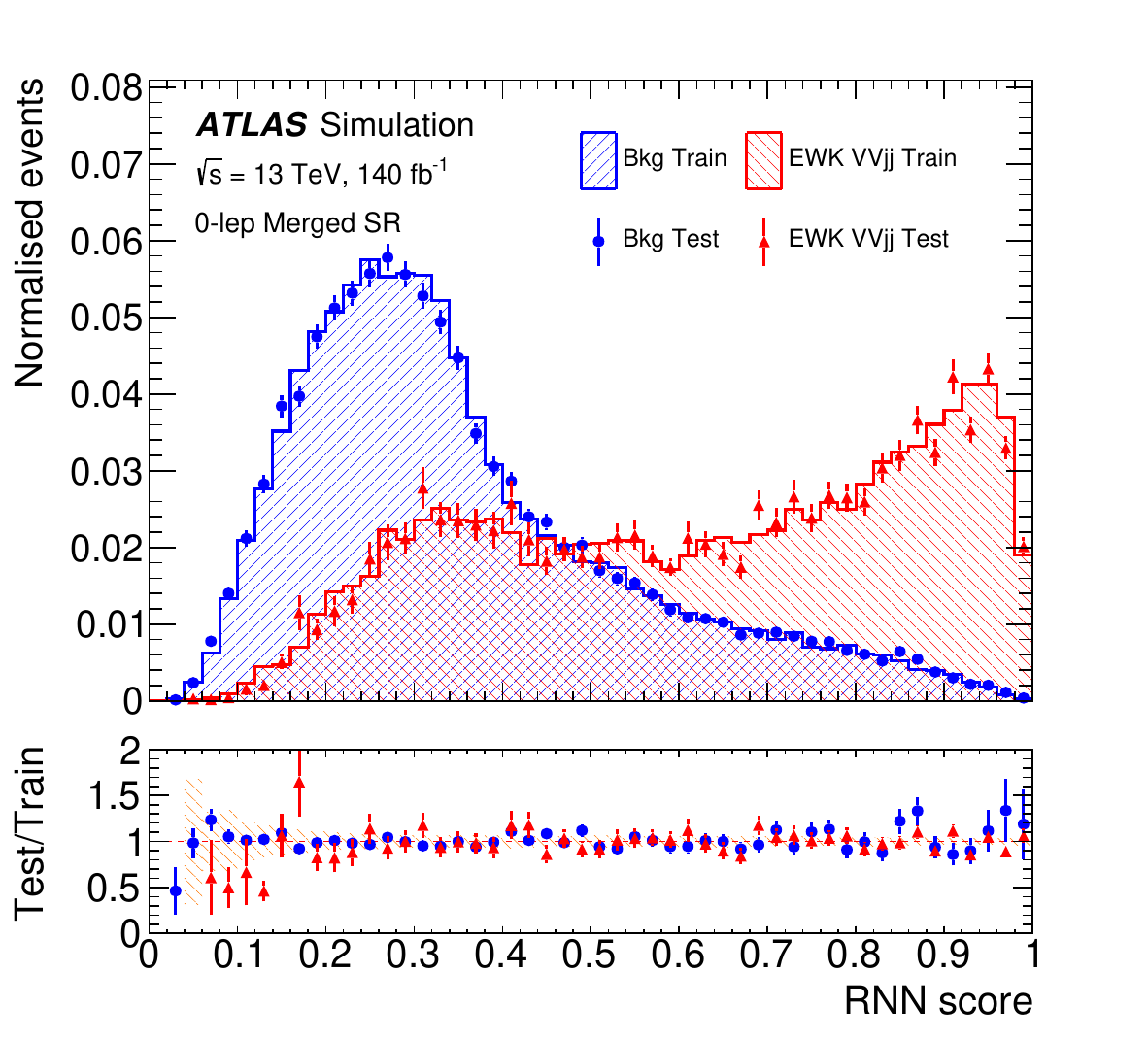}}
\subfigure[]{\includegraphics[width=0.4\textwidth]{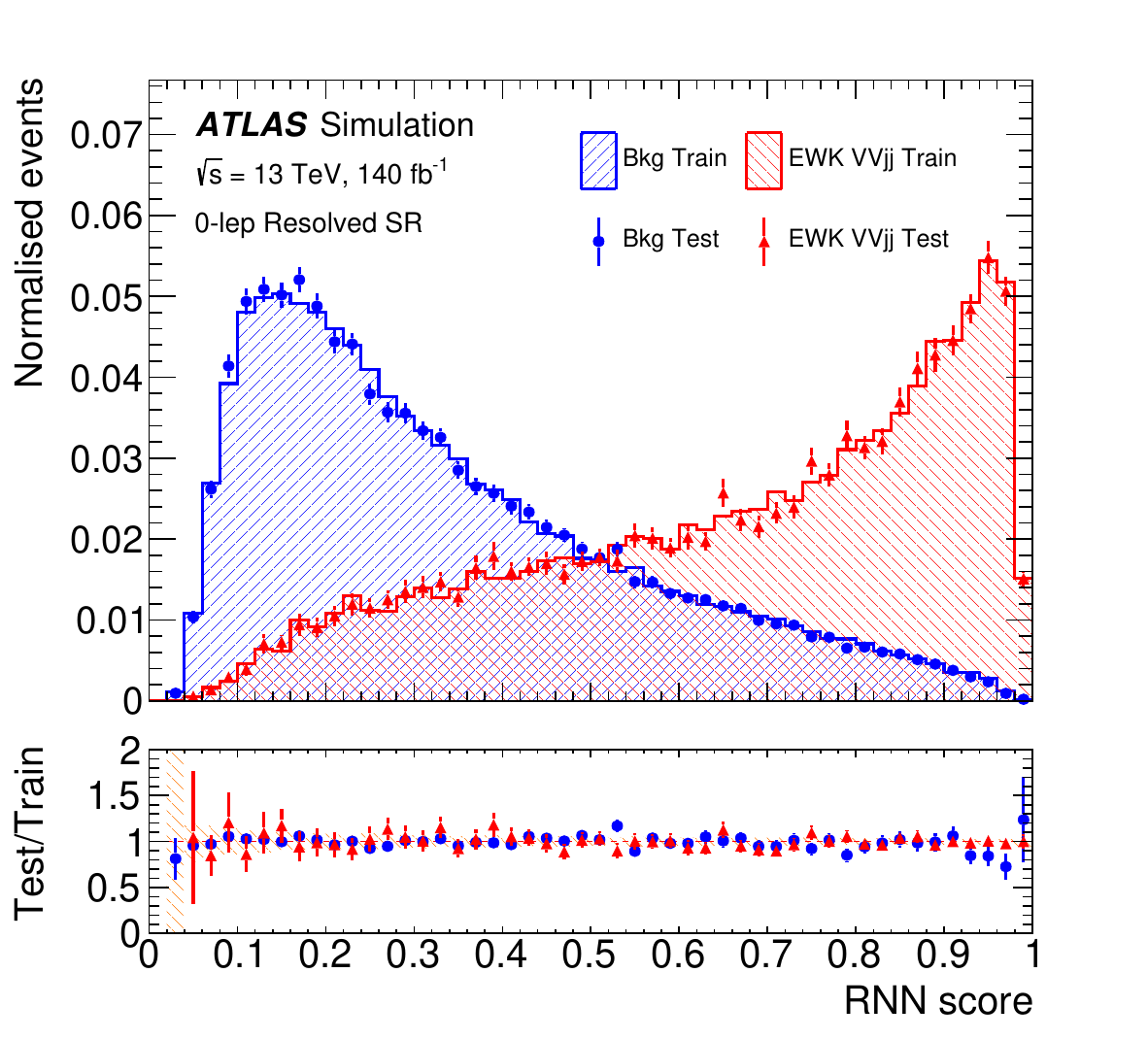}}    \\
\subfigure[]{\includegraphics[width=0.4\textwidth]{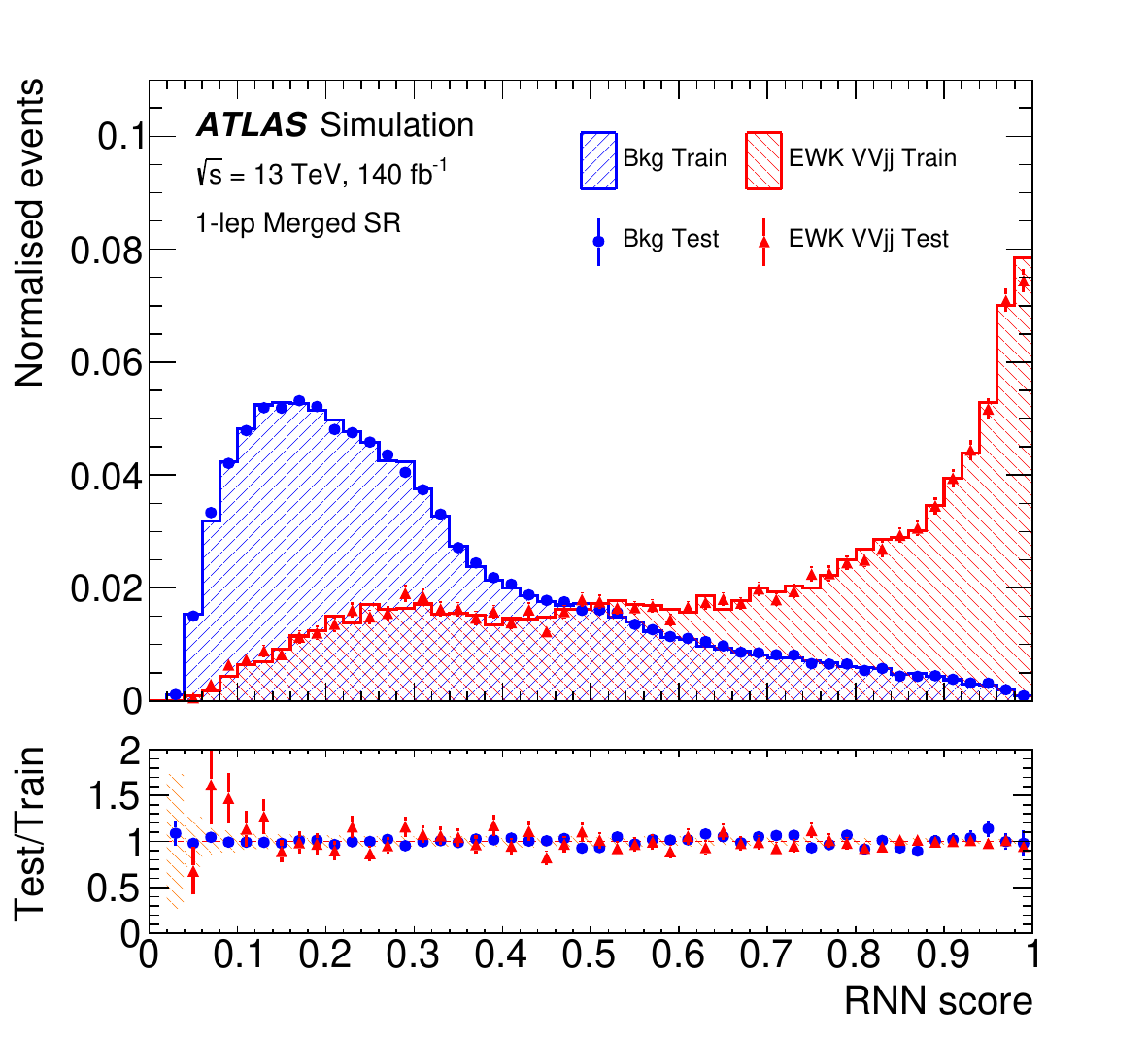}}
\subfigure[]{\includegraphics[width=0.4\textwidth]{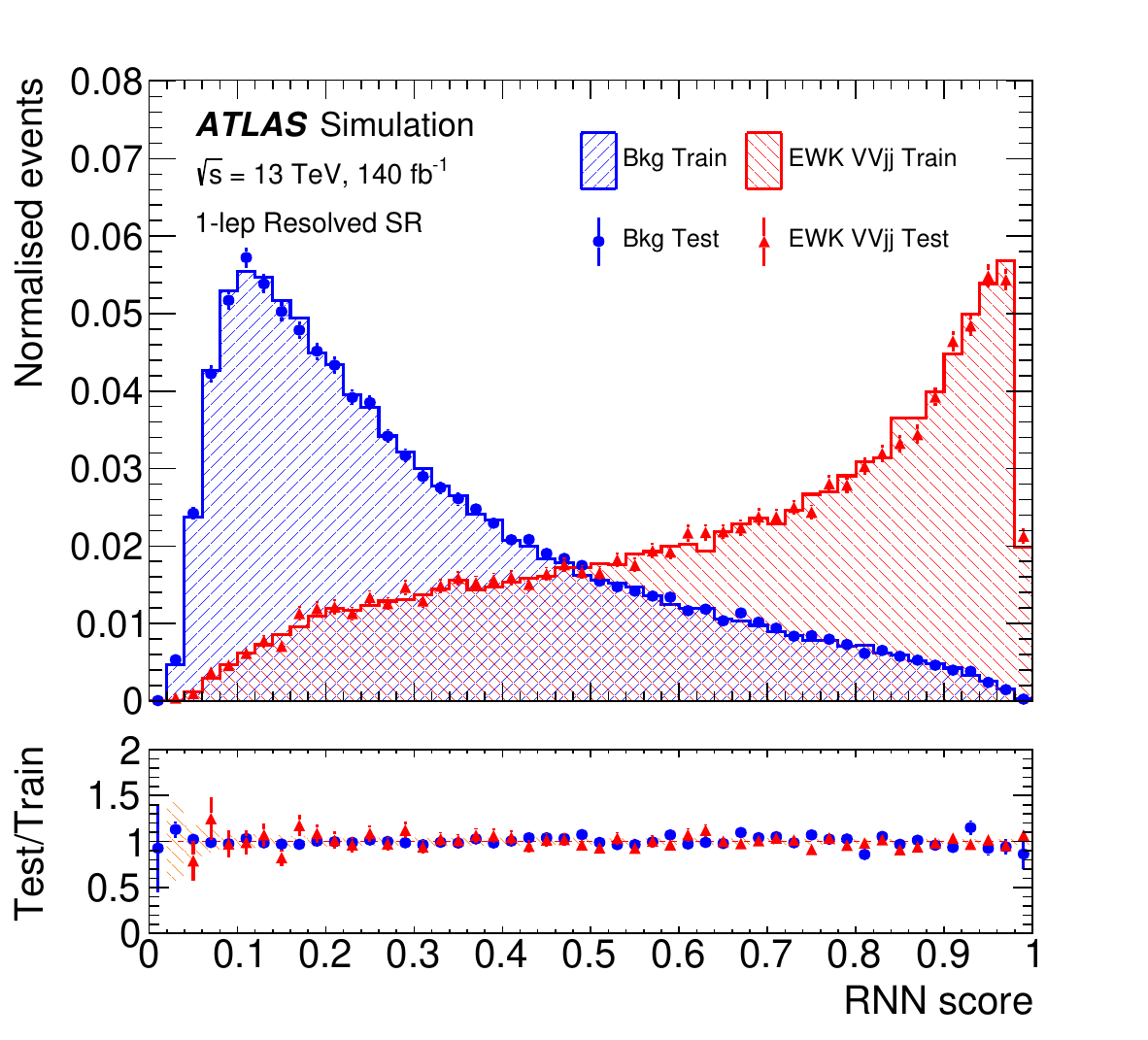}} \\
\subfigure[]{\includegraphics[width=0.4\textwidth]{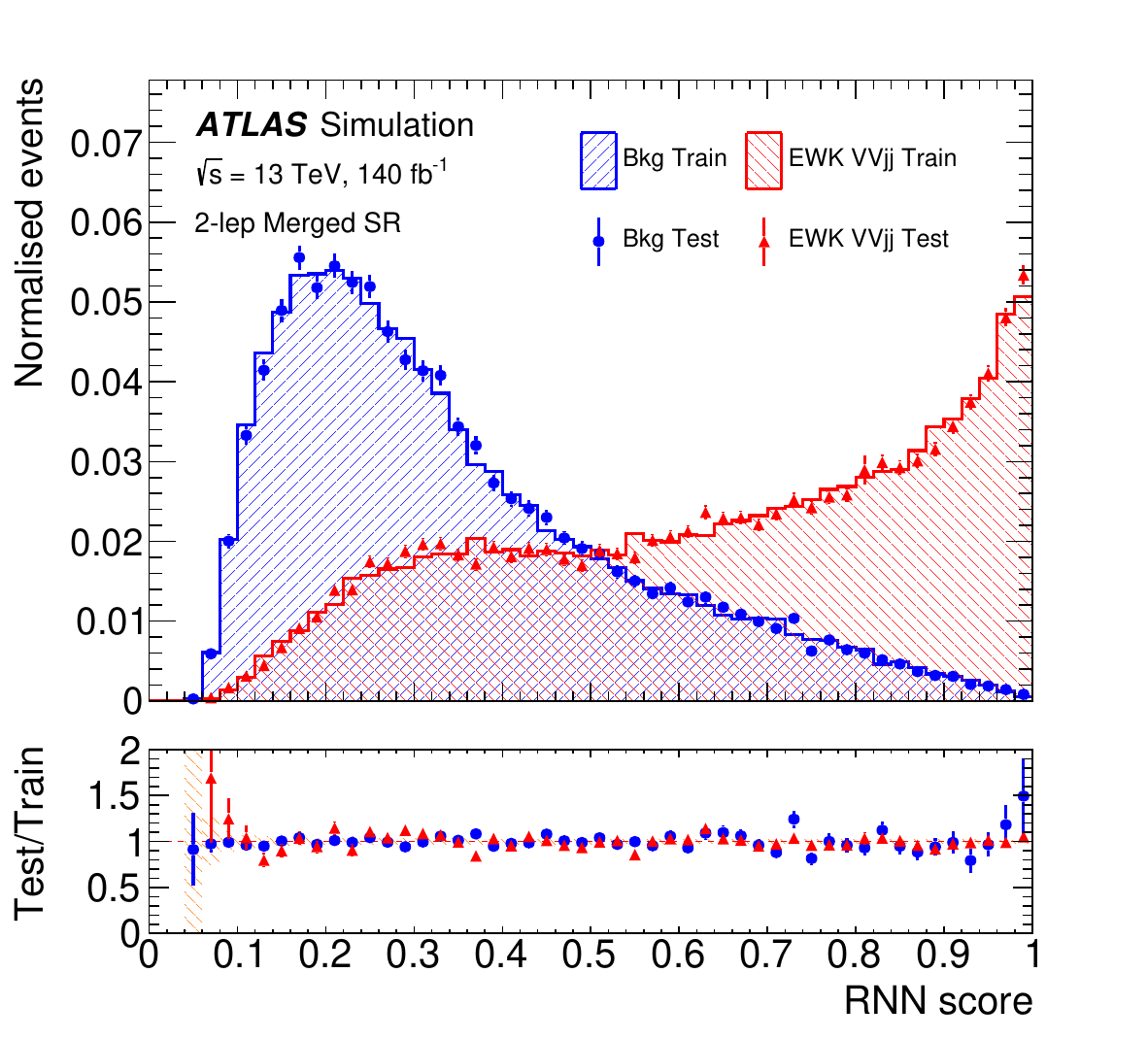}}
\subfigure[]{\includegraphics[width=0.4\textwidth]{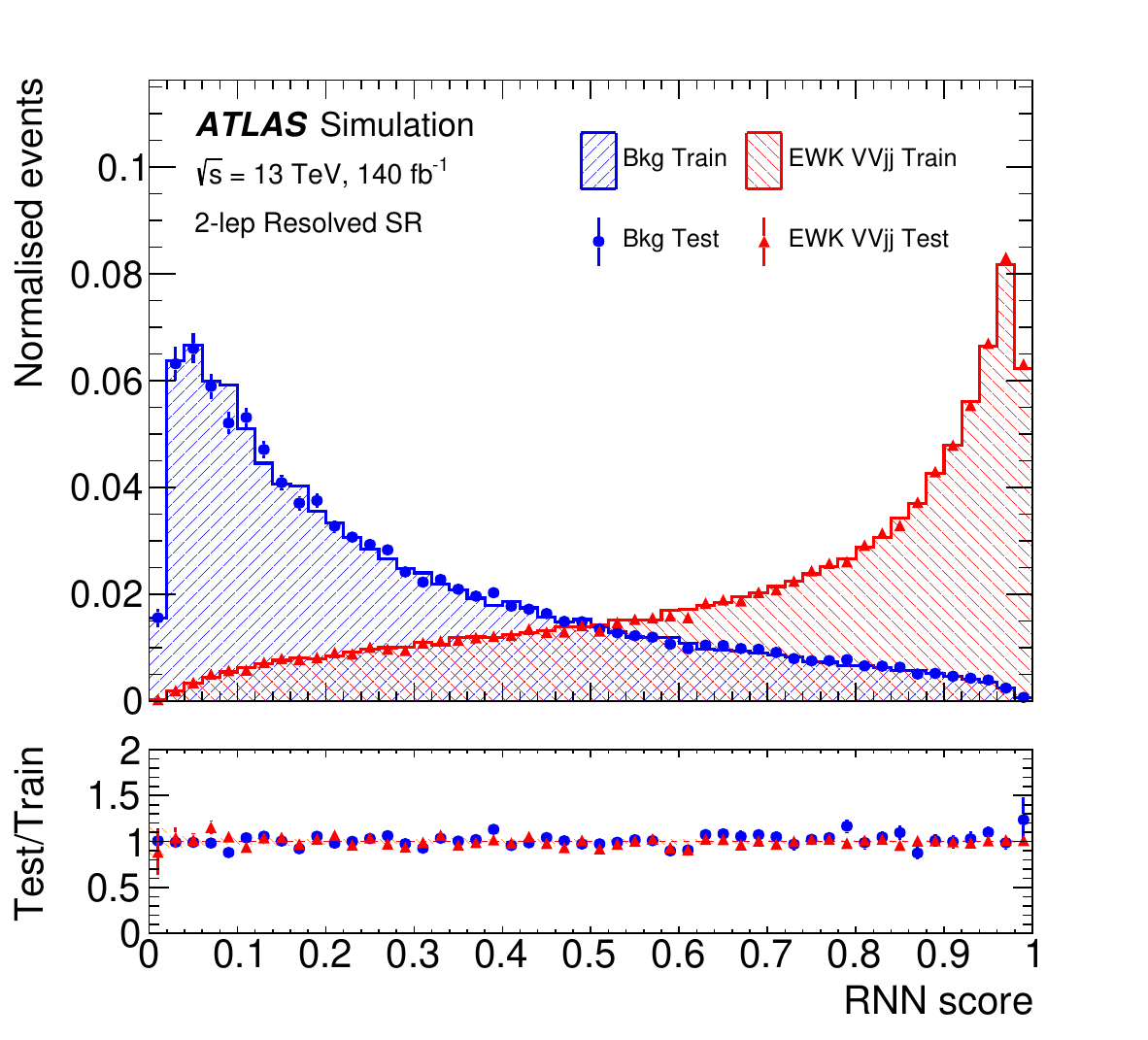}}
\caption{
RNN score distributions for signal and background samples showing the training (histograms) and test (markers) datasets for the (a, b) 0-, (c, d) 1- and (e, f) 2-lepton, (a, c, e) merged and (b, d, f) resolved SRs.
The ratio of the test and training RNN scores in the signal and background samples is shown in the bottom panel.
}
\label{fig:RNN_Score}
\end{figure}

The RNN architecture has already been exploited for a few applications by the ATLAS Collaboration; it was first used in the context of heavy-flavour identification \cite{ATL-PHYS-PUB-2017-003} and then deployed for a vector-boson-fusion classification task in diboson resonant searches \cite{HDBS-2018-10}.
Since it is an architecture designed to handle input features organized in sequences of variable length, it is a good architecture to deal with final states with a variable number of jets, such as the one probed here.

The RNN model is trained to separate the signal from the SM background processes and the output scores are used as final discriminants in the analysis.
To fully benefit from the deep learning regime, low-level input variables, i.e. the four-momentum of the jets and the track multiplicity (\pt, $\eta$, $\phi$, $E$ and
\ensuremath{n_\mathrm{Tracks}}\xspace)
are used.
Specifically, the RNN implementation available in Keras \cite{chollet2015keras} is used and the Tensorflow package \cite{abadi2016tensorflowlargescalemachinelearning} is used as backend for the mathematical computation.
A maximum of five jets is considered for each event; this upper value is optimized by checking the final sensitivity of the analysis;
including less jets would mean missing some hard-scatter jets
when some pile-up jets have a higher \pt than one of the forward VBS jets,
while including more jets would introduce further systematic uncertainties in the result.
A dedicated training is needed in each lepton channel due to the different background composition. Moreover, for the merged regime, one global training without the HP/LP separation is applied to maximize the statistics for the training.

To construct the RNN discriminant, the architecture is designed to leverage recurrent features through two hidden RNN layers and an output layer with a single node. The training process is set for a maximum of 200 epochs but terminates if there are 10 consecutive iterations without improvement. A dropout rate of 0.3 is applied, and the Adam \cite{kingma2017adammethodstochasticoptimization} optimizer is utilized. The two long short-term memory (LSTM) \cite{Sherstinsky_2020} layers use the hyperbolic tangent activation function, while the output layer employs a sigmoid activation function \cite{dubey2022activationfunctionsdeeplearning}.
The MC sample is divided into three parts: $56\%$ for training, $30\%$ for testing, and $14\%$ for validation.
During training, the signal and background events are weighted according to their physical cross sections and MC weights.
For the \Vjets process, the additional reweighting derived to correct the \mjj distribution, as discussed in Section~\ref{sec:eventselection}, is applied;
this ensures a more accurate description of the background events during training, and therefore leads to a more optimal model.
Finally, a global normalization factor is applied to ensure that the signal and background samples contain the same total number of events,
providing a balanced representation of both classes during training.

For the merged regime, the large-$R$ jet candidate information is added to further enhance the VBS signal-to-background separation completing the set of information related to the \Vhad decay;
specifically, the four-momentum components (\pt, $\eta$, $\phi$, $E$) of the large-$R$ jet candidate are used as additional input variables.
The architecture in this case is extended with an additional branch, based on a set of dense layers,
that processes the information of the large-$R$ jet
and combine it with the information of the small-$R$ jets sequences processed through the RNN layers.
The use of an extended architecture allows the exploitation of the correlation across the tagging jets and the \Vhad; for simplicity, this extended architecture will also be referenced to as RNN in the following.

Figure \ref{fig:RNN_Score} shows the RNN scores for the signal and background samples, including the training and test MC samples for the 0-, 1- and 2-lepton merged and resolved SRs. The score distributions show how well the RNN distinguishes between signal and background; the ratio of RNN scores for the test and training samples indicates a good agreement between them and demonstrates the absence of overtraining.
Moreover, a k-Fold cross-validation test is performed;
the input samples are split into training and test events sliding the test sample $k=5$ times, each time training and testing on independent samples.
The performance is found to be consistent within the statistical uncertainty of the sample, confirming that no overtraining is caused from the input sample.

\FloatBarrier


\section{Fiducial phase-space definition}

\label{sec:fiducial}

\begin{table}[t]
\begin{center}
\caption{Summary of the object and event selection performed at the truth level to define the fiducial regions.}
\begin{tabular}{l c c c}
\toprule
\multicolumn{4}{c}{Object selection} \\ \hline
Leptons      & \multicolumn{3}{c}{$\pt > 27$~GeV, $|\eta| < 2.5$ }\\
Small-$R$ jets & \multicolumn{3}{c}{$\pt > 20$~GeV  if $|\eta| < 2.5$ and $\pt > 30$~GeV if $2.5 < |\eta| < 4.5$}\\
Large-$R$ jets & \multicolumn{3}{c}{$\pt > 200$~GeV, $|\eta| < 2.0$ }\\
\bottomrule
\toprule
\multicolumn{4}{c}{Event selection} \\
\midrule
Selection &  0-lepton & 1-lepton & 2-lepton \\
\midrule
\multirow{2}{*}{$V \to \nu\nu/\ell\nu/\ell\ell$}
&  $\met > 200$~GeV & 1 lepton       & 2 leptons \\
&                   & $\met > 80$~GeV  &      \\
\midrule
\multirow{2}{*}{VBS topology tagging jets}
& \multicolumn{3}{c}{Largest \mjj pair with $\eta_{\mathrm{tag}\ j_1} \cdot \eta_{\mathrm{tag}\ j_2} < 0$, } \\
& \multicolumn{3}{c}{$\pt^{tag\ j_{1,2}} > 30$~\textrm{GeV}, $m_{jj}^{tag}$>400~\textrm{GeV}}  \\
\midrule
\multirow{2}{*}{$V \to J$}
& \multicolumn{3}{c}{Leading \pt large-$R$ jet} \\
& \multicolumn{3}{c}{$64 < m_{J} < 106$~\textrm{GeV} } \\
\midrule
\multirow{4}{*}{$V \to jj$}
& \multicolumn{3}{c}{Two leading \pt small-$R$ jets} \\
& \multicolumn{3}{c}{$\pt^{j_{1}}>$40~GeV, $\pt^{j_{2}}>$20~GeV } \\
& \multicolumn{3}{c}{$64 < m_{jj} < 106$~\textrm{GeV} } \\
& \multicolumn{3}{c}{$m_{jjj} > 220~\GeV$ } \\
\midrule
\multirow{1}{*}{Additional $b$-jet veto }
& No & Yes & No \\
\bottomrule

\end{tabular}
\label{tbl:vbs_fid_sel}
\end{center}
\end{table}

To minimize the extrapolation uncertainty, the cross section is measured with respect to a truth fiducial region which closely mimics the reconstructed region.
The fiducial phase space of the measurement is defined using stable final-state particles~\cite{ATL-PHYS-PUB-2015-013}.
The full fiducial selection is summarized in Table~\ref{tbl:vbs_fid_sel}.

Truth charged leptons (electrons or muons) are required not to originate from hadron decays and to have $\pt>27$~\gev\ and $|\eta|<2.5$.
The lepton four-vector is adjusted to include photon radiation within $\Delta R<0.1$.
The truth-level \met is calculated from the four-vector sum of all neutrinos in the event.
Small-$R$ and large-$R$ jets at the truth level are clustered using the same algorithms as the reconstruction level jets using all particles except for prompt charged leptons, neutrinos, and photons from $W/Z/H/\tau$ decays.
Truth large-$R$ jets are required to have $\pt>200$~\gev\ and $|\eta|<2.0$.
Similarly, truth small-$R$ jets are required to have $\pt>20$~\gev\ if $|\eta|<2.5$, and $\pt>30$~\gev\ for $2.5<|\eta|<4.5$.
Jets containing $b$-hadrons are labeled as truth $b$-jets.

The VBS topology is identified by the requirement of two truth small-$R$ jets with $\pt>30$~\gev,
$\eta_{\mathrm{tag}\ j_1} \cdot \eta_{\mathrm{tag}\ j_2} <0$
and dijet mass $m_{jj}^{tag}>400$~\gev.
If more than one dijet pair satisfies these conditions, the pair with highest mass is retained.
The \Vhad candidate jets are selected in the merged and resolved regions as either the leading large-$R$ jet or the two remaining highest-\pt small-$R$ jets, respectively.

The truth fiducial region is further subdivided into categories which align with the corresponding signal regions.
The 0-, 1-, 2-lepton regions are defined by requiring the corresponding number of truth charged leptons.
Additionally, the 0-lepton (1-lepton) region requires truth $\met>200\, (80)$~\gev.
The truth merged regions are required to contain a truth large-$R$ jet passing the above requirements
with $\Delta R(J,j)>1.4$ with respect to the VBS topology jets
and the jet mass passing the $64<m_{J}<106$~\gev requirement.
No $W/Z$-tagging is employed at the truth level so the merged regions are not separated into HP/LP regions as the reconstruction level is.
The acceptance effect due to different jet substructure requirements between the truth fiducial region, which is substructure-inclusive, and the reconstruction level $W/Z$-tagging is evaluated to be negligible.
Resolved events are selected from the remaining events failing the merged selection if the leading (subleading) \pt of the \Vhad small-$R$ jet has $\pt>40\,(20)$~\gev, and the mass of the pair satisfies $64<m_{jj}<106$~\gev.
The resolved regions also require the three-jet mass closest to the top-quark mass to have $m_{jjj}>220$~\gev, following the selection at the reconstruction level
to suppress EWK \ttbar production in the fiducial phase space.
Additionally, in the 1-lepton region, the \Vhad candidate jets are required not to be truth $b$-jets.

%


\section{Systematic uncertainties}

\label{sec:systematics}

Systematic uncertainties result in both changes to the overall predicted yields of processes, as well as shape effects on the kinematic observables.
Broadly, the systematic uncertainties can be categorized as either experimental uncertainties arising from the detector and reconstruction performance, and theory modelling uncertainties in the background and signal simulation.
The uncertainties quoted in this section are the relative uncertainties on the specified quantity, and not the propagated uncertainty in the analysis sensitivity.

\subsection{Experimental uncertainties}

The uncertainty in the combined 2015--2018 integrated luminosity is 0.83\%~\cite{DAPR-2021-01}, obtained using the LUCID-2 detector~\cite{LUCID2} for the primary luminosity measurements.
An additional uncertainty in the re-weighting procedure for the pile-up simulation is included to cover the uncertainty in the ratio of the predicted and measured inelastic cross sections~\cite{STDM-2015-05}.

The uncertainties associated to the lepton triggers are found to be negligible as the efficiency for the selected signal events is high.
The modelling of the electron and muon reconstruction, identification, and isolation efficiencies is evaluated using tag-and-probe methods with $Z\to\ell\ell$ events in both data and simulation~\cite{EGAM-2018-01,MUON-2018-03,PERF-2015-10}.
Corrections of the order of 1\% are applied to the simulation to better model the performance seen in data.
Uncertainties in the electron (muon) energy (momentum) scale and resolution are also considered
\cite{EGAM-2021-02,MUON-2022-01}.
Uncertainties in the small-$R$ jet energy scale and resolution are determined using MC simulation and data-driven techniques~\cite{JETM-2018-05}.
For central jets, the total relative uncertainty in the jet energy scale varies in the range 1–5\% for $\pt > 20$~\GeV.
The uncertainty in the small-$R$ jet energy resolution ranges from 25\% for jets with a $\pt$ of 20~\GeV~to less than 5\% for jets with $\pt > 300$~\GeV.
Since the \VBS topology is enriched in quark-like jets, the \qg fraction is estimated in MC in the signal regions as functions of the jet \pt and $\eta$ and is propagated to the relevant components of the uncertainties to accommodate for potential different responses of quark and gluon initiated jets.
Uncertainties in the efficiencies for successfully tagging $b$-jets and for mis-tagging light-flavour jets are determined from
studies in \ttbar enriched samples~\cite{FTAG-2018-01, FTAG-2019-02}.
Uncertainties in the tracks associated to jets are measured in pile-up events using random triggers \cite{ATL-PHYS-PUB-2015-051} and the uncertainty in their multiplicity is evaluated from dijet events \cite{STDM-2015-12}.

The uncertainties in the large-$R$ jet energy scale and resolution are determined using a similar methodology as for the small-$R$ jets and are found to be of the order of 1\% and 15\%, respectively, in the region of interest \cite{JETM-2018-02}. In addition, the uncertainty in the jet mass scale and resolution is also measured and found to be 1-5\% and 10-20\%, respectively, for jets near the $W/Z$ mass \cite{JETM-2018-02,ATLAS-CONF-2020-022}.
A dedicated calibration for the boson tagger is included. The strategy foresees to calibrate the efficiency working points for $W/Z$-initiated signal jets and for quark/gluon-initiated background jets that are used in the simulation of the various processes considered.
Calibrations and uncertainties in MC are evaluated in data CRs \cite{ATL-PHYS-PUB-2020-017}. Signal-like jets are selected in a \ttbar enriched region for $\pt<600~\gev$, and extrapolated to higher \pt via alternative MC modelling samples, while background-like jets are evaluated from dijet and $\gamma+$jet events. The uncertainty in the signal efficiency is approximately 20\% for the 50\% efficiency working point, and the uncertainty in the background efficiency can vary from 3\% to 10\% as a function of \pt.

Similarly to the lepton triggers, the \met trigger uncertainties are evaluated to be negligible.
Uncertainties in the lepton and jet resolution are directly propagated to the \met uncertainty.
An additional uncertainty ~\cite{JETM-2020-03} in the soft radiation term, defined in Section \ref{sec:objectreconstruction}, is also considered.

\subsection{Modelling uncertainties}

Theoretical uncertainties for \VVjj signal processes include the choice of PDF and QCD scale, as well as the parton-shower modelling. They are accounted for as acceptance effects in bins of the final RNN distribution. The PDF uncertainties are estimated using the \NNPDF[3.0nlo] prescription \cite{Ball:2014uwa},
and the QCD scale uncertainties are estimated by varying the nominal choices of the renormalisation, $\mu_{\text{R}}$, and factorisation, $\mu_\text{F}$, scales independently by factors of two and one-half and considering their envelope.
The resulting signal modelling acceptance uncertainty due to QCD scale and PDF choices is at maximum 10\% and 5\%, respectively, across all signal regions.
An additional uncertainty in the parton-showering modelling of the signal process is also assessed by comparing the nominal MC generator prediction to an alternative sample showered with \Herwig[7] ~\cite{Bahr:2008pv, Bellm:2015jjp}. This uncertainty can be as large as 20\% in the high RNN region.

Due to the fact that the EWK and QCD \VVjj processes are generated separately, the interference term is not included. The impact of the missing term is estimated at the fiducial level by comparing an inclusive \VVjj calculation with the separate QCD and EWK calculations, using the same configuration as the nominal EWK generation.
The ensuing difference is used to evaluate the size of the interference term, which is included as an uncertainty in the nominal EWK \VVjj MC prediction, parameterized as a function of the \mjjtag variable defined at truth level. At reconstructed level, this results in a 5\% (20\%) uncertainty in the RNN distribution in the resolved (merged) SRs.
This uncertainty is found to be small compared to other theory uncertainties.

The normalizations of the dominant \Wjets, \Zjets, and \ttbar background processes are evaluated directly in the final simultaneous fit to the SRs and CRs, as detailed in Section ~\ref{subsec:fit}.
A modelling uncertainty on the RNN shape for the \Wjets and \Zjets processes is estimated from a comparison to an alternative MC generator choice, \MGNLO+\PYTHIA.
An additional modelling uncertainty on the \Vjets processes to account for the \mjjtag reweighting procedure discussed in Section~\ref{sec:eventselection} is also evaluated conservatively as the difference between the shapes before and after the reweighting.
This reweighting uncertainty is evaluated bin-by-bin in the final RNN observable, and is found to vary with respect to the nominal prediction from a few percent to up to 50\% in the 1- and 2-lepton channels, and up to 20\% in the 0-lepton channel.
Similarly, the shape uncertainty for the \ttbar process is obtained by a comparison to an alternative
\HERWIG 7.04 sample using the H7UE set of tuned parameters~\cite{Bellm:2015jjp} and the MMHT2014LO PDF set~\cite{Harland-Lang:2014zoa}.

The cross section for the \VVQCD and \singletop contributions are estimated from the best theoretical cross section available as detailed in Section ~\ref{subsec:mc}.
An uncertainty in the modelling of the \VVQCD process in this phase space
is estimated by comparing to an alternative MC generator choice, \POWHEGBOX+\PYTHIA.
A normalization uncertainty of 20\% in \singletop is assumed with no further shape variation~\cite{TOPQ-2015-16}.
Similarly to the signal process, the uncertainties
from PDF and QCD scale choices for all background processes are calculated separately, and found to be smaller than the other sources of background modelling uncertainty.


\section{Results}

\label{sec:results}

\subsection{Statistical analysis}
\label{subsec:fit}

The statistical analysis is based on the RooFit~\cite{RooFit} and RooStats~\cite{moneta2011roostatsproject} frameworks to define a binned likelihood function, and the profile-likelihood ratio~\cite{Cowan:2010js} is used as test statistic to classify the level of agreement between a model and the observed data.
The analysis performs a fit across all SRs and CRs detailed in Section~\ref{sec:eventselection} simultaneously. In the CRs, the \mjjtag observable is used over several bins to help constrain the uncertainties associated to the background modelling and the \mjjtag reweighting procedure. The observable in the SRs is the RNN distribution, binned using a dedicated procedure detailed in Refs.~\cite{HIGG-2013-23,HIGG-2016-29} to ensure a fine binning in the signal sensitive regions while maintaining a low statistical uncertainty in the expected background prediction.
Sources of uncertainty are treated as additional nuisance parameters in the likelihood function.
In particular, the \Wjets, \Zjets, and \ttbar normalizations in each of the merged and resolved regions are unconstrained floating  parameters, and allowed to converge to the values which provide optimal agreement with the data.
All experimental systematic uncertainties are correlated amongst all regions and constrained by additional Gaussian terms in the likelihood as parametrized by external measurements.
Modelling uncertainties are treated similarly except for the fact that
they are uncorrelated across the lepton channels;
the \mjjtag\ reweighting uncertainties are decorrelated
across the lepton channels
and between the merged and resolved regions.
Each component of the signal uncertainties is individually included in the fit and
it is correlated across all the analysis regions.

\subsection{Inclusive measurement}

\begin{figure}[t]
\centering
\includegraphics[width=.8\linewidth]{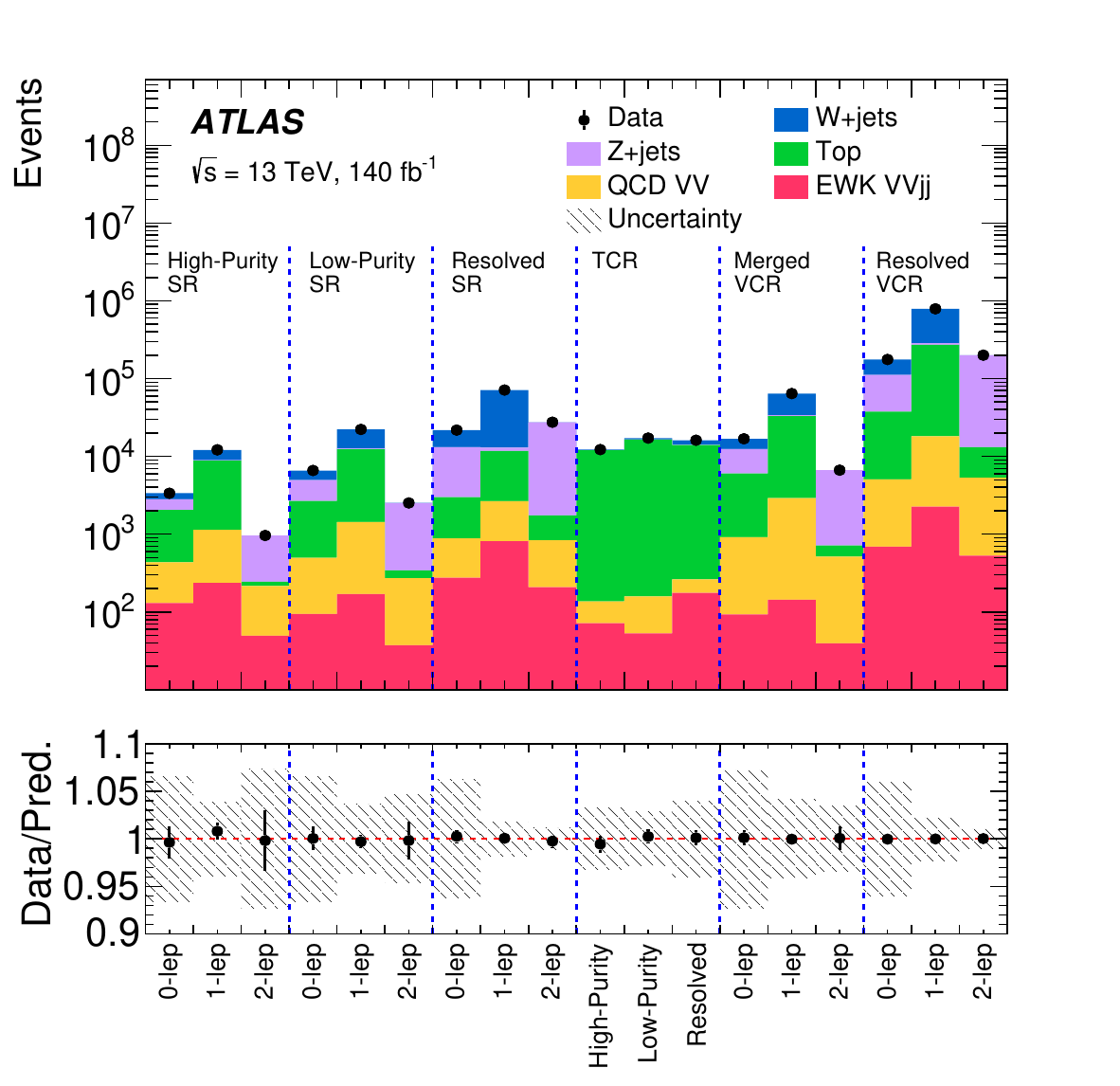}
\caption{
Data and stacked post-fit predicted yields in the various regions.
The systematic uncertainty in the prediction is shown as a hatched band.
}
\label{fig:YieldTable}
\end{figure}

To measure the EWK \VVjj cross section, the parameter of interest is the normalization of the \VVjj process cross section relative to to the SM prediction $\mu=\frac{\sigma_{\text{obs}}}{\sigma_{\text{exp}}}$, which is allowed to float to its optimal value in the fit.
The extracted yields in each region of each of the processes considered are shown in
Figure~\ref{fig:YieldTable}.
The post-fit RNN score distributions of the signal-plus-background fit are shown in Figure~\ref{fig:PostfitPlots}.
In general, good agreement of the model with data is found when including the EWK \VVjj contribution.
The description of the background model in the analysis regions is
checked on other variables not used directly in the fit model,
such as the invariant mass of the reconstructed diboson system \mVV
and \mjjtag, finding a similarly good description.

The measured EWK \VVjj signal strength \muobs is found to be
\begin{equation*}
\muobs= 1.28^{+0.23}_{-0.21} = 1.28 \pm0.09\mathrm{(stat)}^{+0.20}_{-0.19}\mathrm{(syst)}.
\end{equation*}
This is within $1.5\sigma$ agreement of the predicted value, with a $p$-value of 0.16. The measured signal strength is comparable to
those also measured by the ATLAS Collaboration in leptonic \VVjj analyses ~\cite{STDM-2018-35,STDM-2018-32,STDM-2022-06,STDM-2017-19}.
The background-only hypothesis is excluded with a significance of $7.4\sigma$, while the expected significance is $6.1\sigma$. %

The \muobs value for each of the individual lepton channels and for the resolved and merged regimes
is measured by performing a fit with, respectively, three and two parameters to extract;
the results are shown in Figure ~\ref{fig:signalStrengths}.
It can be seen that the 1 and 2-lepton channels are in good agreement
with the SM prediction while the 0-lepton channel
measures a higher EWK VV j j contribution,
with a $2.3\sigma$ compatibility with the inclusive measurement.
The measured signal strength for the \olep channel is $1.02^{+0.26}_{-0.24}$ and this result is compatible with the
value measured by the CMS Collaboration \cite{CMS-SMP-20-013}.

Table~\ref{tab:np_ranking_muobs} shows the contributions of all experimental and theory uncertainties to the fitted signal strength. The analysis is primarily dominated by systematic uncertainties, mainly arising from the theory and modelling factors, while experimental uncertainties related to the reconstructed jets also play a significant role.
The background modelling and \mjjtag\ uncertainties represent the leading sources constrained by the fit,
primarily due to the constraining power of the CRs.

An additional fit is also performed allowing both the EWK and QCD \VVjj contributions to be simultaneously extracted. The negative log-likelihood contours of this fit are shown in Figure ~\ref{fig:EWKvsQCDmu}. The measured value of the EWK \VVjj signal strength in this fit is found to be consistent
with the result obtained when the QCD \VVjj normalization is constrained to the generator prediction, indicating low correlation between these observables. A slight tension with respect to the MC generator predictions is found in these fits, which is similarly seen in other \VBS-enhanced EWK \VVjj topologies~\cite{STDM-2018-35, CMS-SMP-20-013}, and may be indicative of a poor generator modelling of the QCD $VV$ spectrum at high jet multiplicity.

\begin{figure}[t]
\centering
\subfigure[]{\includegraphics[width=.3\linewidth]{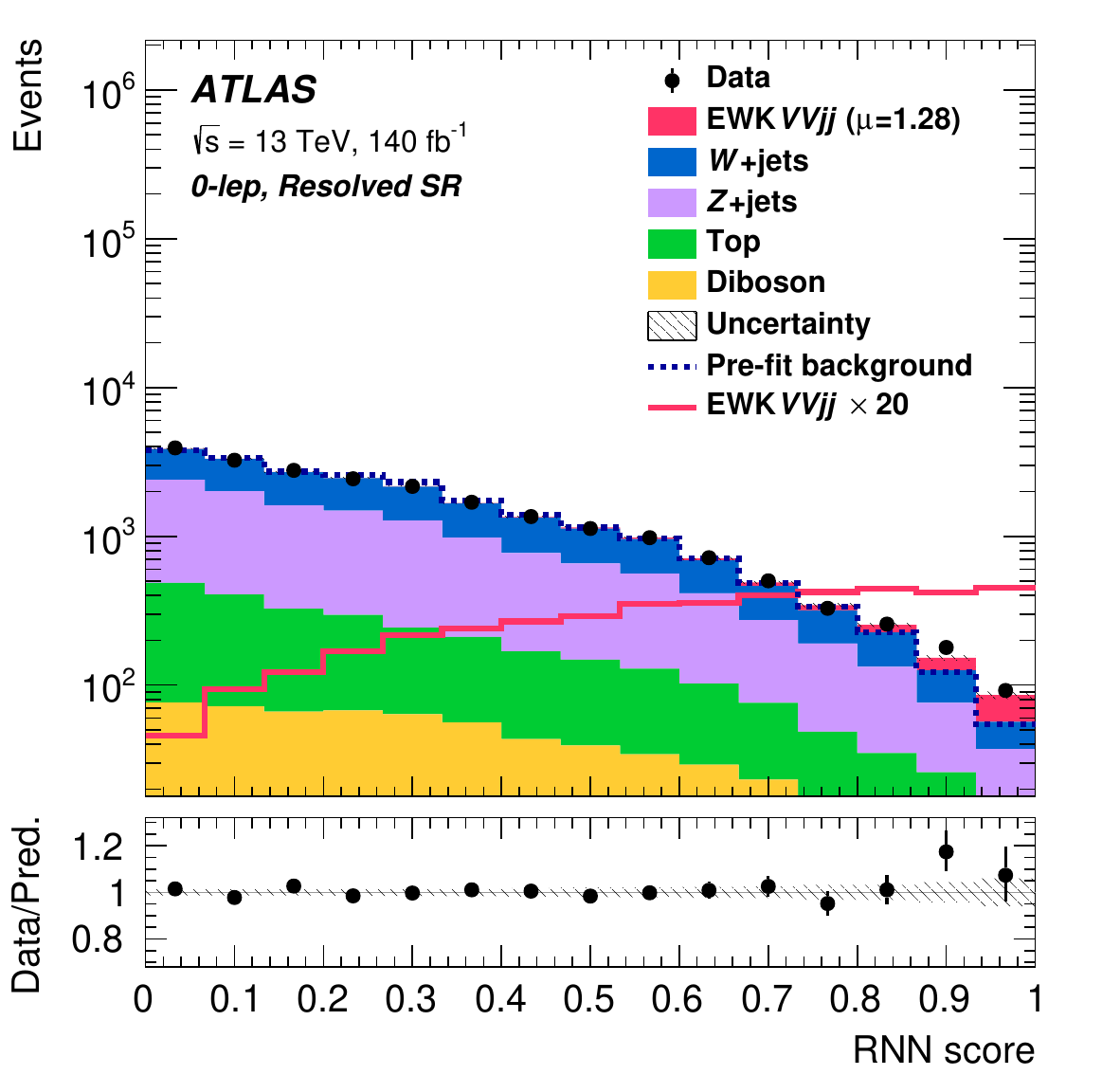}}
\subfigure[]{\includegraphics[width=.3\linewidth]{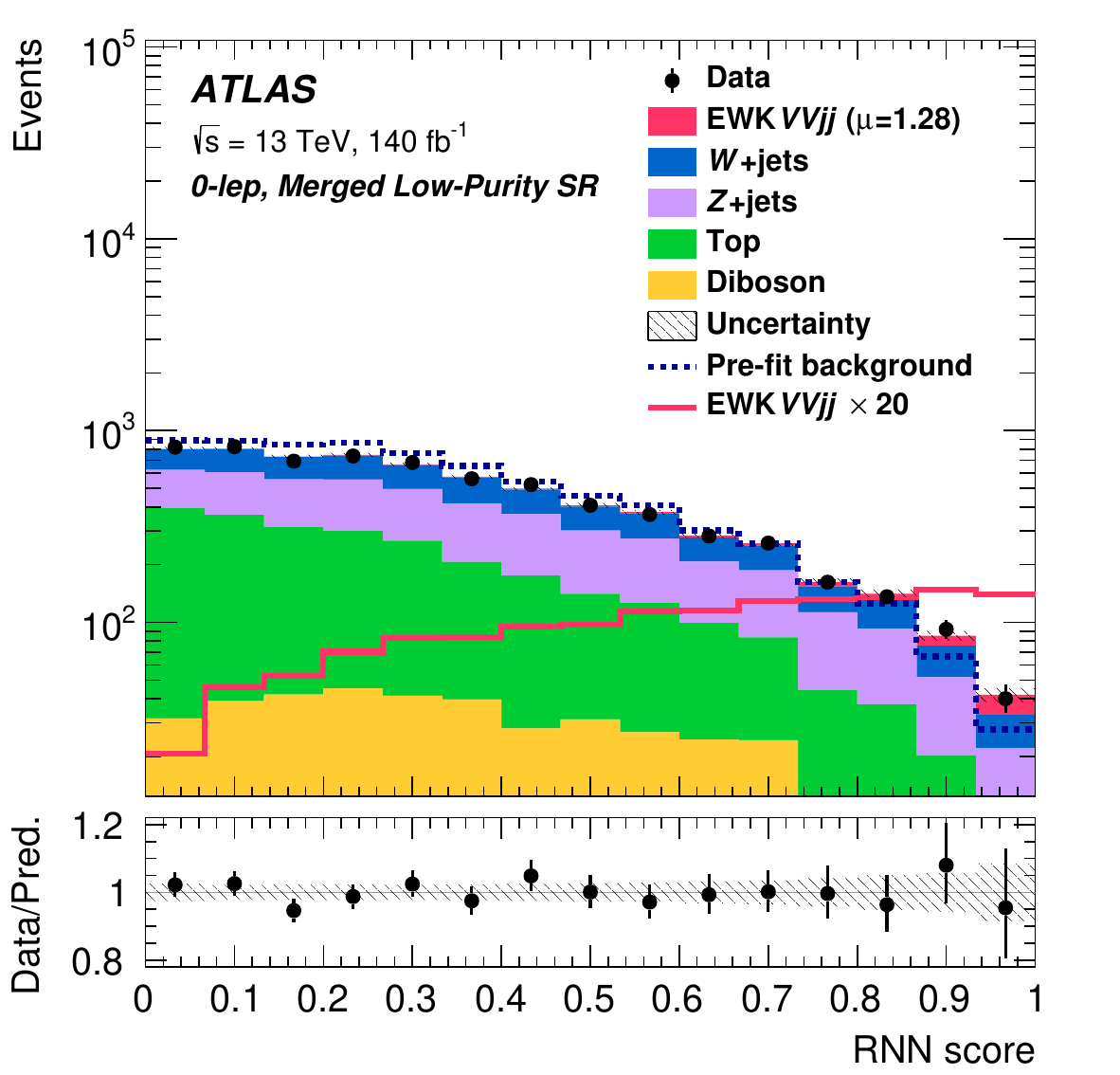}}
\subfigure[]{\includegraphics[width=.3\linewidth]{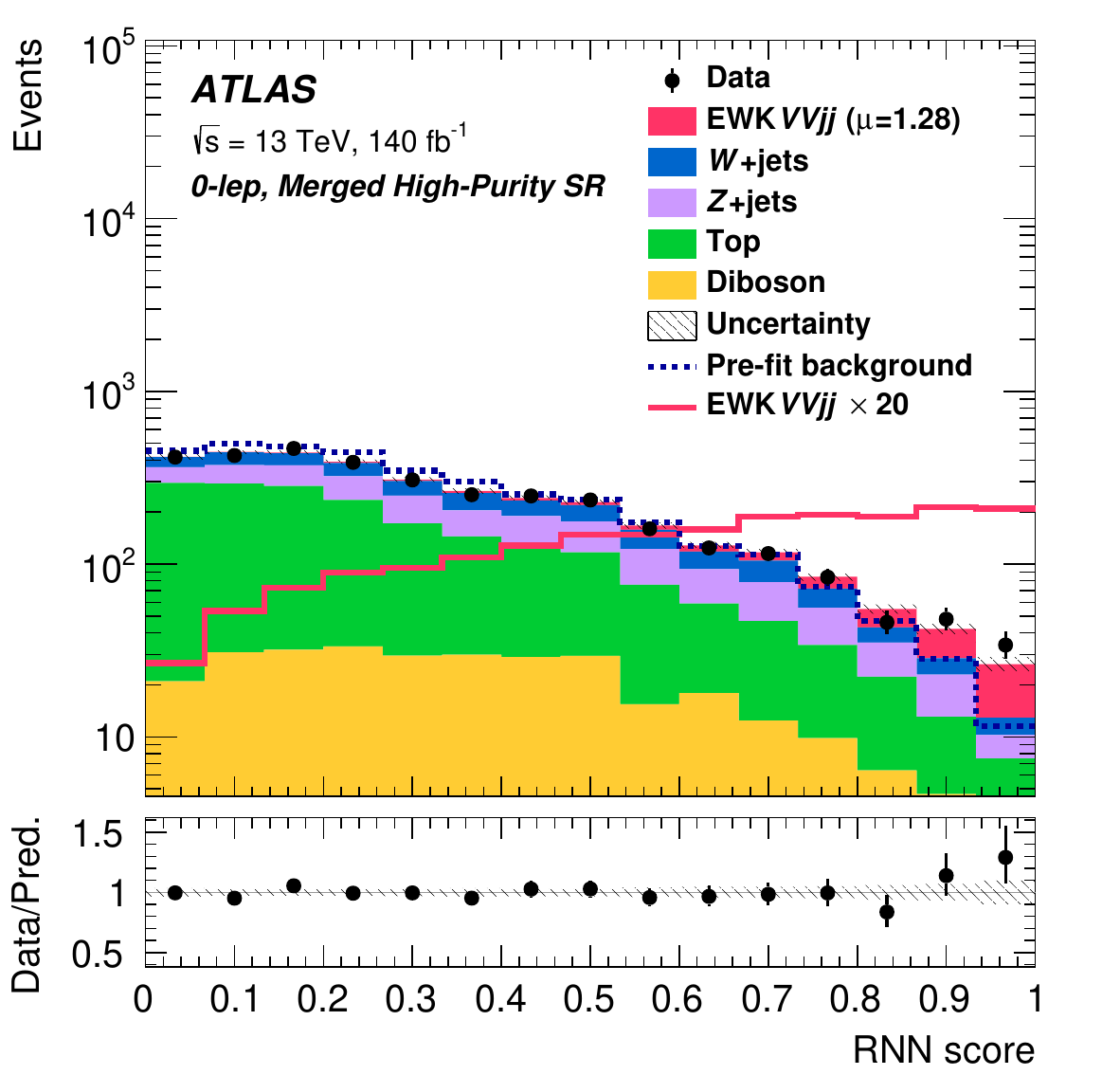}} \\
\subfigure[]{\includegraphics[width=.3\linewidth]{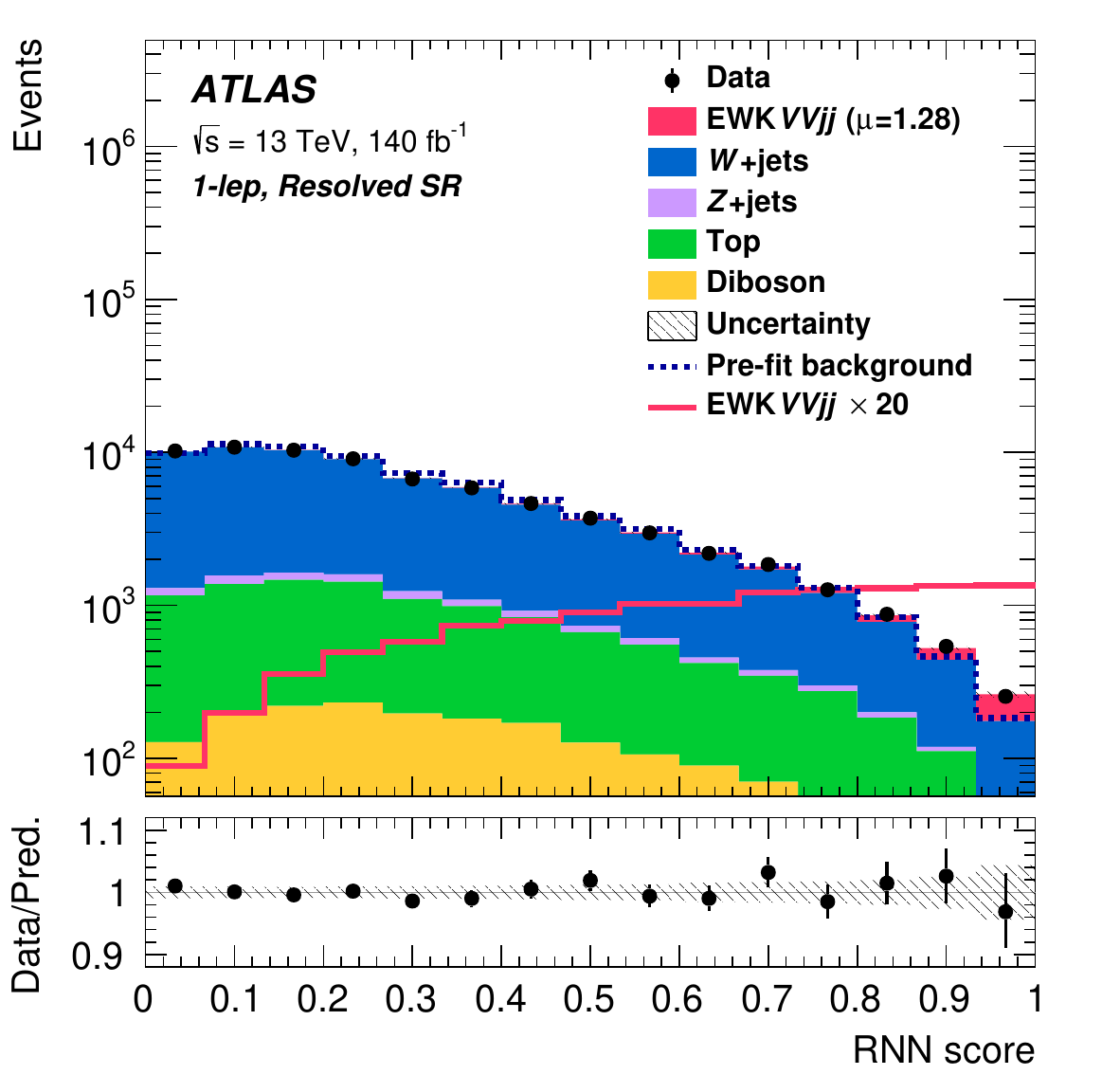}}
\subfigure[]{\includegraphics[width=.3\linewidth]{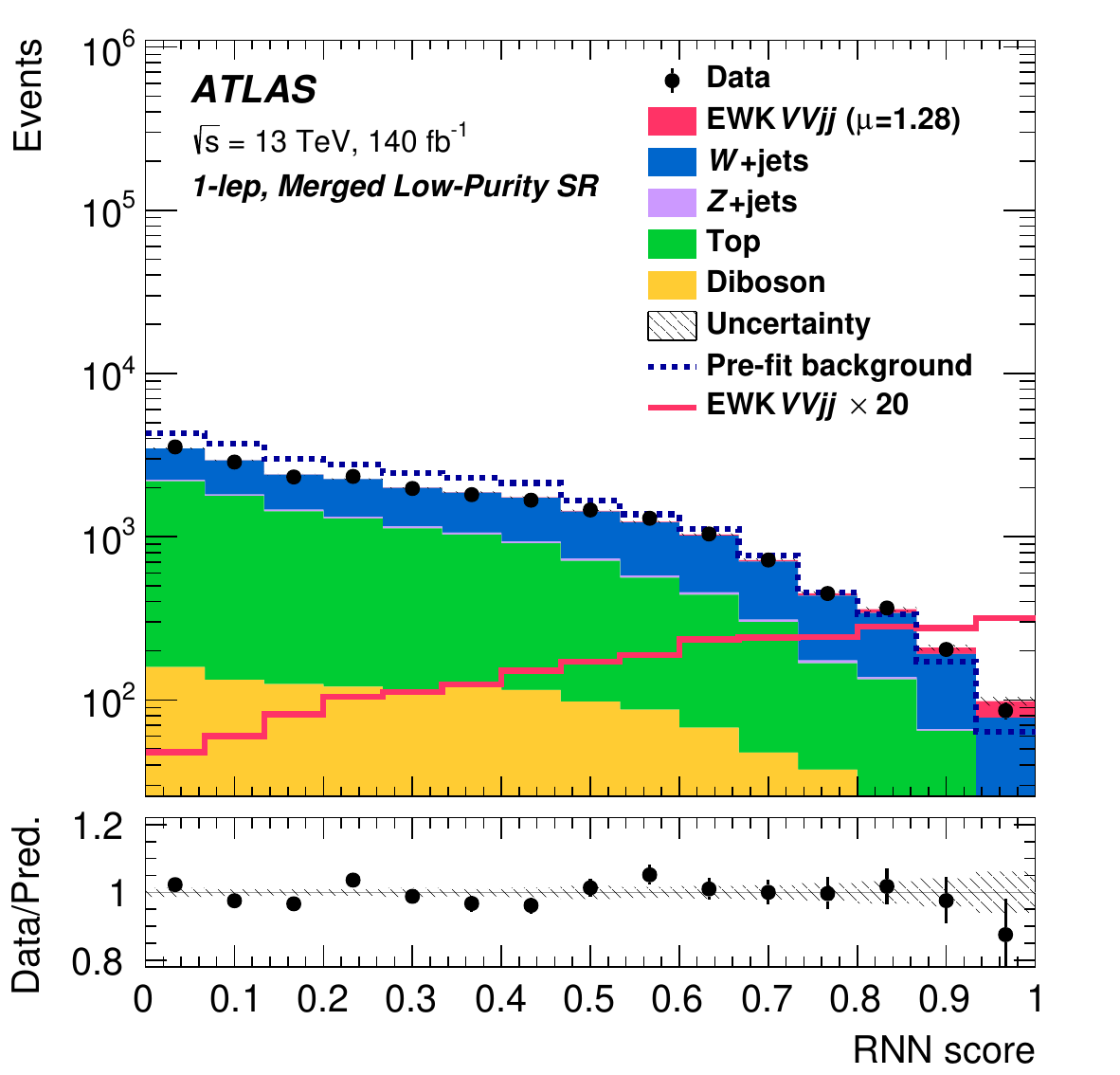}}
\subfigure[]{\includegraphics[width=.3\linewidth]{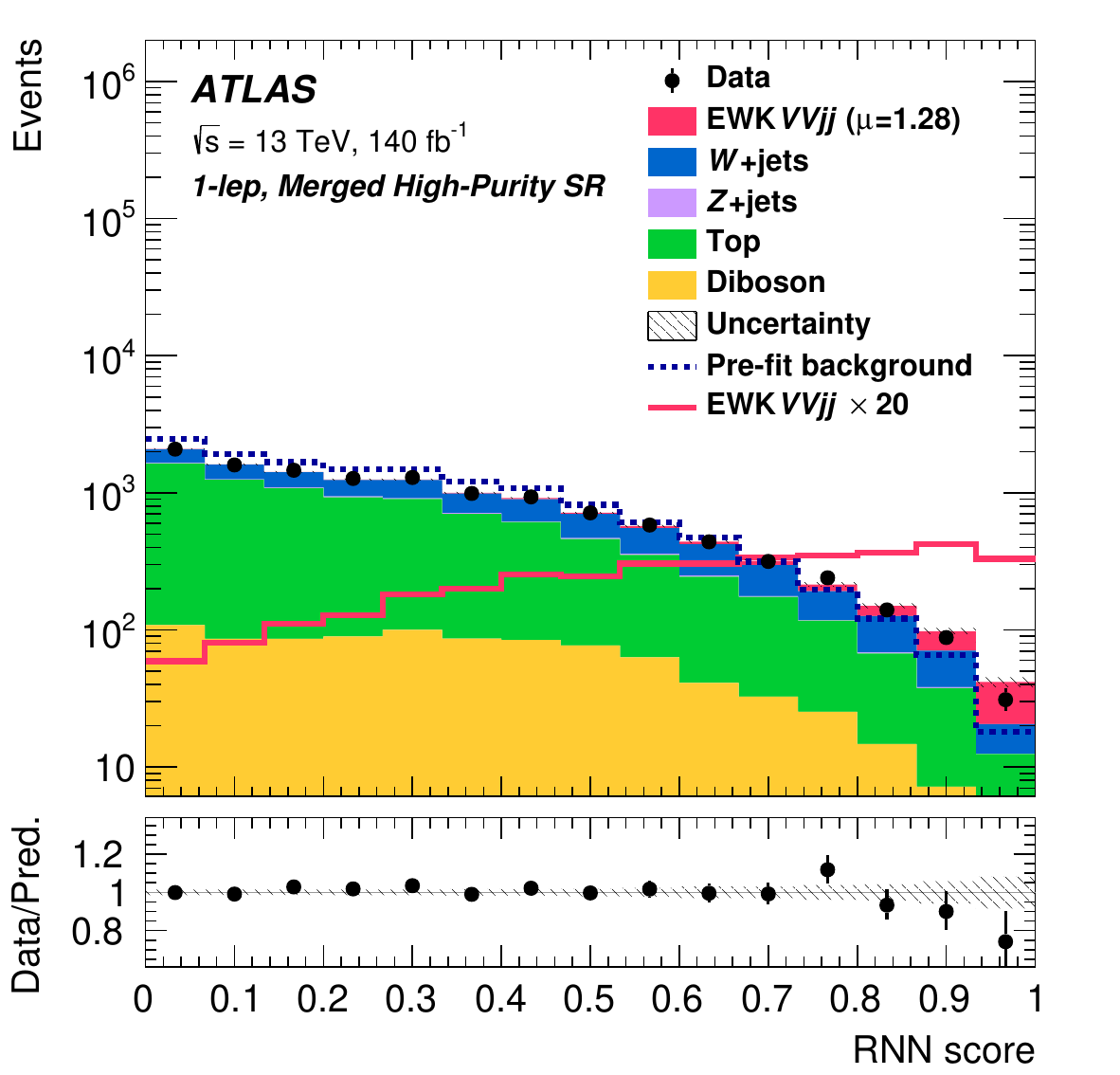}} \\
\subfigure[]{\includegraphics[width=.3\linewidth]{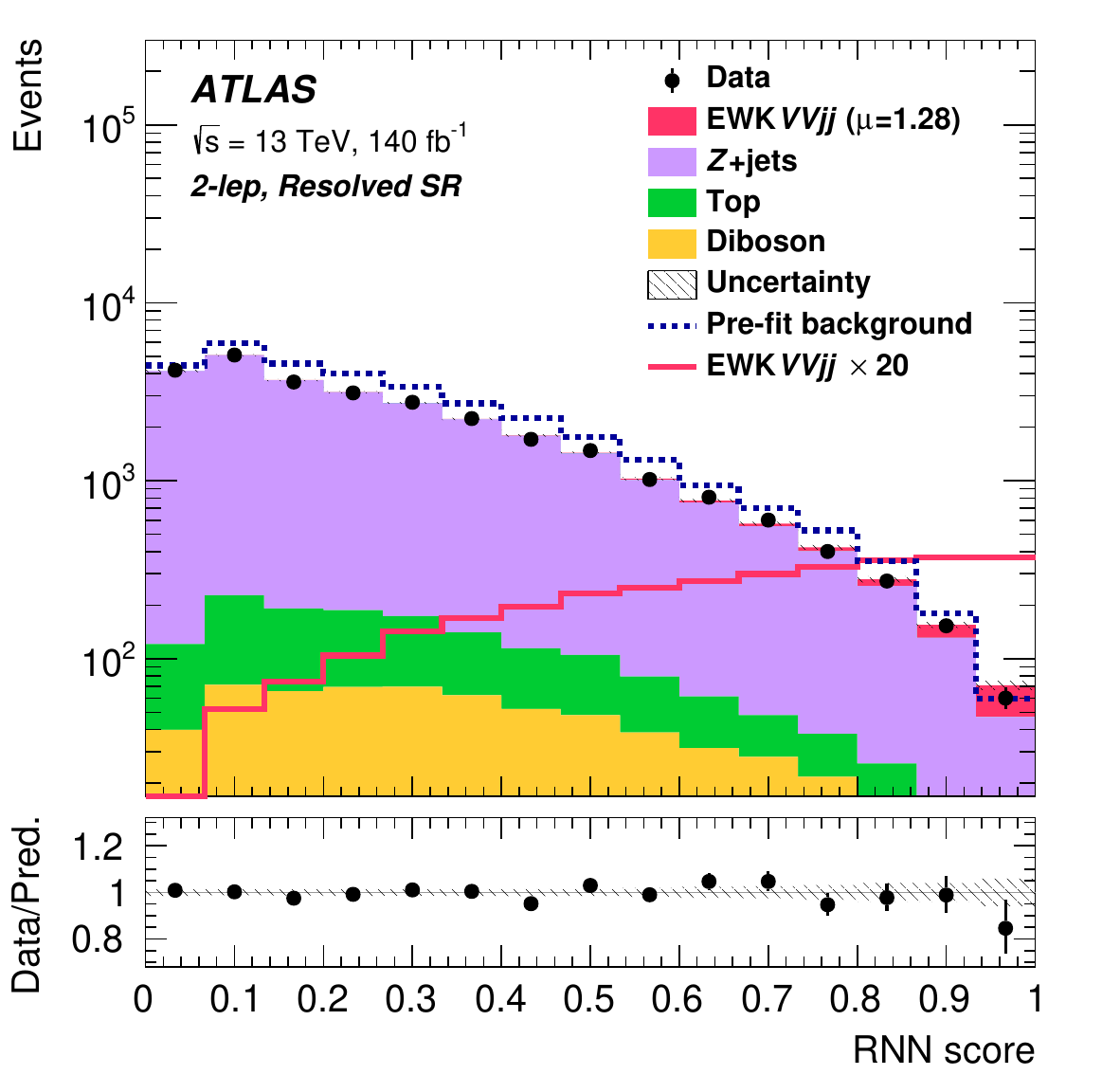}}
\subfigure[]{\includegraphics[width=.3\linewidth]{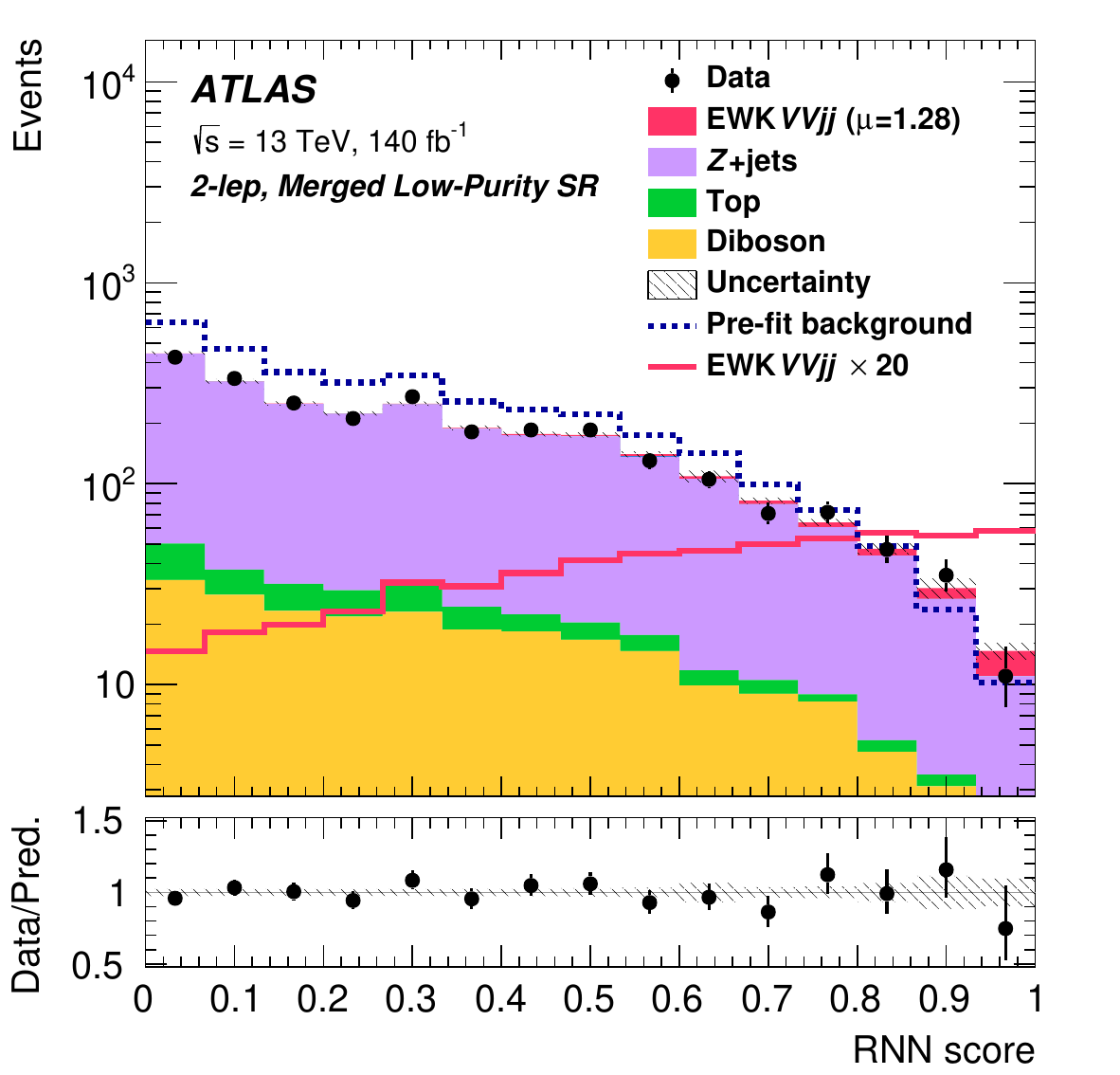}}
\subfigure[]{\includegraphics[width=.3\linewidth]{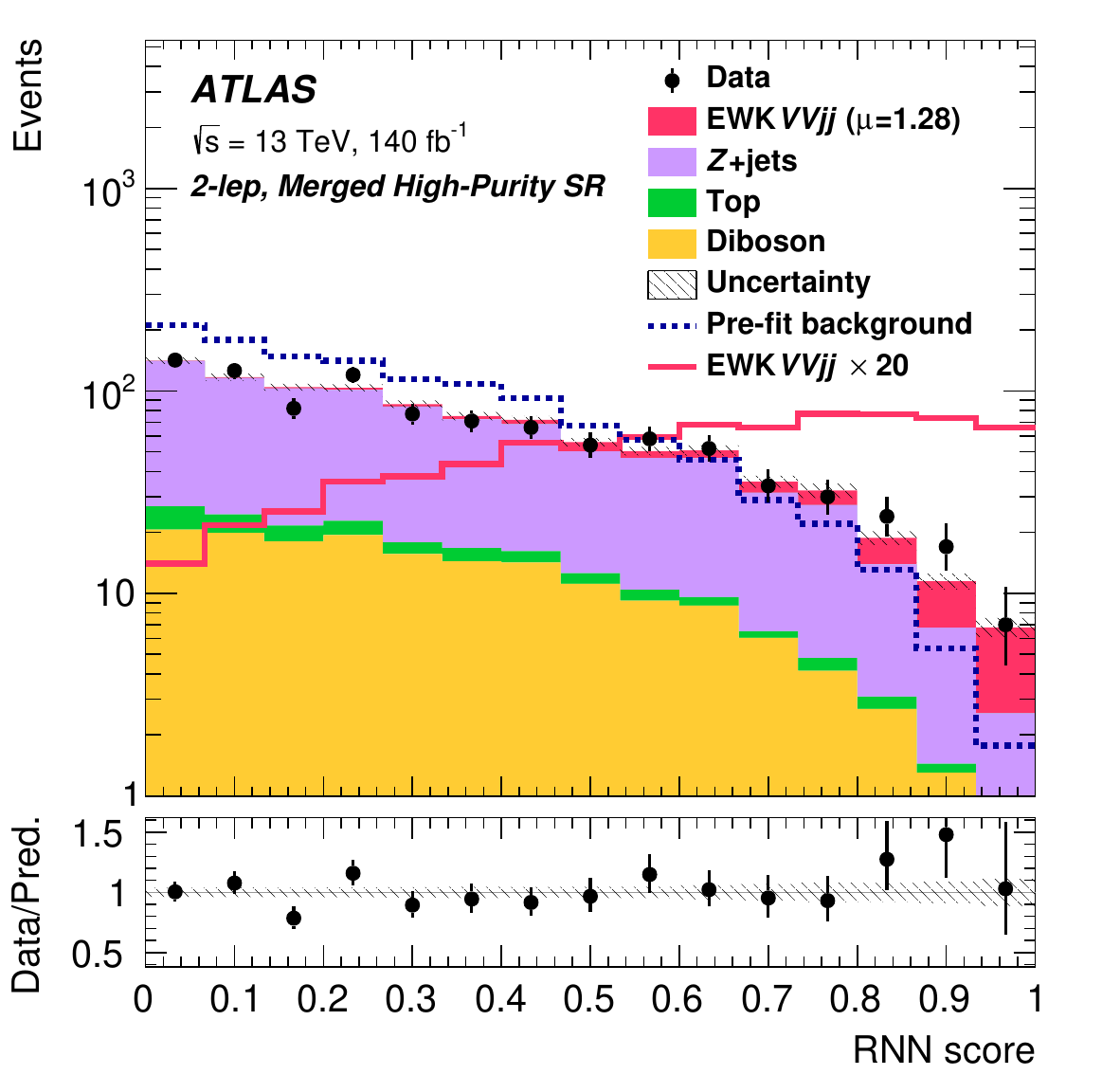}}
\caption{Distribution of the RNN score for data (dots) and post-fit prediction separated by process (stacked histograms).
The EWK \VVjj signal is also shown separately scaled for visibility by a factor of 20.
For comparison, the pre-fit background distribution is overlayed as a dotted histogram.
The total uncertainty including all statistical, experimental, and modelling uncertainties is overlayed as a hatched area. The bottom panel shows the ratio of data to the fitted signal-plus-background model.}
\label{fig:PostfitPlots}
\end{figure}

\begin{table}[t]
\caption{
The symmetrized uncertainty in the extracted signal-strength parameter $\mu$ from different sources. The systematic sources of uncertainty are separated into theory modelling uncertainties and experimental uncertainties, which are further sub-divided into different groups according to their origin.
\label{tab:np_ranking_muobs}}
\centering
\begin{tabular}{lc}
\toprule
Uncertainty source    & $\sigma_{\mu}$ \\  \midrule
Total                            & 0.22  \\
Statistical                      & 0.09  \\
Systematic                       & 0.20   \\
\midrule
\multicolumn{2}{c}{Theory and modelling uncertainties}\\
\midrule
Floating normalizations          & 0.04   \\
\Zjets                           & 0.06   \\
\Wjets                           & 0.07   \\
\ttbar                           & 0.02   \\
QCD \VVjj                      & 0.05   \\
Single-top                       & 0.01   \\
\mjjtag reweighting              & 0.07  \\
Signal modelling                 & 0.13   \\
MC statistics                    & 0.07  \\
\midrule
\multicolumn{2}{c}{Experimental uncertainties}\\
\midrule
Jets and \met                    & 0.09 \\
Tracking                         & 0.03 \\
Leptons                          & $<0.01$   \\
$b$-tagging                      & $<0.01$   \\
Luminosity                       & 0.01   \\
\bottomrule
\end{tabular}
\end{table}

\subsection{Fiducial measurement}

The cross sections in the fiducial regions described in Section~\ref{sec:fiducial} are extracted from a fit in which the normalization effects from the theory modelling uncertainty in the signal are removed. All other experimental, statistical, and theory modelling uncertainties are retained.
The fiducial cross section is then obtained from the fitted signal strength multiplied by the predicted generator cross section in the fiducial region and the fiducial-to-reconstructed region acceptance.
A combined measurement is performed using a single signal strength.
Measurements for the individual lepton channels and for the resolved and merged regimes are also performed with separate fits to extract, respectively, three and two parameters.
The effects of migrations between the various fiducial regions into the same fully-reconstructed region are taken into account by a linear transformation of the individual channel signal strengths by the estimated migration effects in simulations.
The prediction of the cross section includes only LO calculations since the EWK NLO corrections are not available for the semileptonic final state.

\FloatBarrier

\begin{figure}[t]
\centering
\includegraphics[width=.8\textwidth]{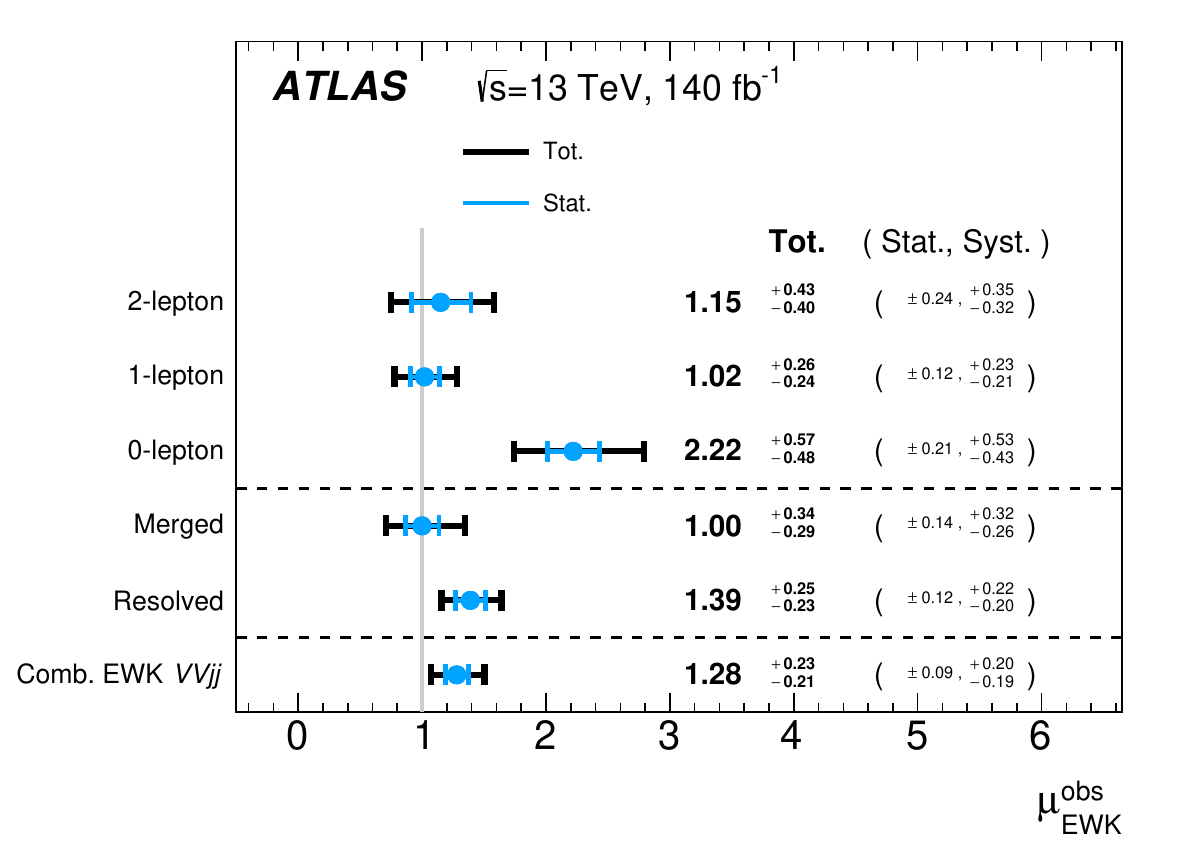}
\caption{Extracted signal strength \muobs from the combined fit, as well as from the fits which allow various signal regions to be uncorrelated. The total, statistical and systematic uncertainties in these values are displayed.\label{fig:signalStrengths}}
\end{figure}

\begin{figure}[t]
\centering
\includegraphics[width=.8\textwidth]{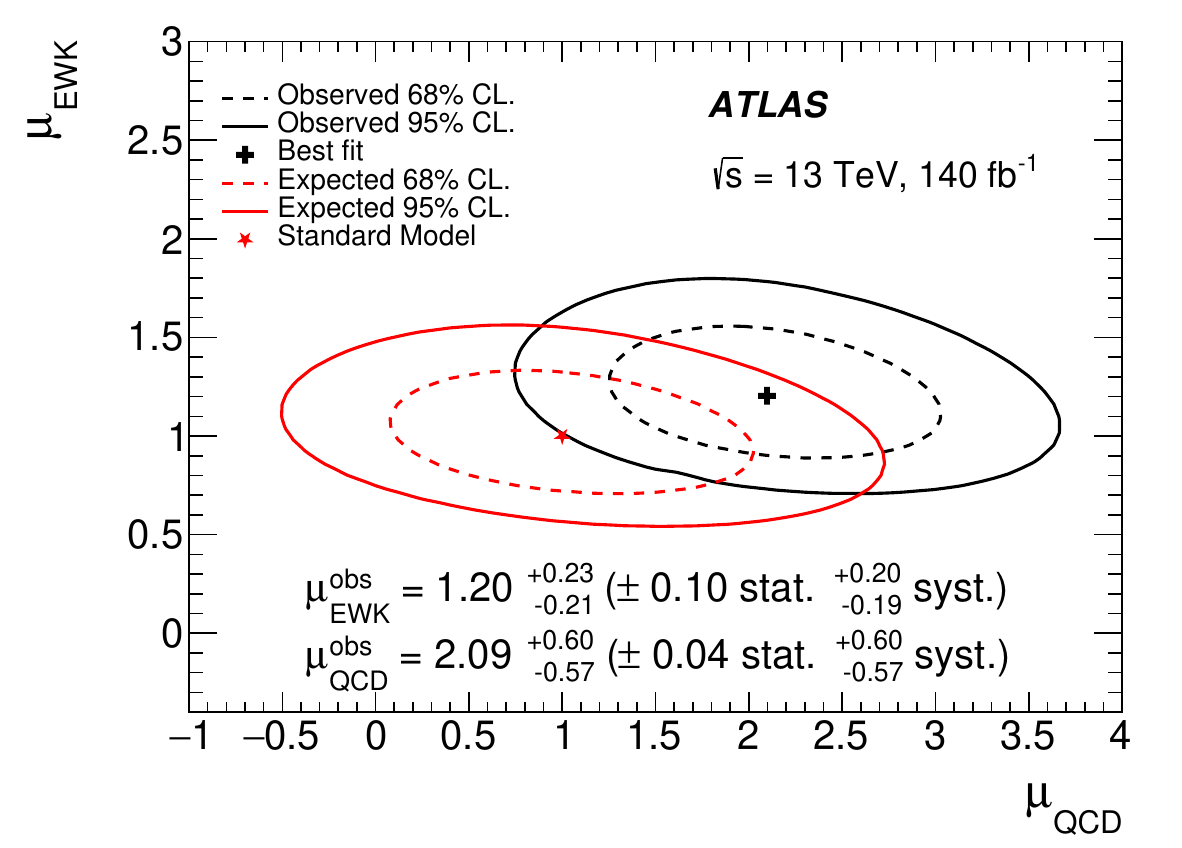}
\caption{The two dimensional 68\% and 95\% negative log-likelihood contours for the simultaneous fit to $\mu_{\mathrm{EWK}}$ and $\mu_{\mathrm{QCD}}$. The SM prediction from the generator is highlighted as a star.\label{fig:EWKvsQCDmu}}
\end{figure}

\FloatBarrier

The total fiducial cross section for the EWK \VVjj production summed across all fiducial regions is found to be $\sigmaobs= 29.2 \pm 4.9$~fb. The measured signal strength is $\muobs = 1.43 \pm 0.24$; it differs with respect to the inclusive signal strength due to the different fit approaches.
The impact of the parton-shower and QCD scale systematic uncertainties is lower in the fiducial measurement fit since only the impact on the shape is evaluated, and not the impact on the expected normalization.
These uncertainties are larger in the 0-lepton channel due to a tighter analysis selection with respect to the other two channels.
This results in a higher relative importance for the SRs of the 0-lepton channel, in which an upward fluctuation is observed.
The results of the fiducial cross-section measurements in different sub-regions are shown in Table ~\ref{tbl:FiducialResult};
in each region, a separate normalization is used in the extraction.
The measured cross sections are in agreement with the SM expectations.
In the fits to extract the signal strength for each lepton channel, only the \zlep channel measures a higher cross section, with a compatibility of 2.1$\sigma$ with the combined measurement.
This measurement complements the one conducted by the CMS Collaboration \cite{CMS-SMP-20-013}
as this result also includes the 0- and \tlep channels that were not considered in that measurement
and it considers a fiducial phase space that is tighter with respect to the one defined in the CMS analysis.

\begin{table}[t]
\caption{Expected and observed fiducial cross section of the EWK \VVjj process. The combined measurement refers to the case in which one signal strength is measured using all the analysis SRs.
The cross sections are also presented in fiducial bins across lepton multiplicity and in the merged and resolved fiducial regions.}
\label{tbl:FiducialResult}
\centering
\resizebox{\textwidth}{!}{\begin{tabular}{ccccccc}
\toprule
& Combined & \zlep & \olep & \tlep & Resolved & Merged \\
\midrule
$\sigmaexp$ & 20.4 $\pm$ 3.5 fb & 7.3 $\pm$ 2.5 fb & 10.3 $\pm$ 2.5 fb & 2.8 $\pm$ 1.1 fb & 11.7 $\pm$ 3.4 fb & 8.7 $\pm$ 2.5 fb \\
$\sigmaobs$ & 29.2 $\pm$ 4.9 fb & 15.7 $\pm$ 2.8 fb & 10.7 $\pm$ 2.8 fb & 3.1 $\pm$ 1.1 fb & 17.9 $\pm$ 4.3 fb & 11.4 $\pm$ 3.4 fb \\
\bottomrule
\end{tabular}}
\end{table}

\subsection{EFT interpretation}

The distribution of events in the SRs are also used to constrain higher-order operators in an EFT framework. In particular, new operators of mass-dimension 8 contribute to aQGCs, which would enhance EWK \VVjj production at high diboson invariant mass, \mVV.

The Eboli model~\cite{eboli2006p} is employed to describe the signal, which introduces 21 new dimension-8 operators satisfying the SM $SU(2)\times U(1)_Y$ symmetry.
These operators can be categorized by the types of coupling; scalar types that only contain covariant derivatives of the Higgs field: $f_{S0}$, $f_{S1}$, $f_{S2}$, tensor types  that only contain field strengths: $f_{T0}$, ..., $f_{T9}$, and mixed operators that exhibit two covariant derivatives of the Higgs field and two field strengths: $f_{M0}$, ..., $f_{M7}$.
As a convention, the whole Lagrangian term for an operator is denoted by its Wilson coefficient i.e. $\mathcal{L}_{n} = \frac{f_{n}}{\Lambda^4} O_{n}$ will be denoted as $f_{n}$.
Due to the fact that the operators $O_{S0}$ and $O_{S2}$ are Hermitian conjugates, they are varied simultaneously with equal coefficient values $f_{S02}$.

The EFT contributions are modelled with MC samples, as described in Section ~\ref{subsec:mc}, using a decomposition of the process amplitude. The total matrix element with the addition of new dimension-8 operators can be written as
\begin{equation*}
|A_{SM}+\frac{f_i}{\Lambda^4}A_i|^2=|A_{SM}^2|+\sum\limits_i \frac{f_i^2}{\Lambda^8}|A_{i}^2|+ \sum\limits_i 2 \frac{f_i}{\Lambda^4} \mathrm{Re}(A_{SM}^\star A_i) +\sum\limits_{i\neq j} \frac{f_i}{\Lambda^4} \frac{f_j}{\Lambda^4} \mathrm{Re}(A_i^\star A_j),
\end{equation*}
where $|A_{SM}|^2$ is the SM matrix element, $|A_{i}^2|$ represents the pure-EFT matrix elements for the $f_{i}$ operator, and $2 \mathrm{Re}(A_{SM}^\star A_i)$ is its corresponding interference term with the SM. The term $\mathrm{Re}(A_i^\star A_j)$ includes the possible interference term between EFT operators $f_{i}$ and $f_{j}$, and will not be considered here.
Individual simulation samples for the pure-EFT and interference terms are generated; to obtain predictions at a specific coupling point, these samples are scaled quadratically and linearly by $f_i$, respectively.

The EFT results are extracted in a similar fit setup as described in Section~\ref{subsec:fit}, where now the EFT coefficients $f_i$, are set as the parameters of interest.
Since the EFT is expected to contribute more at high \mVV, each of the SRs is further split into two (high and low) \mVV bins to improve the sensitivity and the RNN score is used as final discriminant.
The high- and low-\mVV regions are split at $\mVV=1500~\GeV$ in the 1- and 2-lepton channels; in the 0-lepton channel, the transverse mass ($m_\mathrm{T}$),
defined as the invariant mass of the \Vhad candidate and the \met vector in the transverse plane, is used instead with the splitting occurring at $1050~\GeV$ ($1200~\GeV$) in the merged (resolved) region.
The CRs in this fit are the same as those in the inclusive measurement.
The post-fit RNN score distributions in the more aQGC-sensitive high \mVV bin are shown in Figure~\ref{fig:postfitaQGCH};
the binning of the distributions was reduced to account for the lower number of events expected in the high \mVV bin.
The distributions agree reasonably well with the SM predictions.

For the EFT interpretations, the EWK \VVjj normalization
and the signal strength of the EFT process are left free in the fit.
The EWK \VVjj normalization is constrained by the low \mVV bins;
the post-fit value is compatible with the one obtained in the inclusive measurement.
Confidence intervals are set at the
95\% confidence level assuming that the Wald approximation holds ~\cite{Wilks:1938dza,Cowan:2010js}.

To provide interpretable limits which avoid the unitarity violation at high center of mass energy, the clipping method, described in Ref. \cite{CMS-SMP-19-012}, is used. In this method, a clipping scale $E_{\text{clip}}$ is chosen and signal events with truth \mVV greater than this clipping scale are removed. The expected and observed 95\% confidence level limits on the different aQGC operators as functions of the clipping scale are shown in Figures~\ref{fig:limitClippedFT} to \ref{fig:limitClippedFM}.
The theory unitary bounds for each operator are taken from Ref. \cite{da2020unitarity} considering partial-wave unitarity for a single operator at a time.
A summary of the expected and observed limits on the different dimension-8 EFT operators is shown in Table~\ref{tbl:EFTlimits}.

This result improves upon the constraints from other ATLAS searches for aQGC~\cite{STDM-2022-06,STDM-2020-02,STDM-2018-31,STDM-2018-32,STDM-2020-02,STDM-2018-59,STDM-2018-35,STDM-2021-09} for all $f_{S}$ and $f_{M}$ type operators.
Improvements are found for both
the first clipping point that, in most cases, is within the unitarity bound
and for the limits without any clipping scale.
The latter improvement can be as large as a factor
3.3 for $f_{M0}$,
1.8 for $f_{M1}$,
1.5 for $f_{M7}$,
1.5 for $f_{S02}$
and
2.9 for $f_{S1}$
compared to the best limits reported by the ATLAS Collaboration in other diboson channels.
The $f_{T}$ operators are instead found to be more tightly constrained by the measurement of $Z(\rightarrow \nu\nu)\gamma$ in association with two jets~\cite{STDM-2018-59}.

\begin{figure}[t]
\centering
\FloatBarrier
\subfigure[]{\includegraphics[width=.3\linewidth]{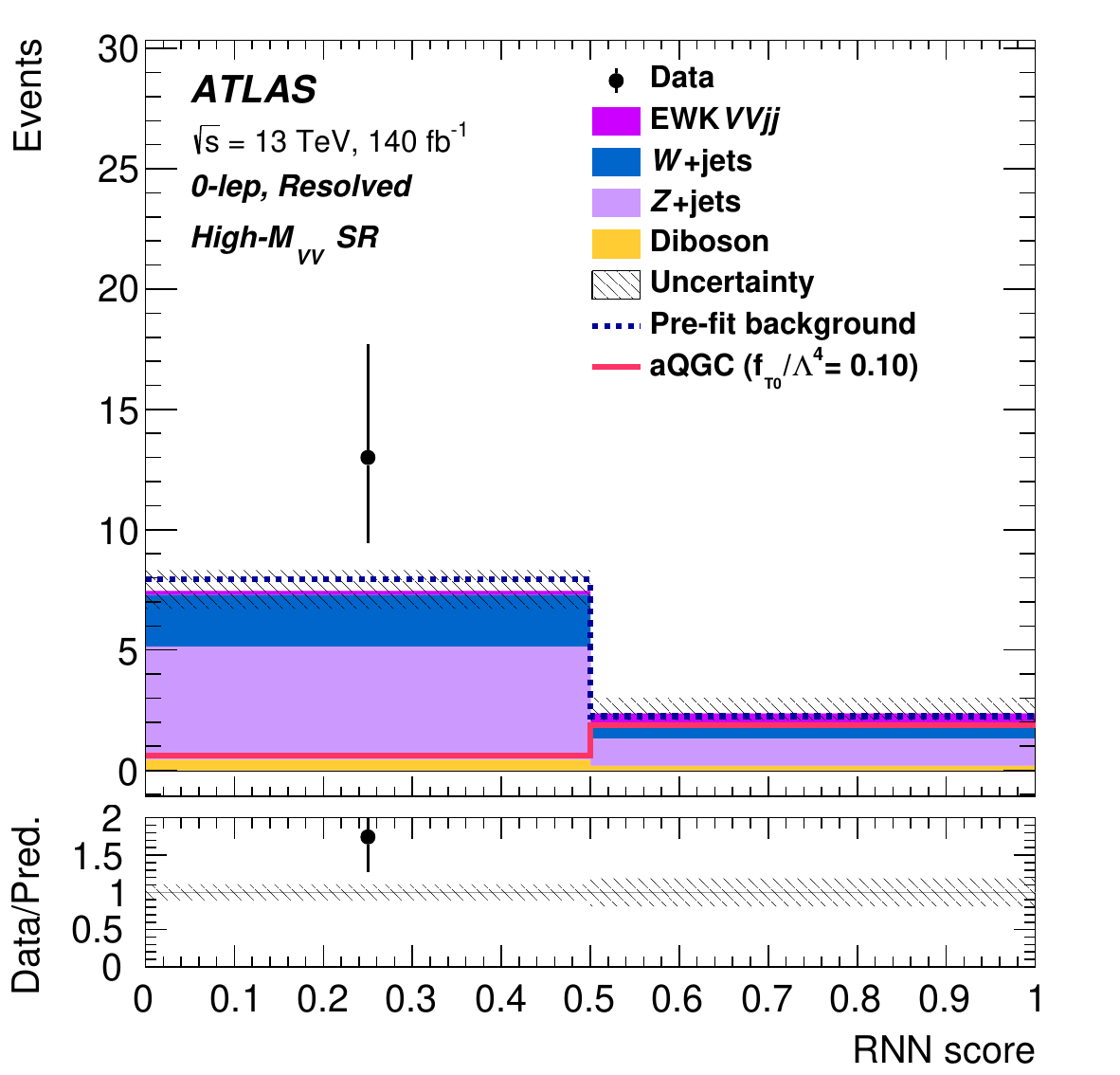}}
\subfigure[]{\includegraphics[width=.3\linewidth]{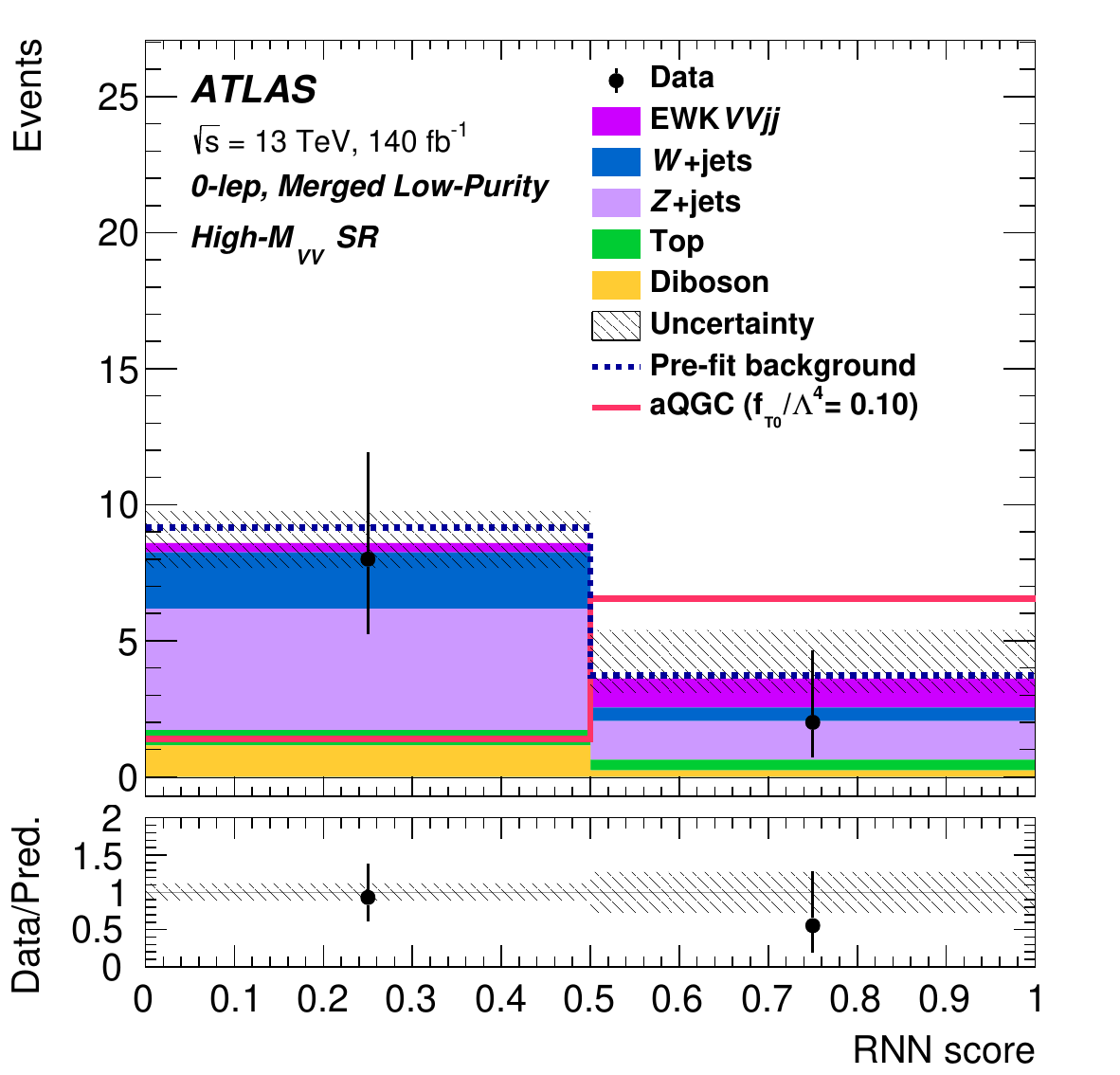}}
\subfigure[]{\includegraphics[width=.3\linewidth]{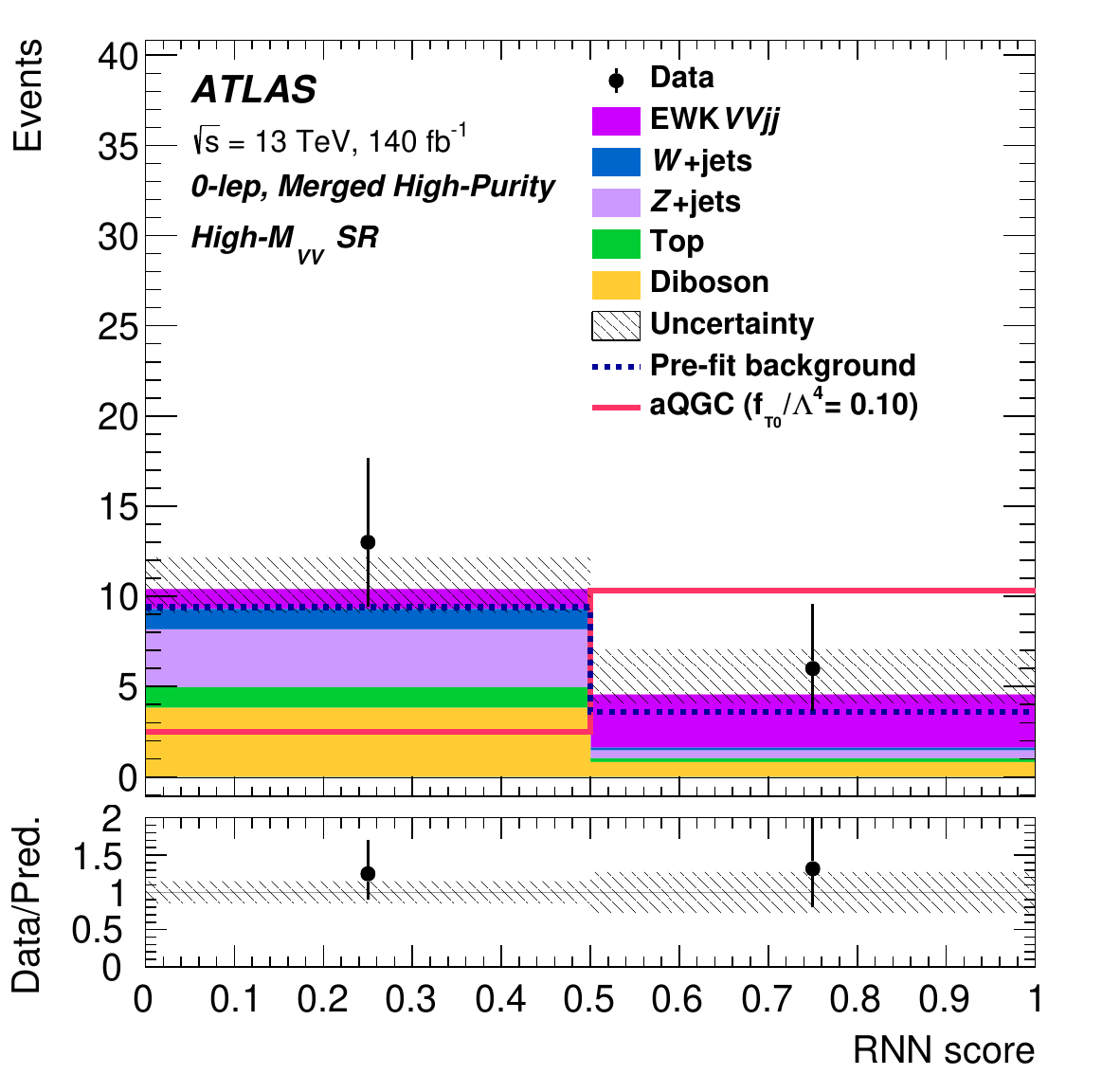}} \\
\subfigure[]{\includegraphics[width=.3\linewidth]{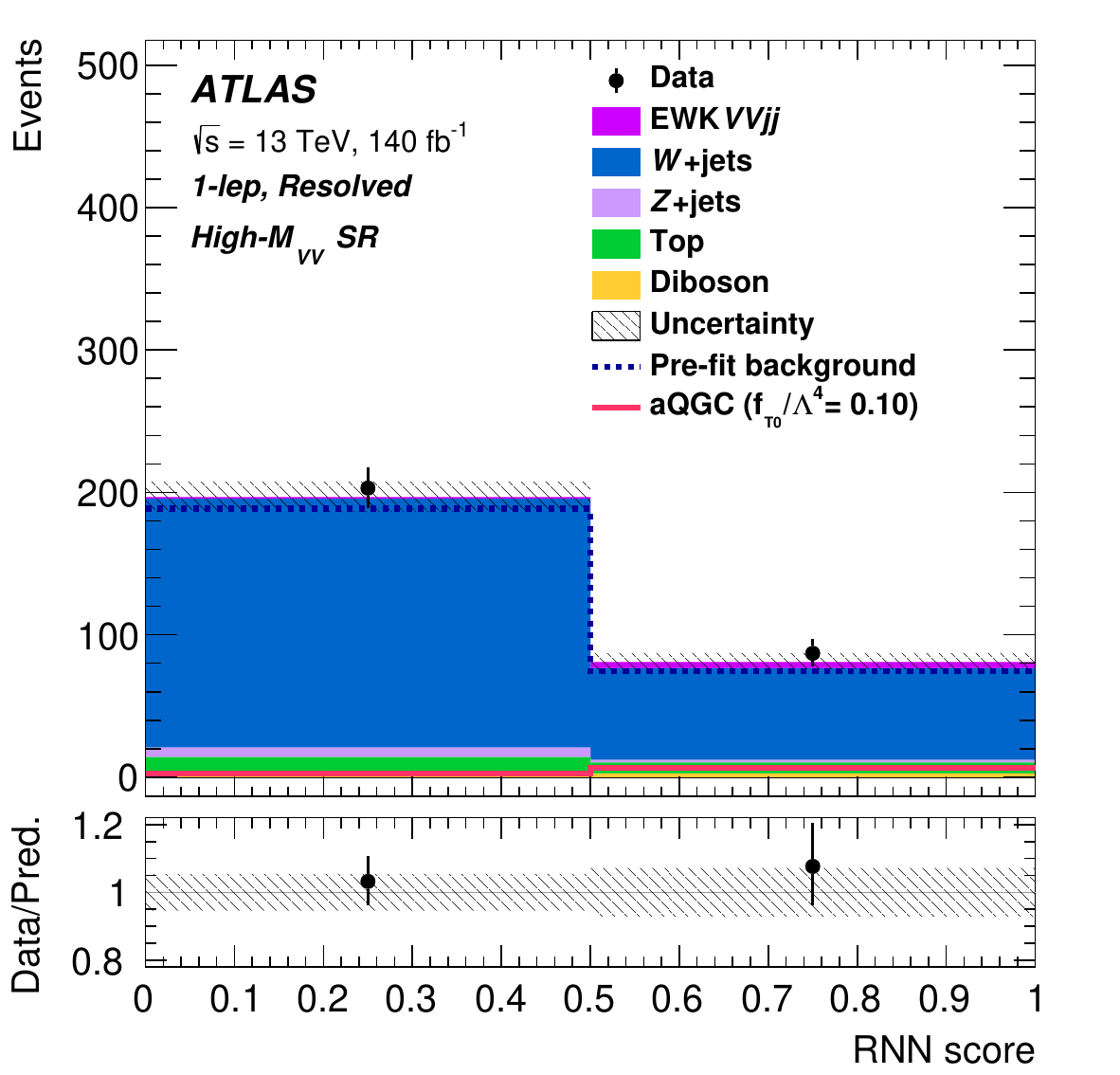}}
\subfigure[]{\includegraphics[width=.3\linewidth]{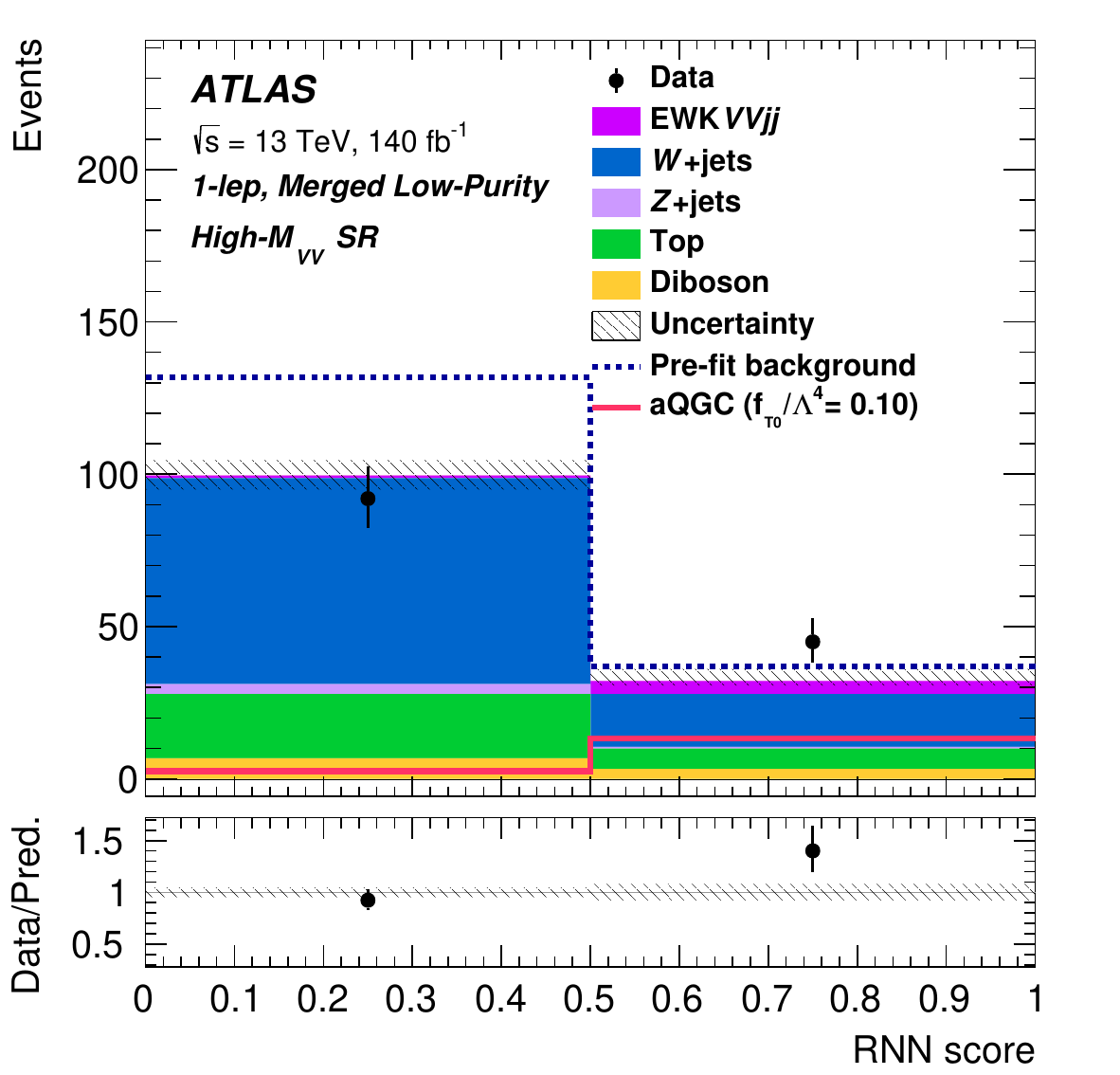}}
\subfigure[]{\includegraphics[width=.3\linewidth]{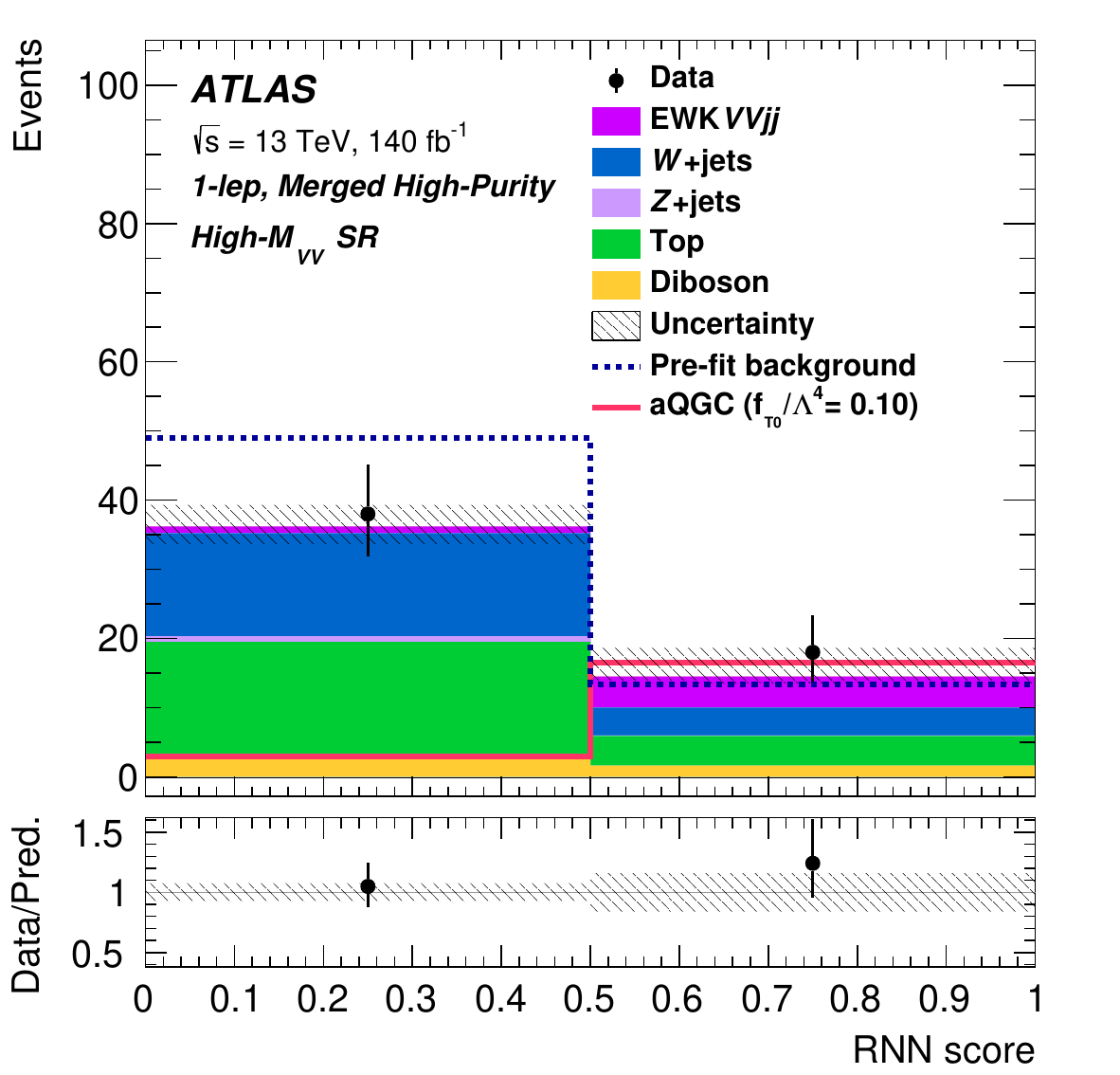}} \\
\subfigure[]{\includegraphics[width=.3\linewidth]{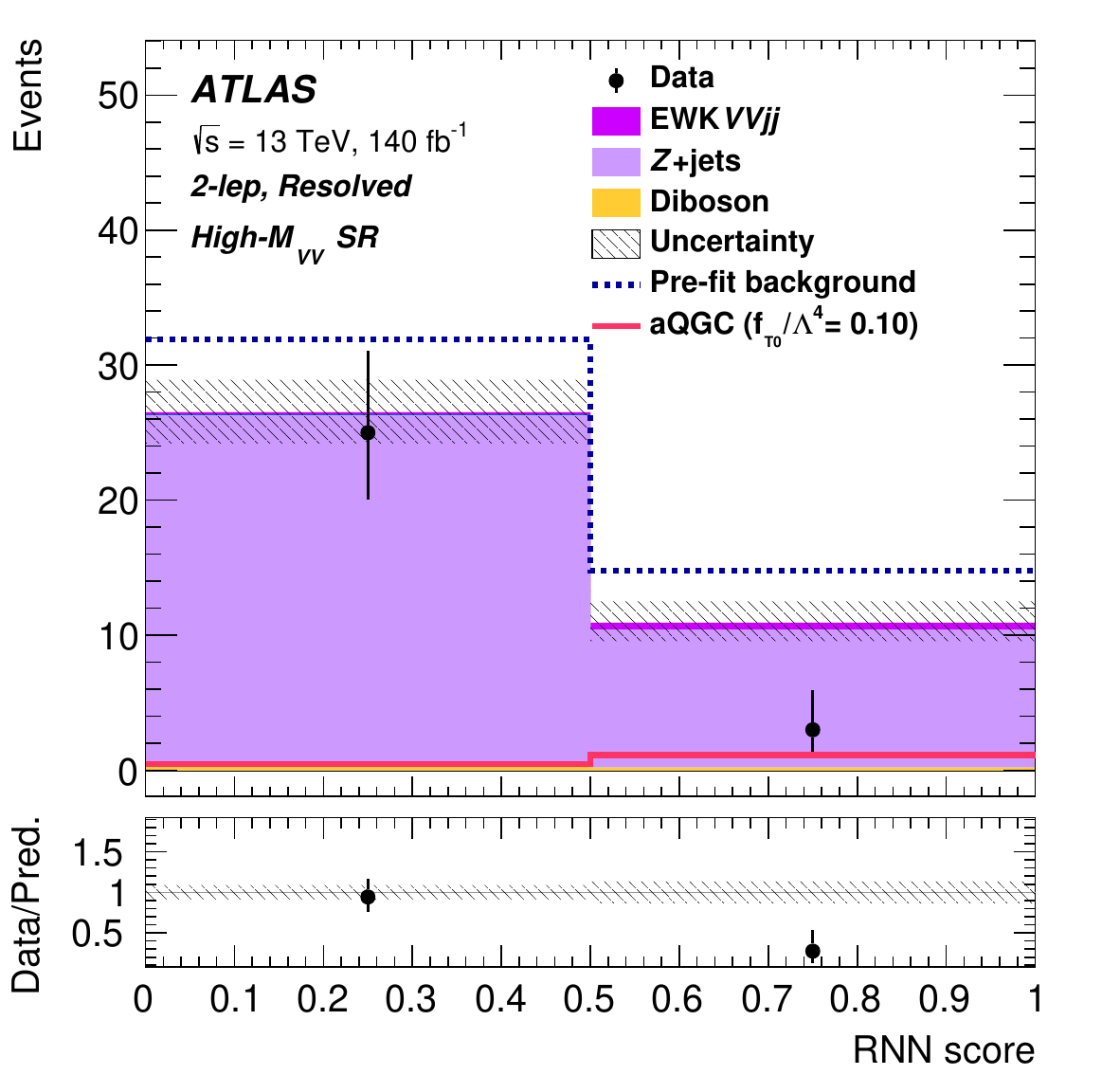}}
\subfigure[]{\includegraphics[width=.3\linewidth]{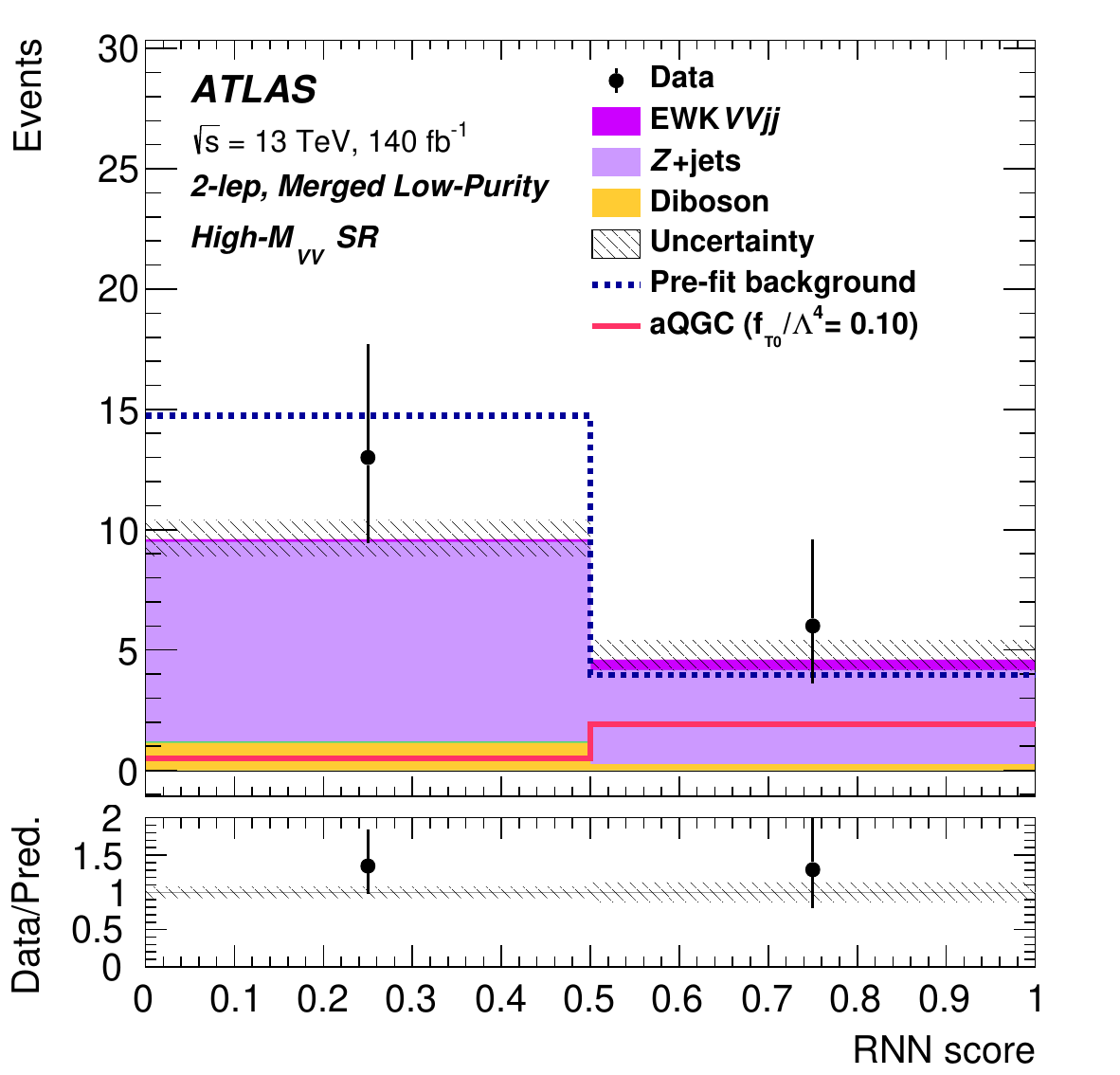}}
\subfigure[]{\includegraphics[width=.3\linewidth]{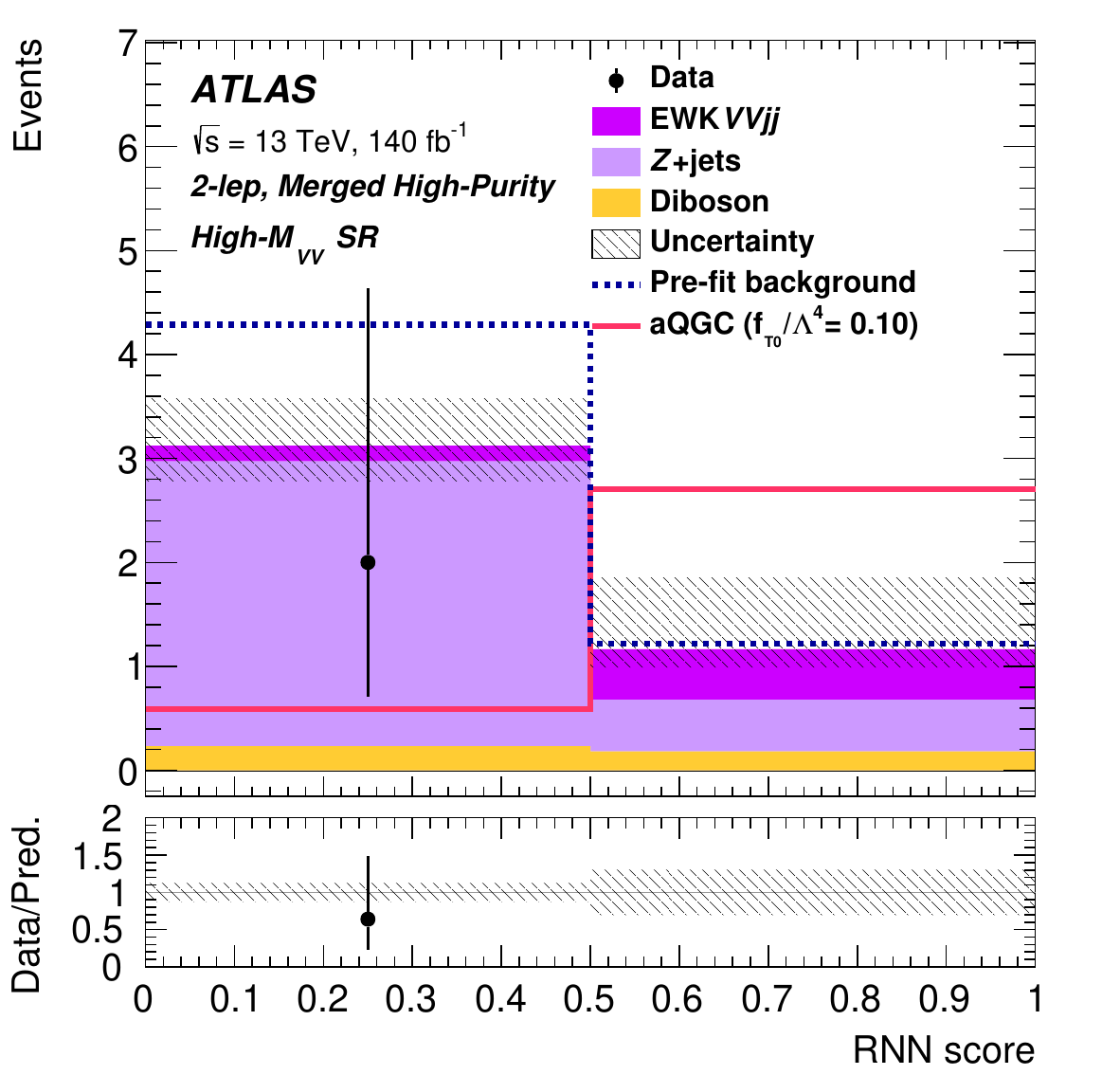}}
\caption{
Distribution of the RNN score for data (dots) and post-fit prediction separated by process (stacked histograms) for the high-$m_{VV}$ aQGC signal regions.
The $f_{T0}$ operator is shown separately with a normalisation corresponding to $f/\Lambda^4=0.10$~$\TeV^{-4}$. For comparison, the pre-fit background distribution is overlayed as a dotted histogram.
The total uncertainty including all statistical, experimental, and modelling uncertainties is overlayed as a hatched area. The bottom panel shows the ratio of data to the fitted signal-plus-background model.
}
\label{fig:postfitaQGCH}
\end{figure}

\begin{figure}[t]
\centering
\subfigure[]{\includegraphics[width=.3\linewidth]{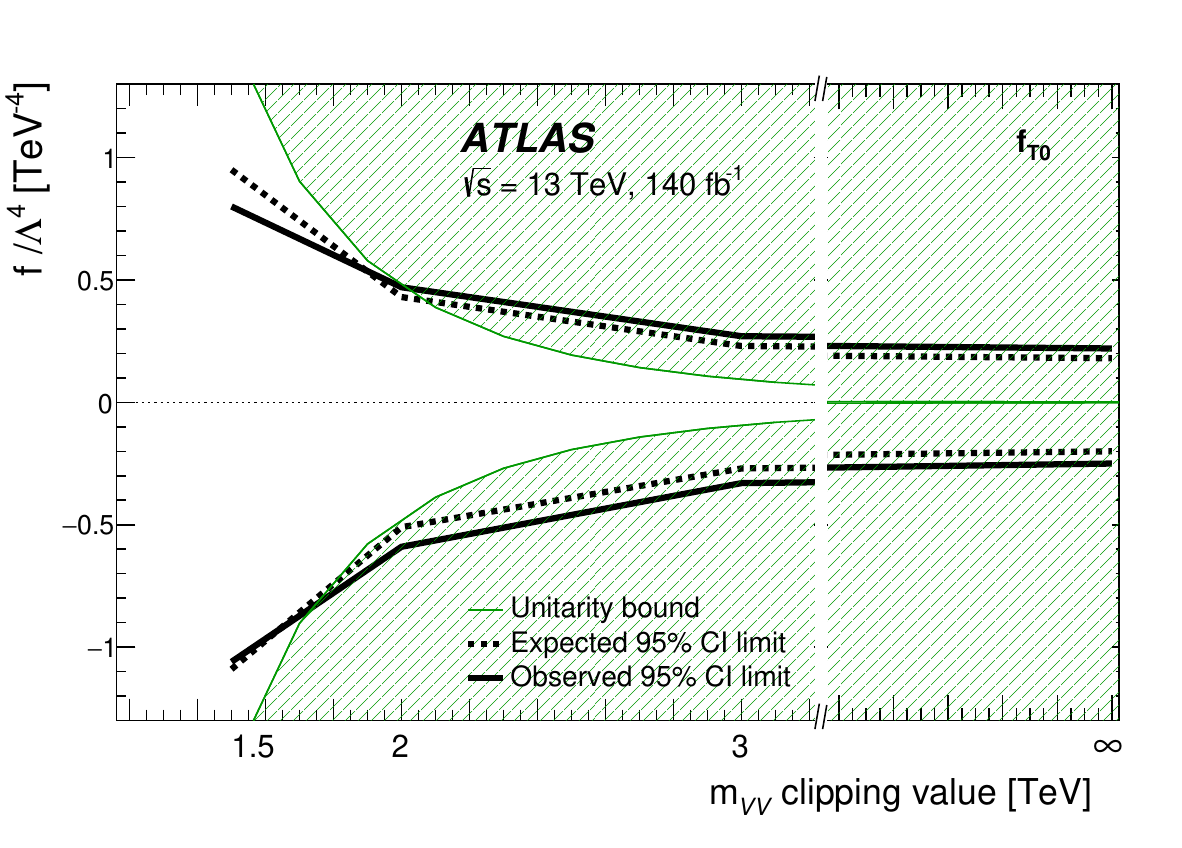}}
\subfigure[]{\includegraphics[width=.3\linewidth]{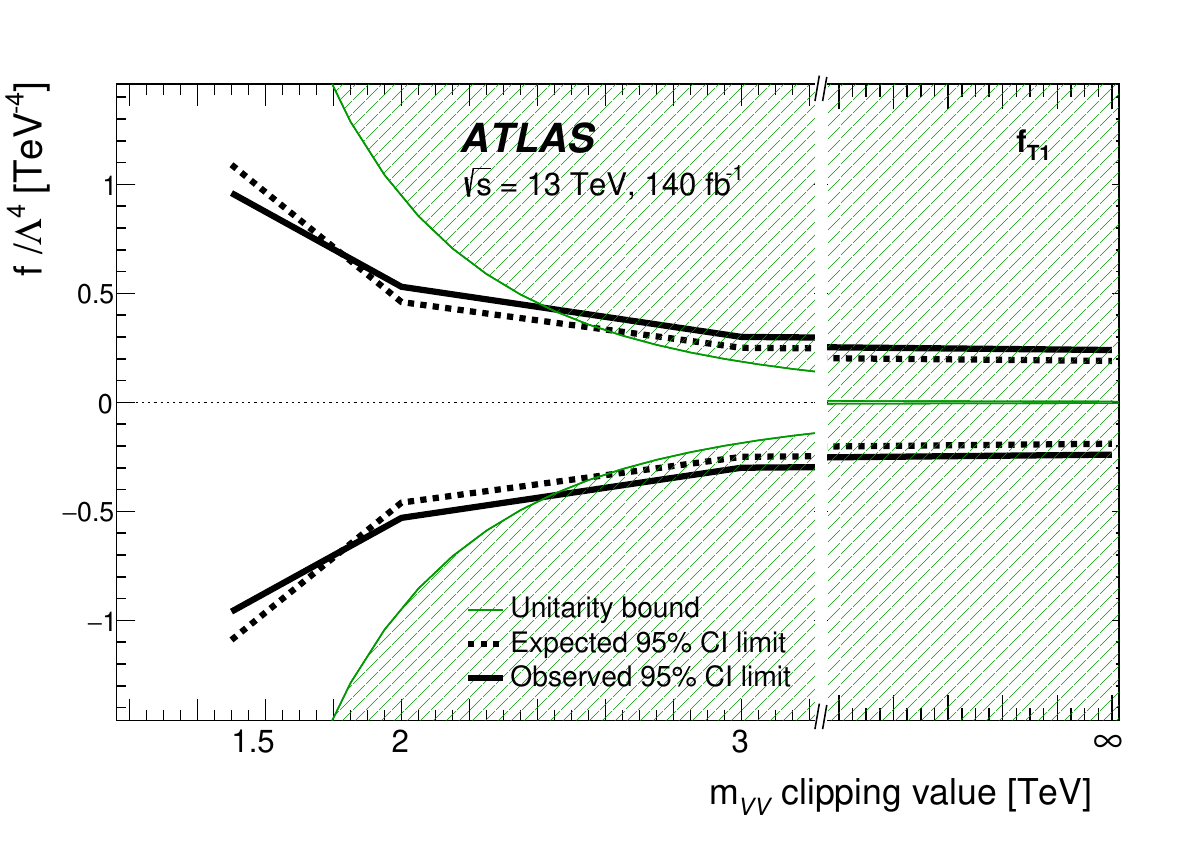}}
\subfigure[]{\includegraphics[width=.3\linewidth]{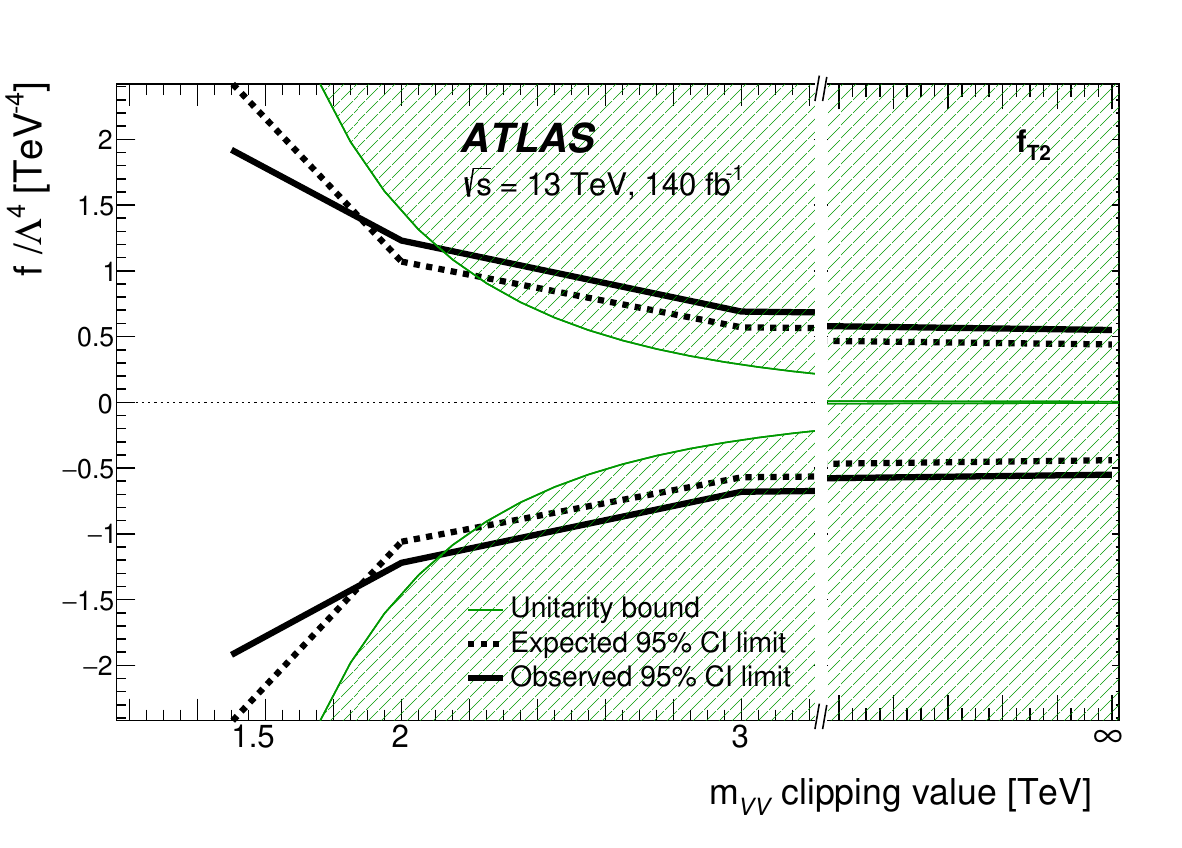}}
\subfigure[]{\includegraphics[width=.3\linewidth]{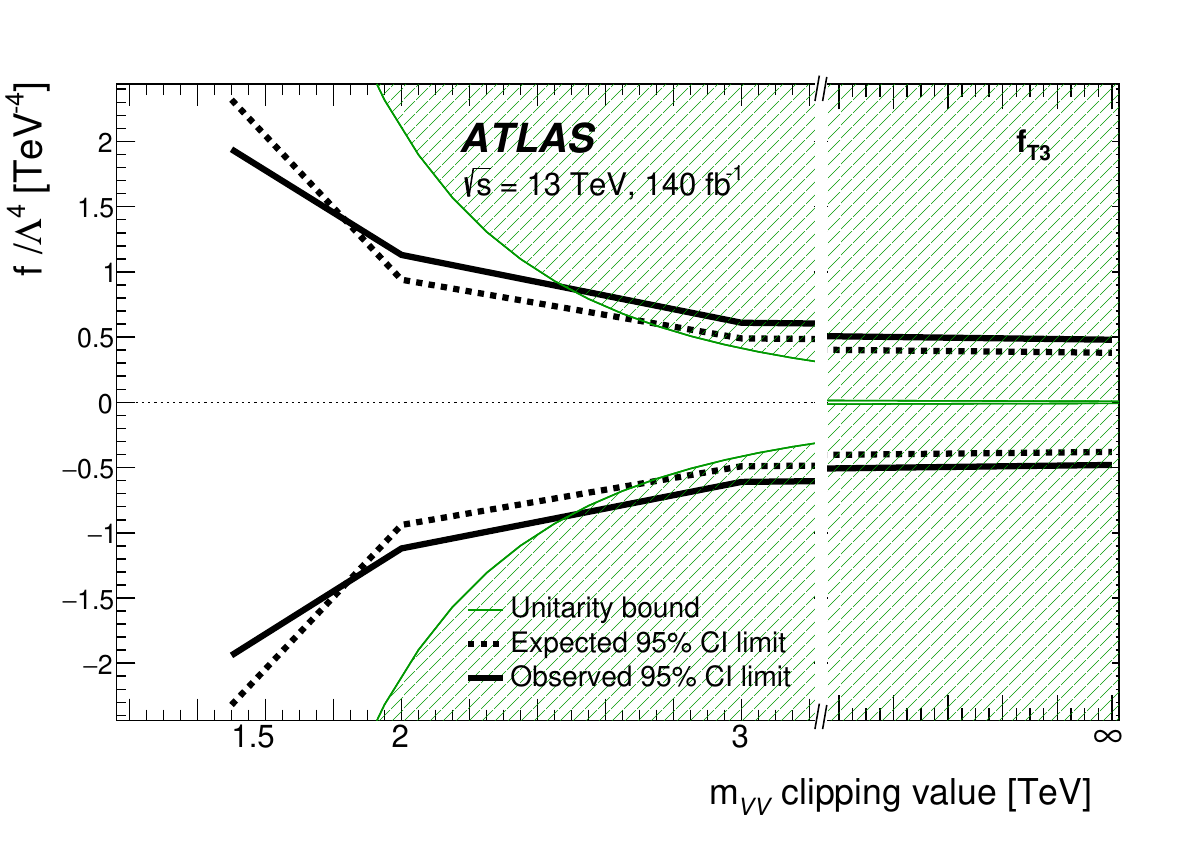}}
\subfigure[]{\includegraphics[width=.3\linewidth]{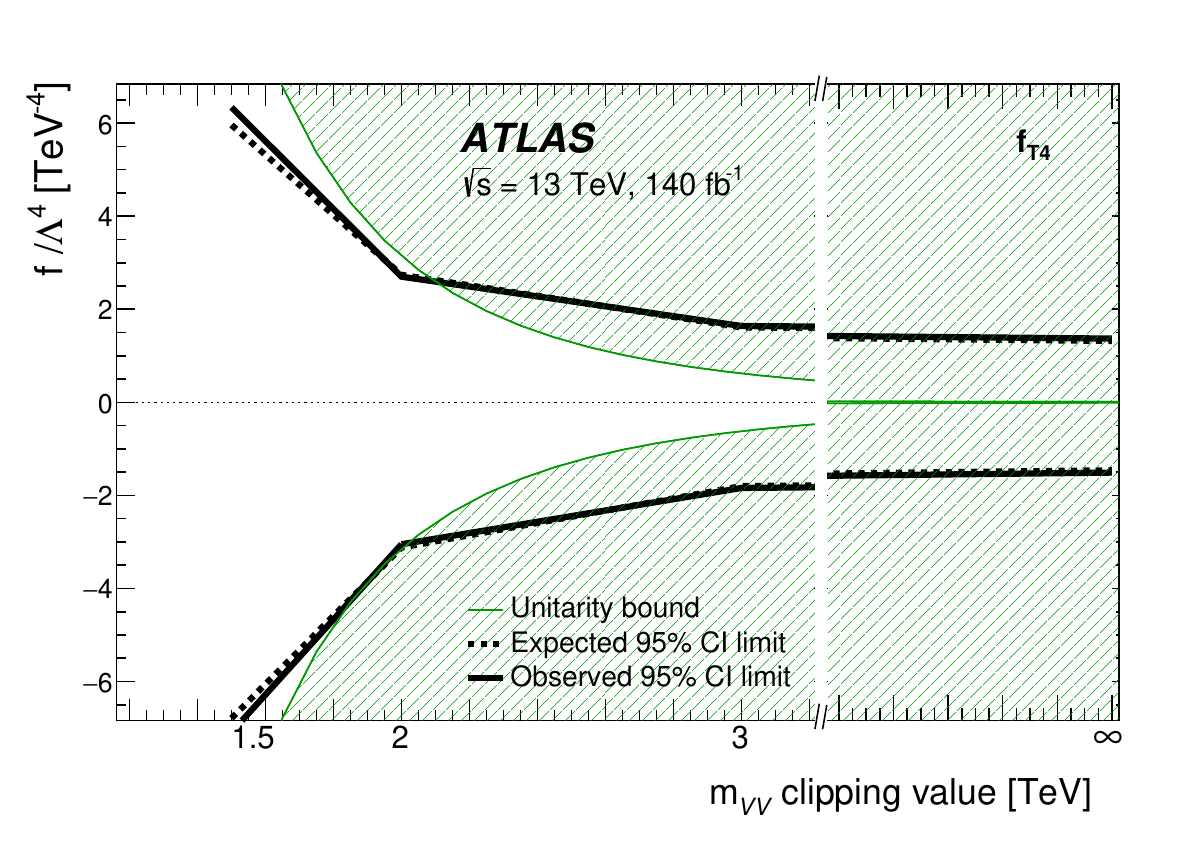}}
\subfigure[]{\includegraphics[width=.3\linewidth]{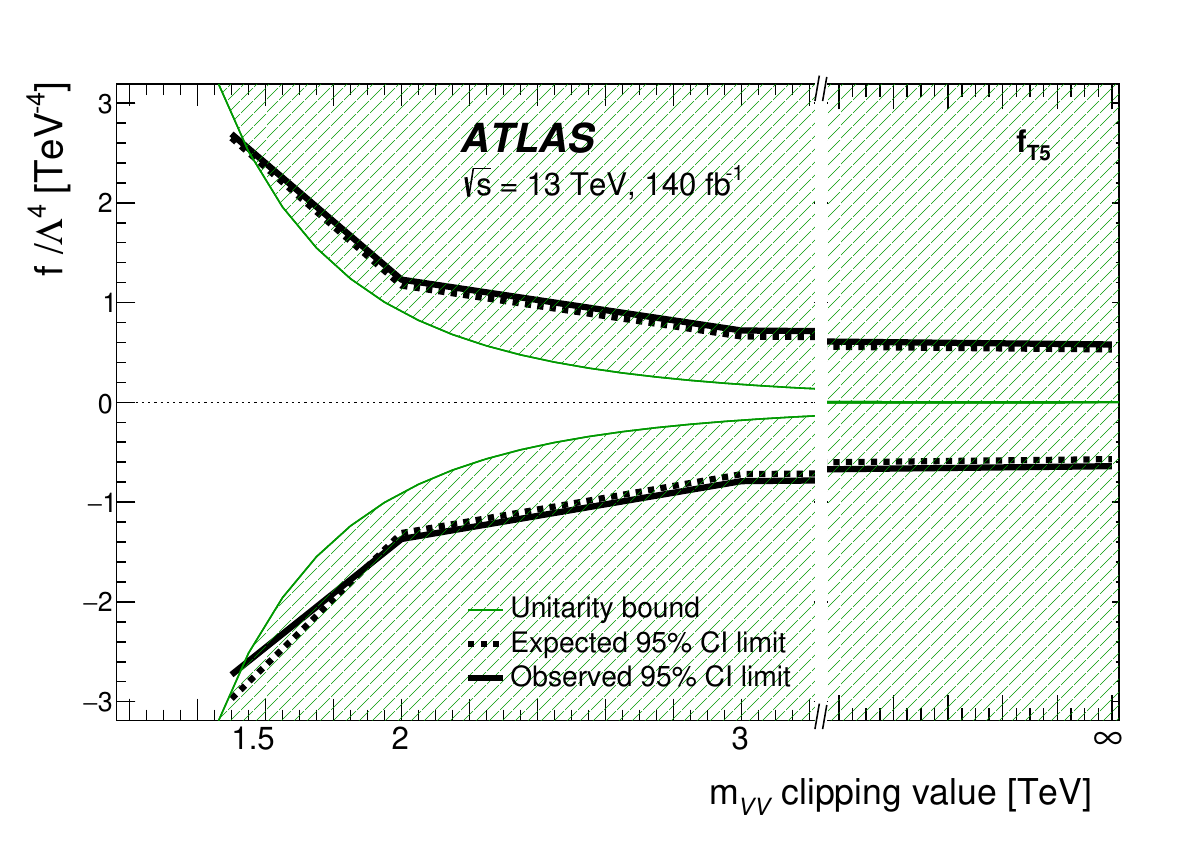}}
\subfigure[]{\includegraphics[width=.3\linewidth]{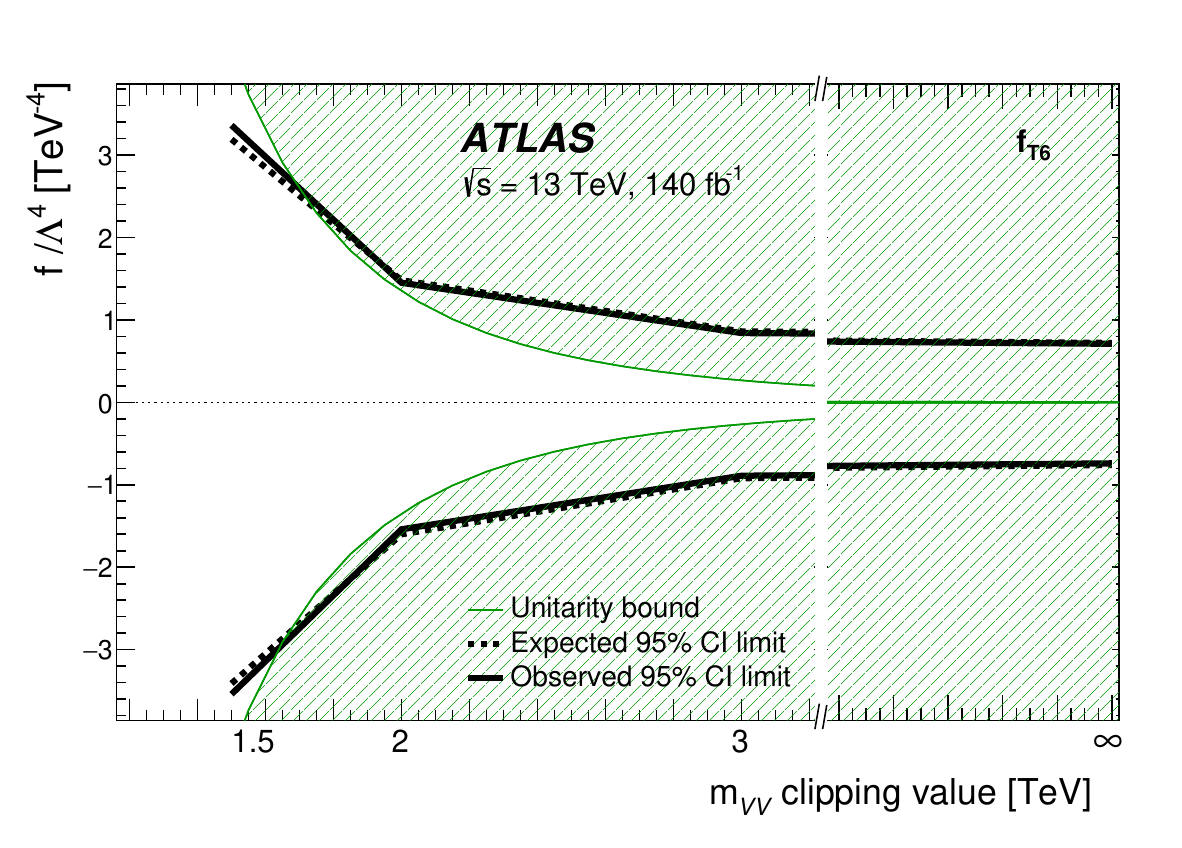}}
\subfigure[]{\includegraphics[width=.3\linewidth]{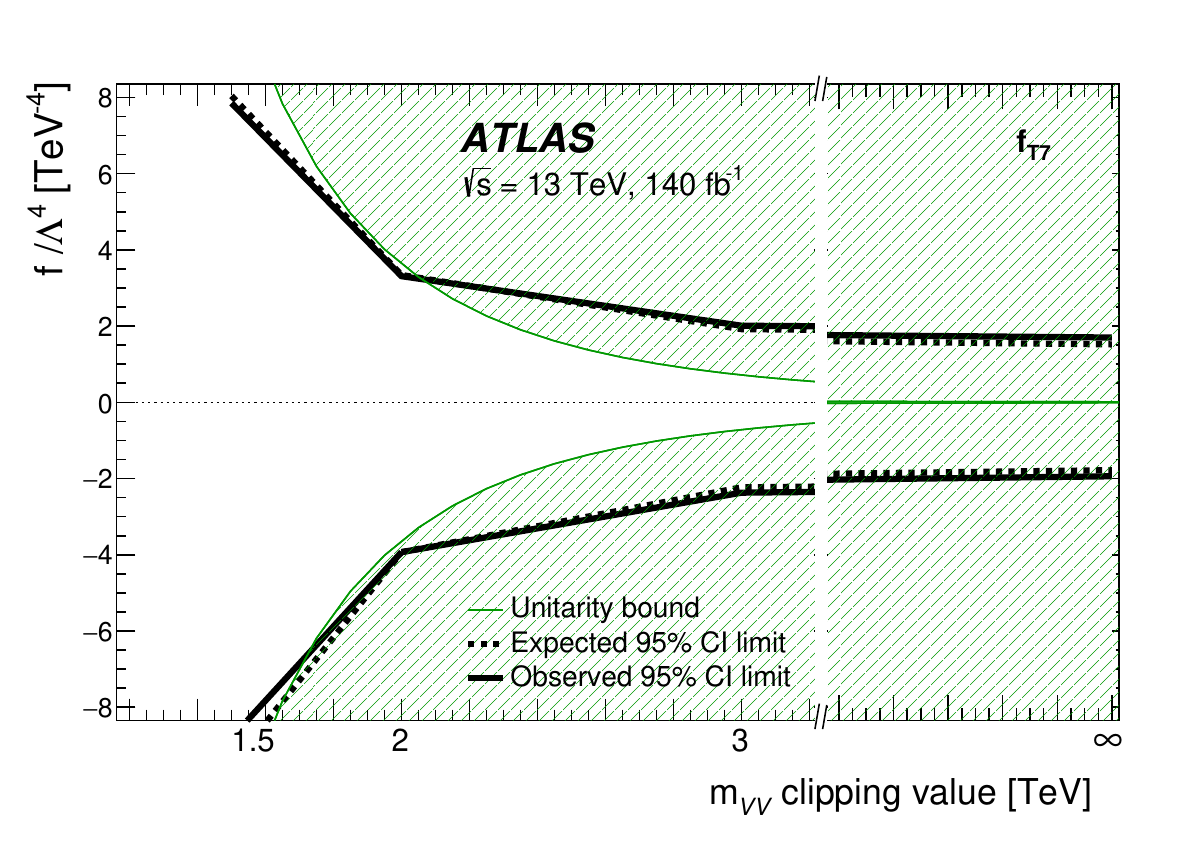}}
\subfigure[]{\includegraphics[width=.3\linewidth]{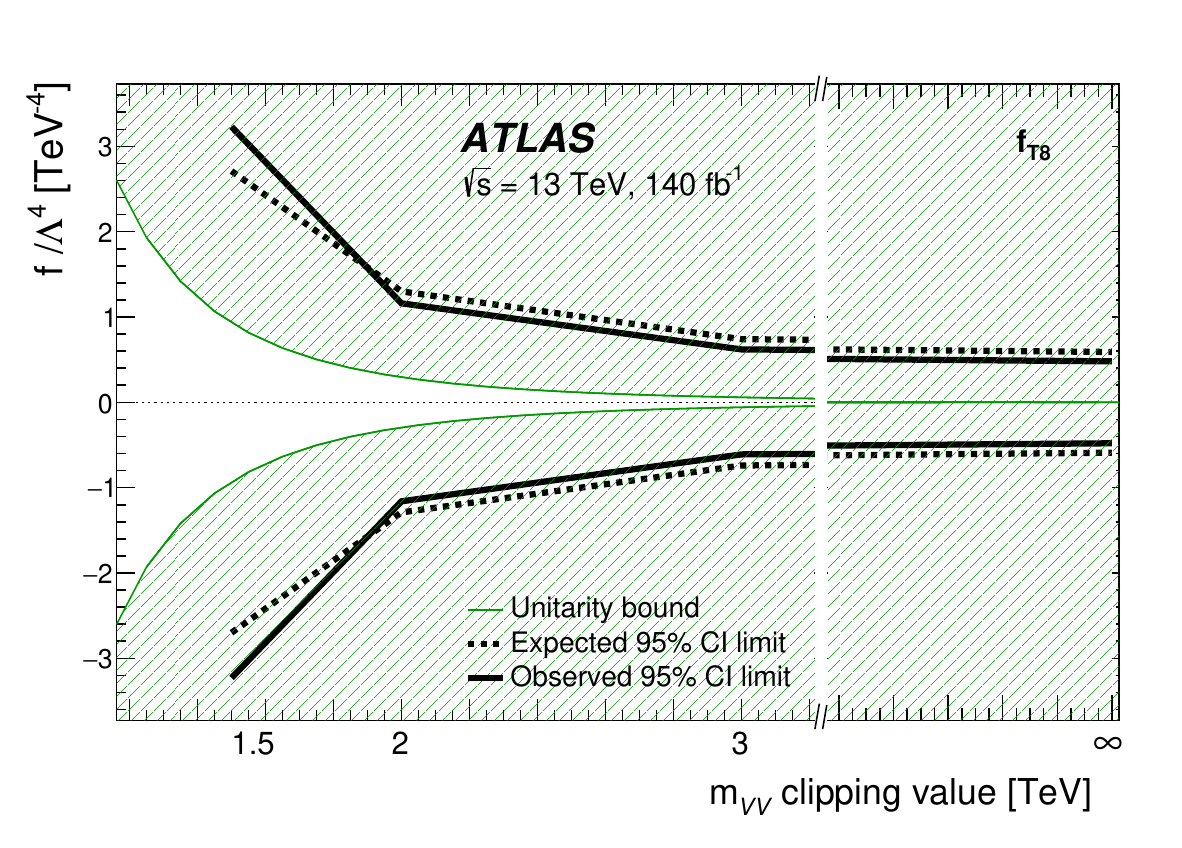}}
\subfigure[]{\includegraphics[width=.3\linewidth]{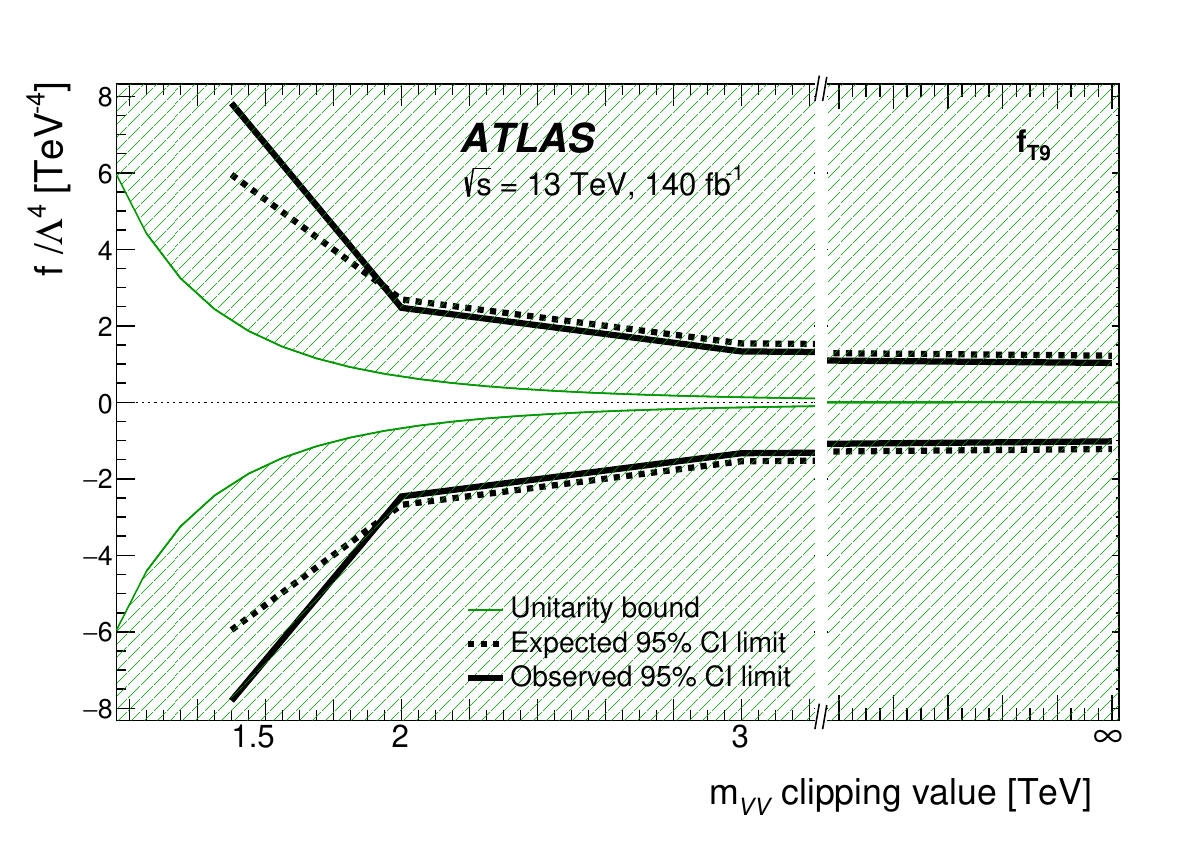}}
\caption{Expected and observed 95\% confidence limits for the dimension-8 $f_{T}$ operators as a function of the clipping scale.
Regions excluded by unitarity constraints are shown as a hatched area.
}
\label{fig:limitClippedFT}
\end{figure}

\begin{figure}[t]
\centering
\subfigure[]{\includegraphics[width=.45\linewidth]{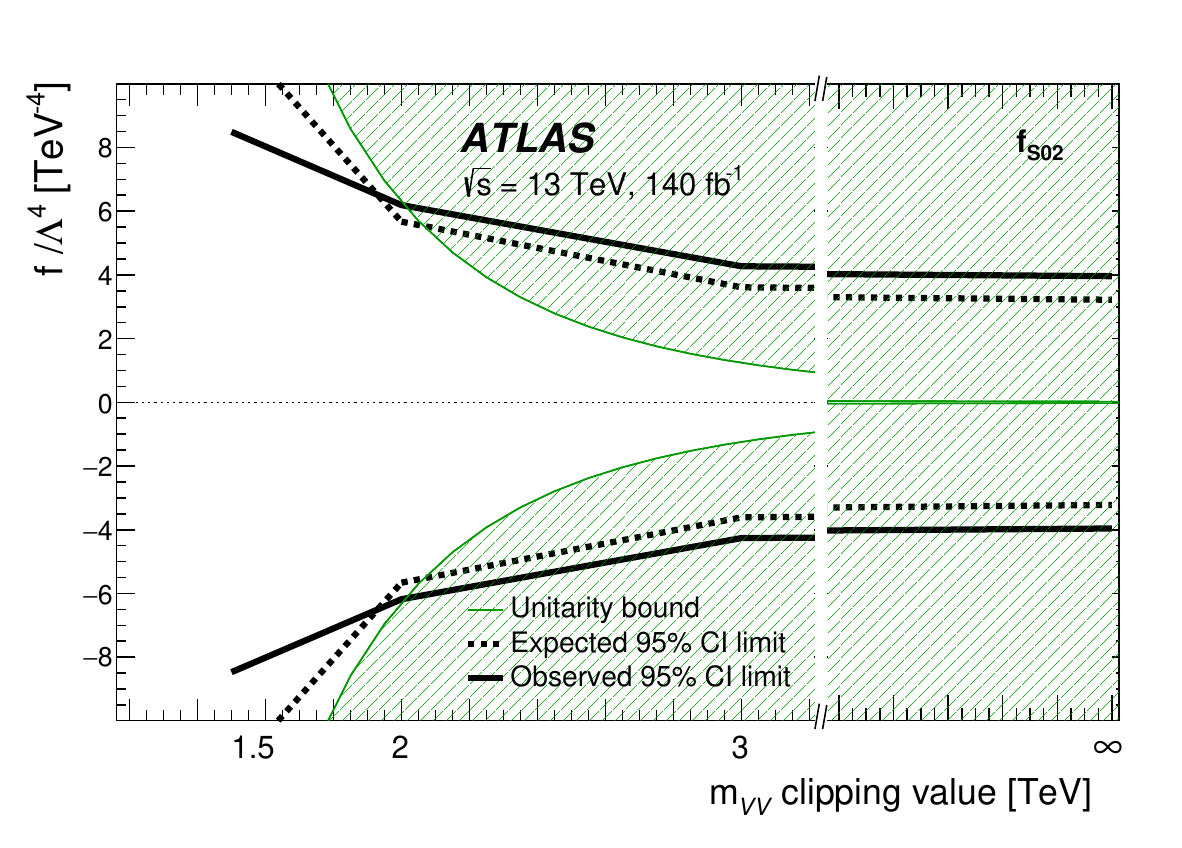}}
\subfigure[]{\includegraphics[width=.45\linewidth]{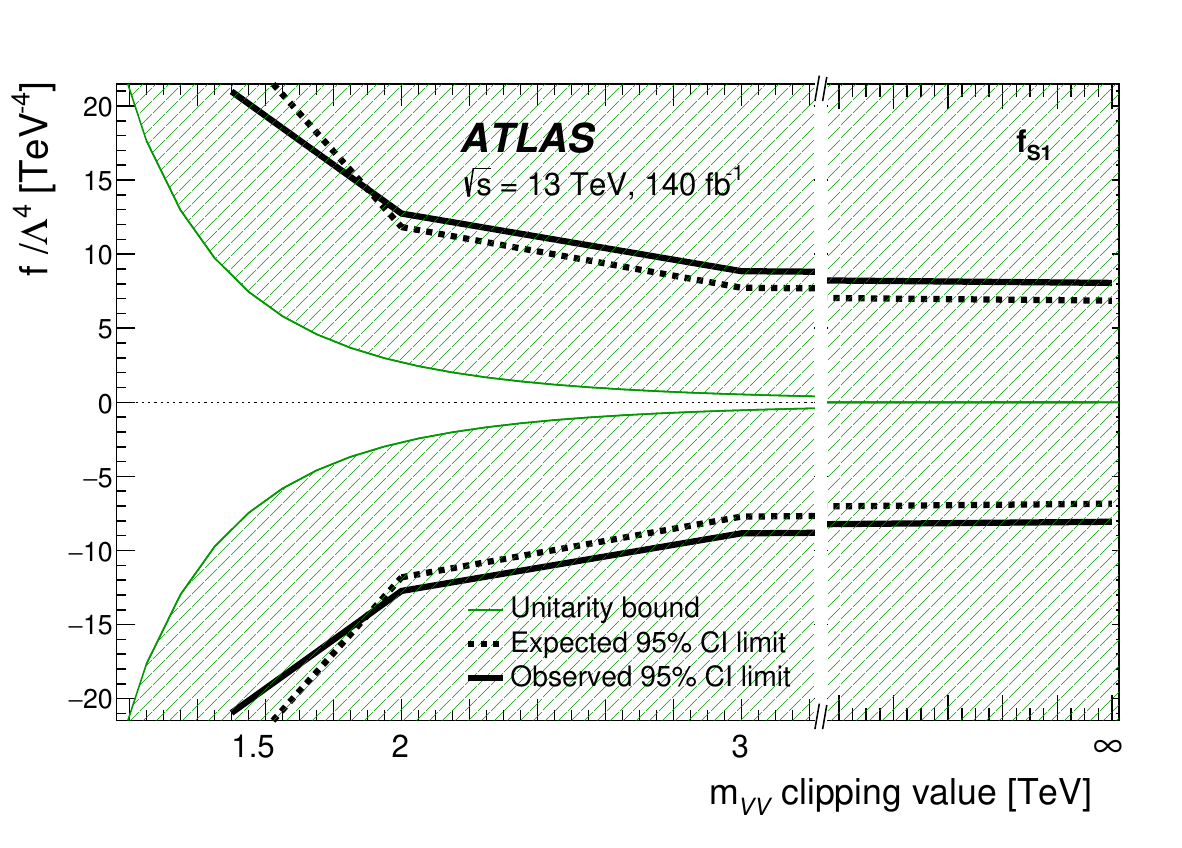}}
\caption{Expected and observed 95\% confidence limits for the dimension-8 $f_{S}$ operators as a function of the clipping scale.
Regions excluded by unitarity constraints are shown as a hatched area.
}
\end{figure}

\begin{figure}[t]
\centering
\subfigure[]{\includegraphics[width=.40\linewidth]{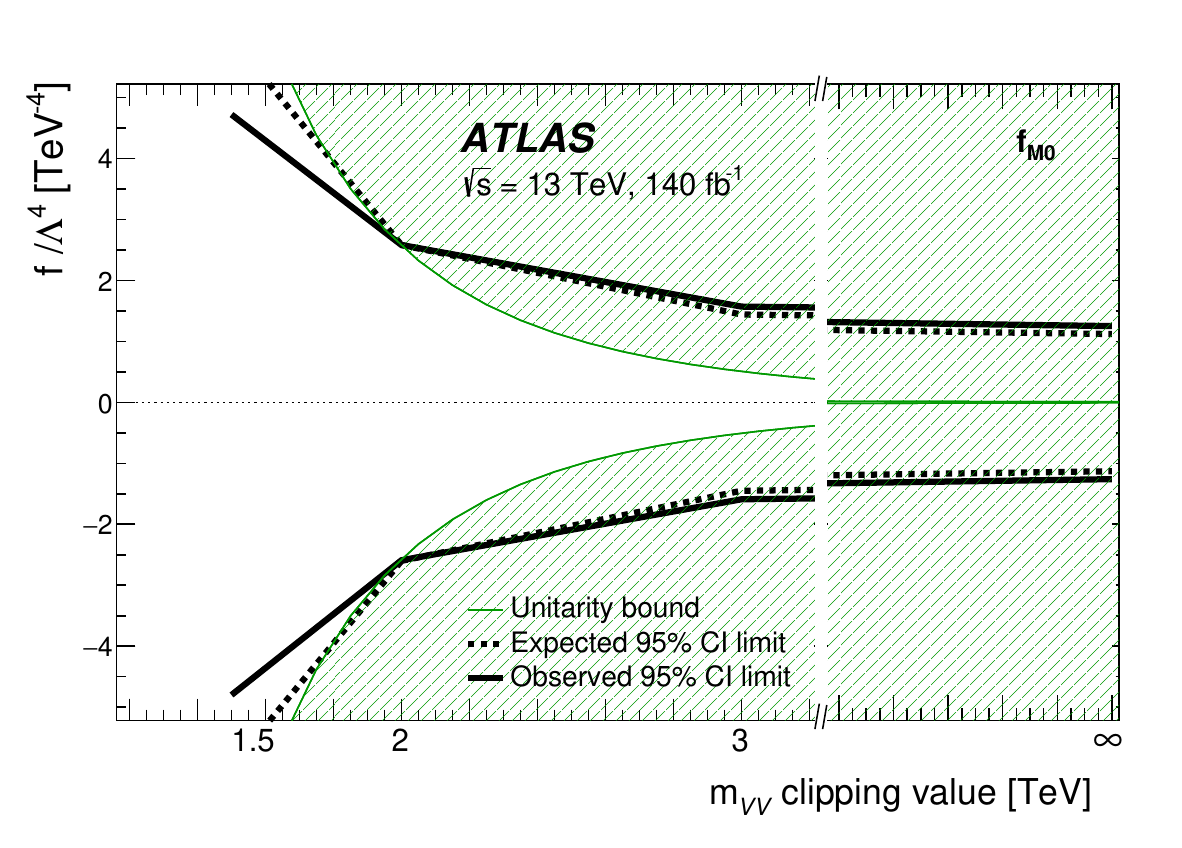}}
\subfigure[]{\includegraphics[width=.40\linewidth]{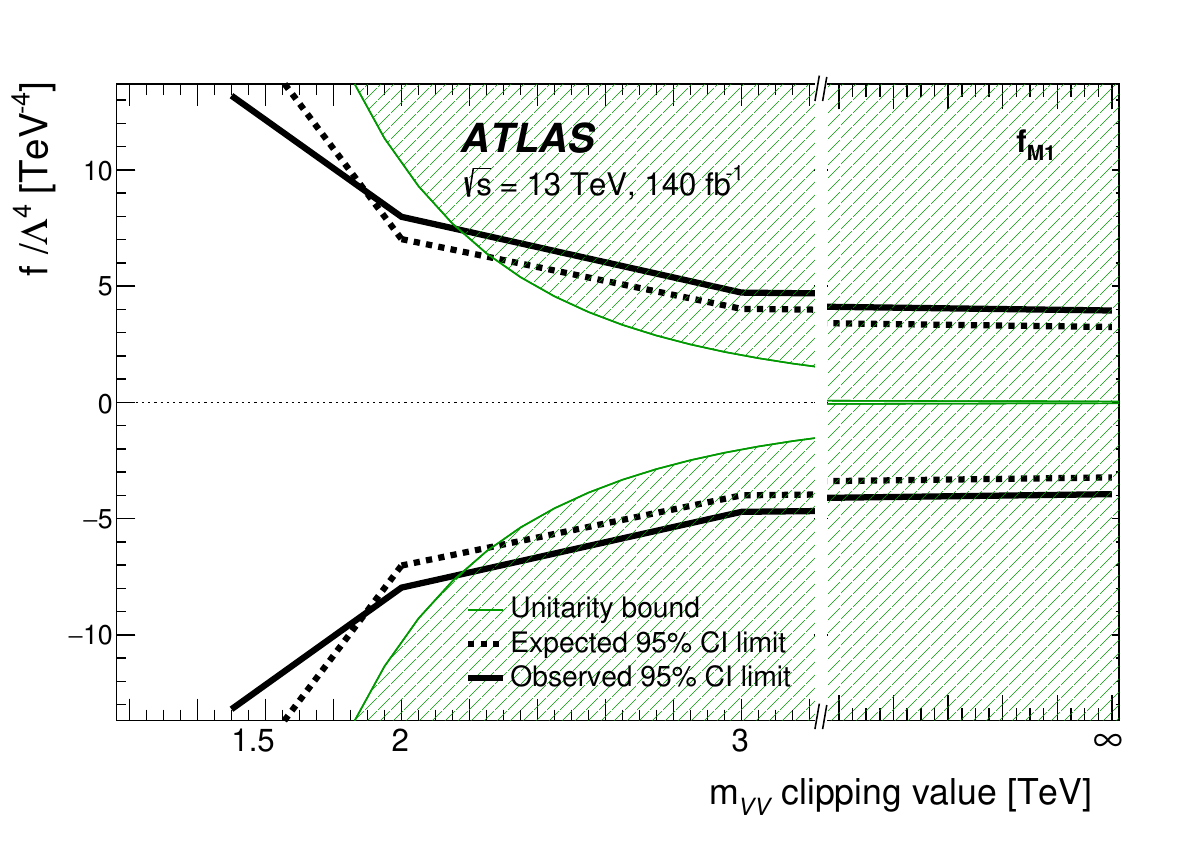}}
\subfigure[]{\includegraphics[width=.40\linewidth]{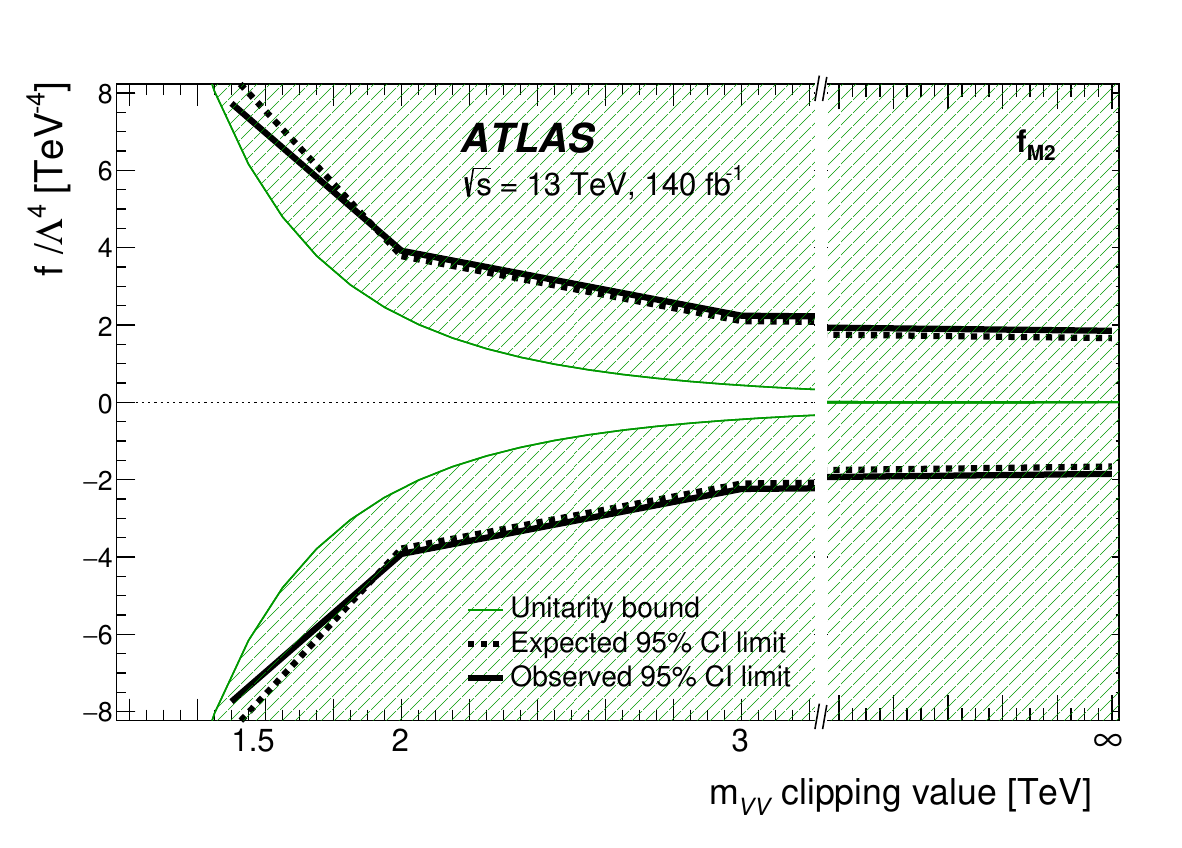}}
\subfigure[]{\includegraphics[width=.40\linewidth]{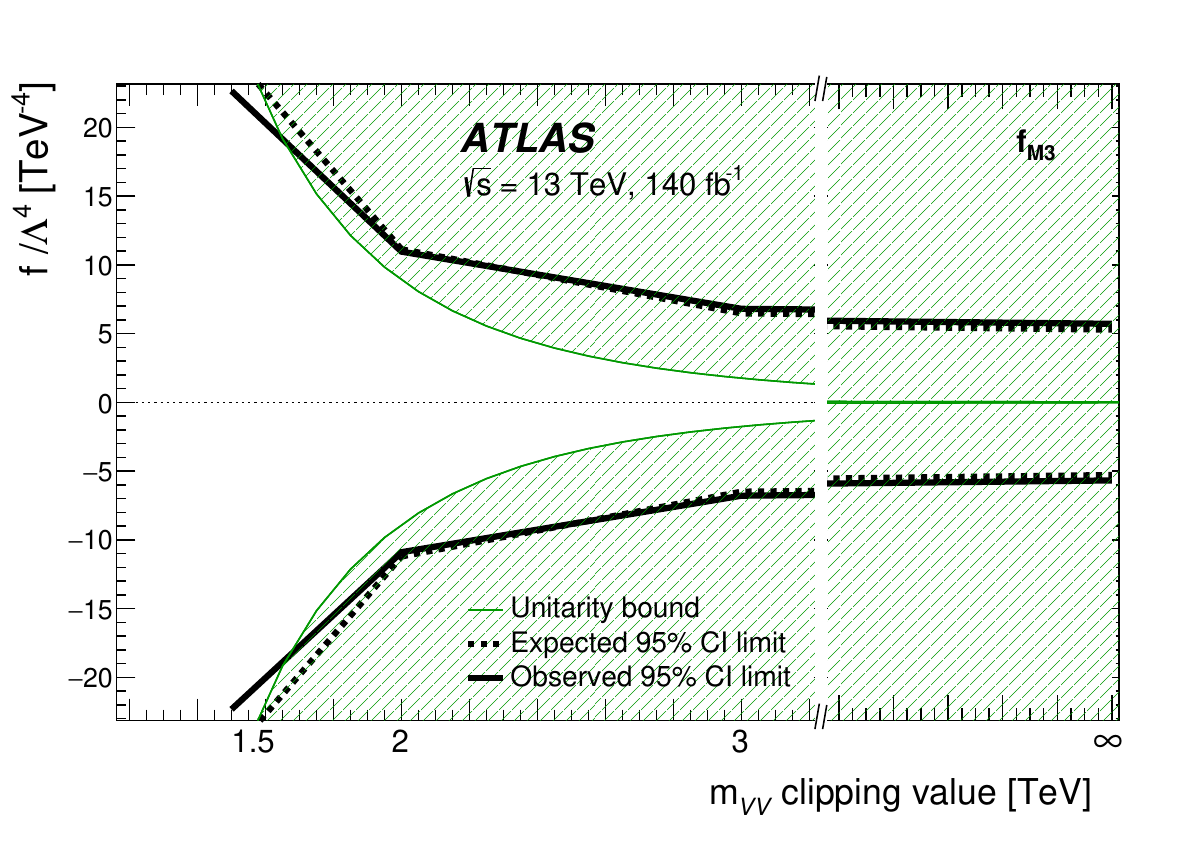}}
\subfigure[]{\includegraphics[width=.40\linewidth]{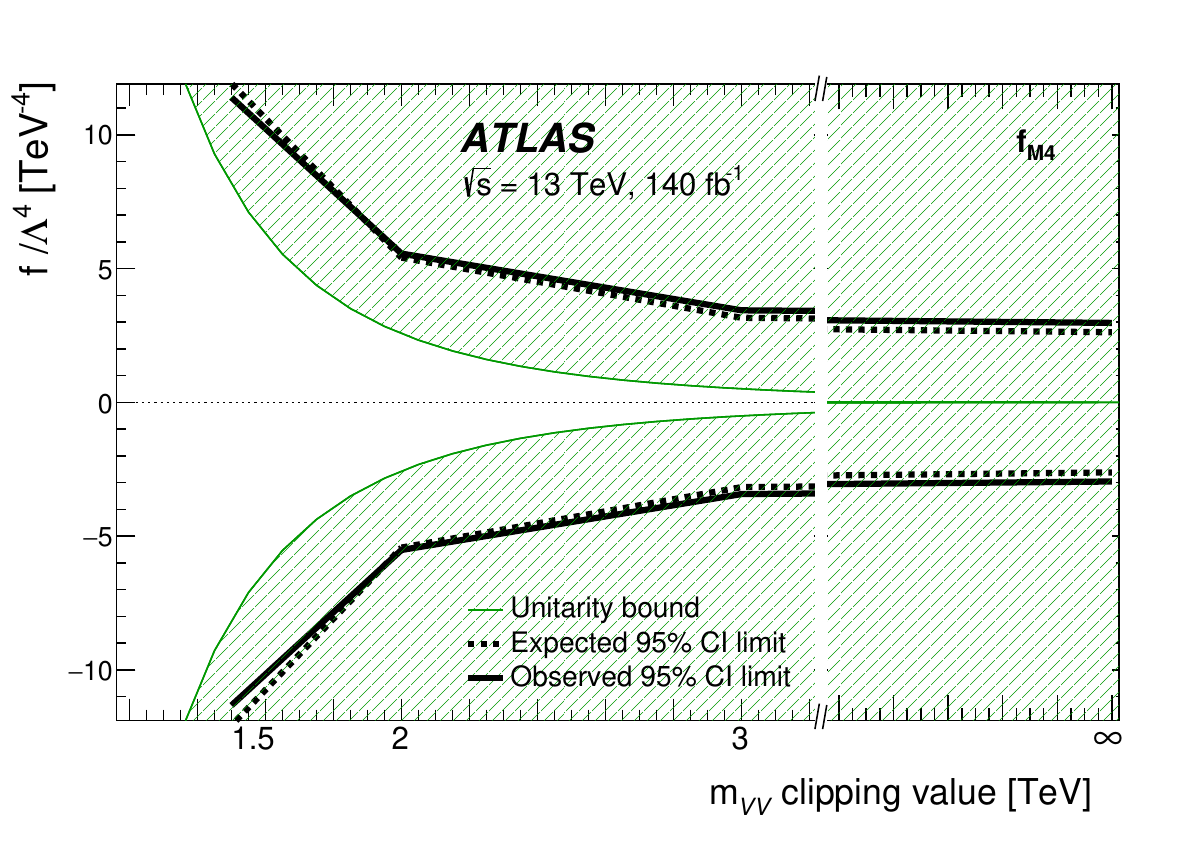}}
\subfigure[]{\includegraphics[width=.40\linewidth]{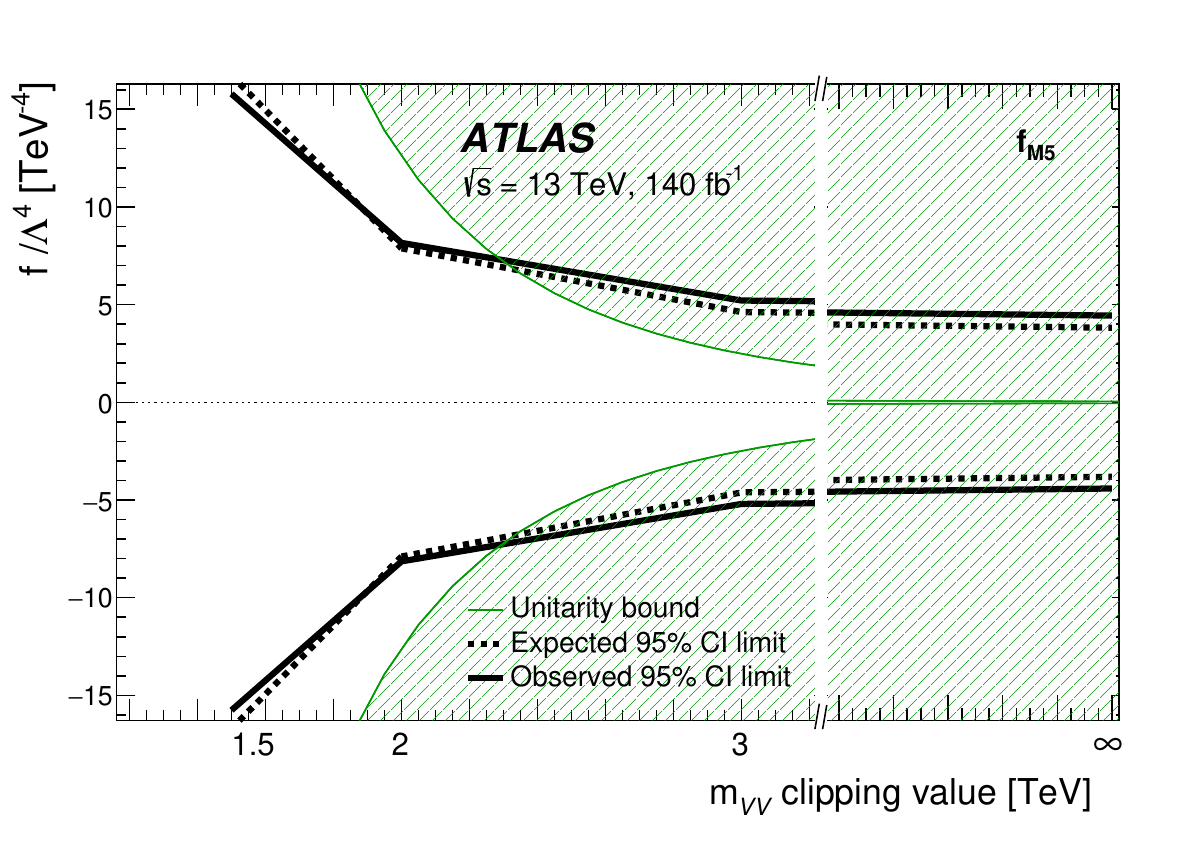}}
\subfigure[]{\includegraphics[width=.40\linewidth]{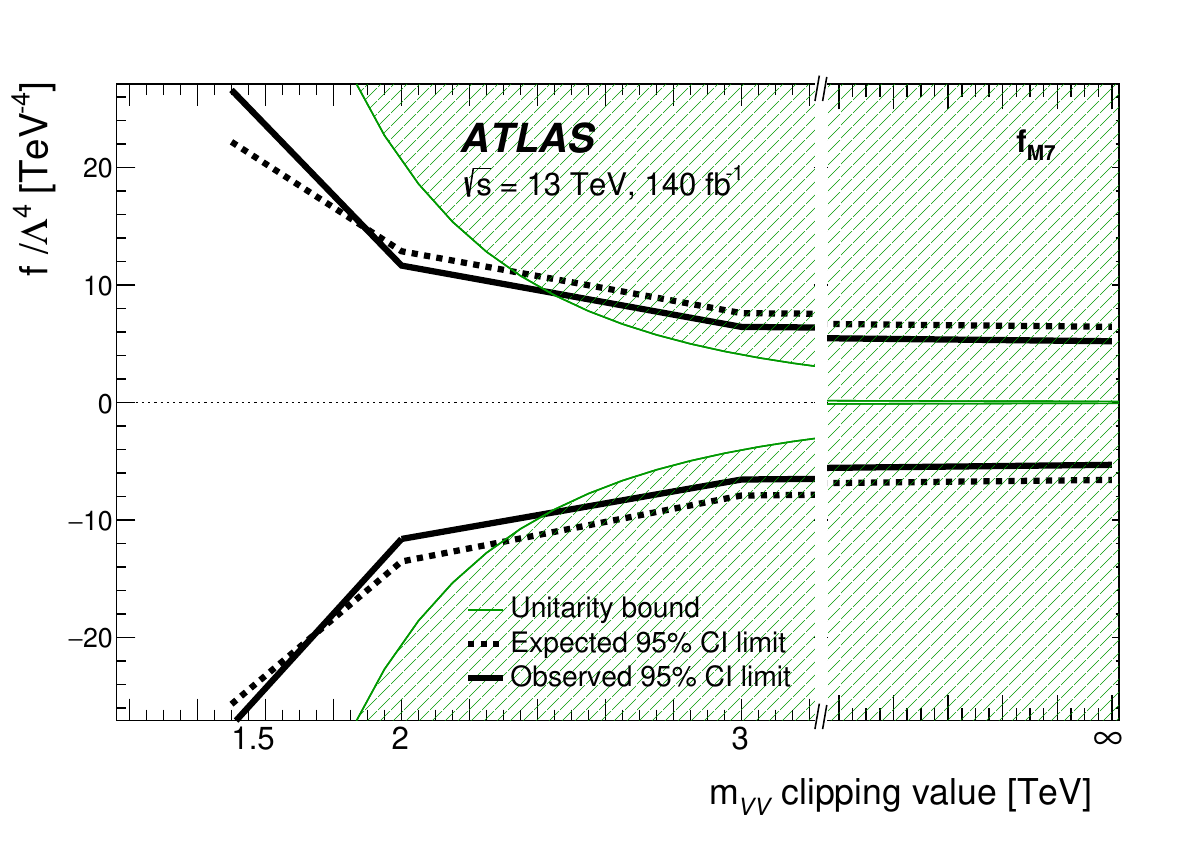}}
\caption{Expected and observed 95\% confidence limits for the dimension-8 $f_{M}$ operators as a function of the clipping scale.
Regions excluded by unitarity constraints are shown as a hatched area.
}
\label{fig:limitClippedFM}
\end{figure}

\FloatBarrier

\begin{table}[th]
\caption{Expected and observed 95\% confidence interval limits on the Wilson coefficients $f_{N}/\Lambda^4$ of various dimension-8 operators.
Limits are obtained considering a single Wilson coefficient to be non-zero at a time.
First, the limit obtained without any clipping scale is shown,
followed by the limit at the clipping scale where the operator crosses the unitarity bounds,
along with the value of the clipping scale itself.
}
\begin{center}
\resizebox{0.9\hsize}{!}{
\begin{tabular}{c c c c c }
\toprule
Wilson coefficient  & Expected limit [$\TeV^{-4}$]  & Observed limit [$\TeV^{-4}$] & Expected limit unitarized [$\TeV^{-4}$] & Observed limit unitarized [$\TeV^{-4}$] \\
\midrule

$f_{T 0} / \Lambda^4$  & [-0.20, 0.18] &[-0.25, 0.22]  & [-0.79, 0.47] at [1.76, 1.96]~TeV   &[-0.85, 0.47] at [1.73, 2.00]~TeV  \\
$f_{T 1} / \Lambda^4$  & [-0.19, 0.19] &[-0.24, 0.24]  & [-0.34, 0.34] at [2.59, 2.59]~TeV   &[-0.43, 0.43] at [2.43, 2.43]~TeV  \\
$f_{T 2} / \Lambda^4$  & [-0.44, 0.44] &[-0.55, 0.55]  & [-0.95, 0.96] at [2.22, 2.22]~TeV   &[-1.16, 1.17] at [2.12, 2.11]~TeV  \\

$f_{T 3} / \Lambda^4$  & [-0.38, 0.38] &[-0.48, 0.48]  & [-0.62, 0.62] at [2.71, 2.71]~TeV   &[-0.88, 0.88] at [2.49, 2.48]~TeV  \\
$f_{T 4} / \Lambda^4$  & [-1.46, 1.32] &[-1.51, 1.37]  & [-3.03, 2.60] at [2.02, 2.09]~TeV   &[-3.03, 2.60] at [2.02, 2.10]~TeV  \\

$f_{T 5} / \Lambda^4$  & [-0.57, 0.53] &[-0.64, 0.58]  & -                                   &[-2.65, 2.57] at [1.53, 1.54]~TeV  \\
$f_{T 6} / \Lambda^4$  & [-0.76, 0.72] &[-0.74, 0.71]  & [-2.82, 2.01] at [1.66, 1.73]~TeV   &[-2.98, 2.62] at [1.64, 1.69]~TeV  \\
$f_{T 7} / \Lambda^4$  & [-1.78, 1.52] &[-1.94, 1.70]  & [-7.88, 4.29] at [1.65, 1.90]~TeV   &[-6.70, 4.11] at [1.72, 1.91]~TeV  \\
$f_{T 8} / \Lambda^4$  & [-0.59, 0.59] &[-0.48, 0.48]  & -                                   & -                                 \\
$f_{T 9} / \Lambda^4$  & [-1.22, 1.22] &[-1.02, 1.03]  & -                                   & -                                 \\
$f_{S 02} / \Lambda^4$ & [-3.22, 3.22] &[-3.96, 3.96]  & [-5.53, 5.54] at [2.07, 2.67]~TeV   &[-6.16, 6.17] at [2.01, 2.01]~TeV  \\
$f_{S 1} / \Lambda^4$  & [-6.84, 6.86] &[-8.06, 8.06]  & -                                   & -                                 \\
$f_{M 0} / \Lambda^4$  & [-1.13, 1.12] &[-1.26, 1.25]  & [-2.61, 2.58] at [2.00, 2.00]~TeV   &[-2.71, 2.65] at [1.97, 1.98]~TeV  \\
$f_{M 1} / \Lambda^4$  & [-3.23, 3.24] &[-3.95, 3.95]  & [-6.22, 6.22] at [2.27, 2.27]~TeV   &[-7.42, 7.43] at [2.17, 2.17]~TeV  \\
$f_{M 2} / \Lambda^4$  & [-1.66, 1.67] &[-1.85, 1.85]  & -                                   & -                                 \\
$f_{M 3} / \Lambda^4$  & [-5.29, 5.29] &[-5.68, 5.71]  & [-23.69, 23.39] at [1.57, 1.57]~TeV &[-18.62, 19.10] at [1.66, 1.65]~TeV \\
$f_{M 4} / \Lambda^4$  & [-2.62, 2.62] &[-2.96, 2.97]  & -                                   & -                                 \\
$f_{M 5} / \Lambda^4$  & [-3.81, 3.82] &[-4.41, 4.44]  & [-6.80, 6.80] at [2.33, 2.33]~TeV   &[-7.28, 7.30] at [2.29, 2.29]~TeV  \\
$f_{M 7} / \Lambda^4$  & [-5.32, 5.20] &[-6.60, 6.43]  & [-9.47, 9.38] at [2.43, 2.43]~TeV   &[-11.91, 11.11] at [2.29, 2.33]~TeV \\ \bottomrule

\end{tabular}
}
\label{tbl:EFTlimits}
\end{center}
\end{table}


%
%
\FloatBarrier

\section{Conclusion}

%
%

A measurement of EWK \VVjj production in proton--proton collisions is performed in the semileptonic final states, using 140~\ifb of data collected at $\sqrt{s}=13$ TeV by the ATLAS experiment at the LHC. The analysis is performed in three decay topologies based on the lepton multiplicity and in two boson \pT regimes.
The measured phase space is designed to be enriched in the \VBS component. The total process is observed with a significance of $7.4\sigma$, while $6.1\sigma$ was expected. The EWK \VVjj production cross section is measured to be $29.2 \pm 4.9$ fb
(expected $20.4 \pm 3.5$) in a fiducial phase space close to the detector and analysis acceptance.
The signal strengths of both electroweak and QCD associated diboson productions are measured in a two-dimensional fit, in agreement with the SM prediction.

Additionally, a search is conducted on possible anomalous quartic gauge couplings, and dimension-8 EFT operator contributions to \VBS production were tested. No significant deviation from the SM prediction is observed, and constraints on the Wilson coefficients
of 19 operators are presented for the first time in this final state by the ATLAS Collaboration.
Some of these constraints are competitive with or better than previous results from other \VBS and triboson measurements.
The observed limits for the S02, T0 and M0 operators are
$(-3.96 < f_{S02} / \Lambda^4 < 3.96)$~$\TeV^{-4}$,
$(-0.25 < f_{T0} / \Lambda^4 < 0.22)$~$\TeV^{-4}$,
$(-1.26 < f_{M0} / \Lambda^4 < 1.25)$~$\TeV^{-4}$.


%
\section*{Acknowledgements}
%

%
%

%
%

We thank CERN for the very successful operation of the LHC and its injectors, as well as the support staff at
CERN and at our institutions worldwide without whom ATLAS could not be operated efficiently.

The crucial computing support from all WLCG partners is acknowledged gratefully, in particular from CERN, the ATLAS Tier-1 facilities at TRIUMF/SFU (Canada), NDGF (Denmark, Norway, Sweden), CC-IN2P3 (France), KIT/GridKA (Germany), INFN-CNAF (Italy), NL-T1 (Netherlands), PIC (Spain), RAL (UK) and BNL (USA), the Tier-2 facilities worldwide and large non-WLCG resource providers. Major contributors of computing resources are listed in Ref.~\cite{ATL-SOFT-PUB-2025-001}.

We gratefully acknowledge the support of ANPCyT, Argentina; YerPhI, Armenia; ARC, Australia; BMWFW and FWF, Austria; ANAS, Azerbaijan; CNPq and FAPESP, Brazil; NSERC, NRC and CFI, Canada; CERN; ANID, Chile; CAS, MOST and NSFC, China; Minciencias, Colombia; MEYS CR, Czech Republic; DNRF and DNSRC, Denmark; IN2P3-CNRS and CEA-DRF/IRFU, France; SRNSFG, Georgia; BMFTR, HGF and MPG, Germany; GSRI, Greece; RGC and Hong Kong SAR, China; ICHEP and Academy of Sciences and Humanities, Israel; INFN, Italy; MEXT and JSPS, Japan; CNRST, Morocco; NWO, Netherlands; RCN, Norway; MNiSW, Poland; FCT, Portugal; MNE/IFA, Romania; MSTDI, Serbia; MSSR, Slovakia; ARIS and MVZI, Slovenia; DSI/NRF, South Africa; MICIU/AEI, Spain; SRC and Wallenberg Foundation, Sweden; SERI, SNSF and Cantons of Bern and Geneva, Switzerland; NSTC, Taipei; TENMAK, T\"urkiye; STFC/UKRI, United Kingdom; DOE and NSF, United States of America.

Individual groups and members have received support from BCKDF, CANARIE, CRC and DRAC, Canada; CERN-CZ, FORTE and PRIMUS, Czech Republic; COST, ERC, ERDF, Horizon 2020, ICSC-NextGenerationEU and Marie Sk{\l}odowska-Curie Actions, European Union; Investissements d'Avenir Labex, Investissements d'Avenir Idex and ANR, France; DFG and AvH Foundation, Germany; Herakleitos, Thales and Aristeia programmes co-financed by EU-ESF and the Greek NSRF, Greece; BSF-NSF and MINERVA, Israel; NCN and NAWA, Poland; La Caixa Banking Foundation, CERCA Programme Generalitat de Catalunya and PROMETEO and GenT Programmes Generalitat Valenciana, Spain; G\"{o}ran Gustafssons Stiftelse, Sweden; The Royal Society and Leverhulme Trust, United Kingdom.

In addition, individual members wish to acknowledge support from Armenia: Yerevan Physics Institute (FAPERJ); CERN: European Organization for Nuclear Research (CERN DOCT); Chile: Agencia Nacional de Investigaci\'on y Desarrollo (FONDECYT 1230812, FONDECYT 1240864, Fondecyt 3240661, Fondecyt Regular 1240721); China: Chinese Ministry of Science and Technology (MOST-2023YFA1605700, MOST-2023YFA1609300), National Natural Science Foundation of China (NSFC - 12175119, NSFC 12275265); Czech Republic: Czech Science Foundation (GACR - 24-11373S), Ministry of Education Youth and Sports (ERC-CZ-LL2327, FORTE CZ.02.01.01/00/22\_008/0004632), PRIMUS Research Programme (PRIMUS/21/SCI/017); EU: H2020 European Research Council (ERC - 101002463); European Union: European Research Council (BARD No. 101116429, ERC - 948254, ERC 101089007), European Regional Development Fund (SMASH COFUND 101081355, SLO ERDF), Horizon 2020 Framework Programme (MUCCA - CHIST-ERA-19-XAI-00), European Union, Future Artificial Intelligence Research (FAIR-NextGenerationEU PE00000013), Horizon 2020 (EuroHPC - EHPC-DEV-2024D11-051), Italian Center for High Performance Computing, Big Data and Quantum Computing (ICSC, NextGenerationEU); France: Agence Nationale de la Recherche (ANR-21-CE31-0022, ANR-22-EDIR-0002, ANR-24-CE31-0504-01); Germany: Baden-Württemberg Stiftung (BW Stiftung-Postdoc Eliteprogramme), Deutsche Forschungsgemeinschaft (DFG - 469666862, DFG - CR 312/5-2); China: Research Grants Council (GRF); Italy: Istituto Nazionale di Fisica Nucleare (ICSC, NextGenerationEU), Ministero dell'Università e della Ricerca (NextGenEU 153D23001490006 M4C2.1.1, NextGenEU I53D23000820006 M4C2.1.1, NextGenEU I53D23001490006 M4C2.1.1); Japan: Japan Society for the Promotion of Science (JSPS KAKENHI JP22H01227, JSPS KAKENHI JP22H04944, JSPS KAKENHI JP22KK0227, JSPS KAKENHI JP23KK0245, JSPS KAKENHI JP24K23939); Norway: Research Council of Norway (RCN-314472); Poland: Ministry of Science and Higher Education (IDUB AGH, POB8, D4 no 9722), Polish National Science Centre (NCN 2021/42/E/ST2/00350, NCN OPUS 2023/51/B/ST2/02507, NCN OPUS nr 2022/47/B/ST2/03059, NCN UMO-2019/34/E/ST2/00393, UMO-2022/47/O/ST2/00148, UMO-2023/49/B/ST2/04085, UMO-2023/51/B/ST2/00920, UMO-2024/53/N/ST2/00869); Portugal: Foundation for Science and Technology (FCT); Spain: Generalitat Valenciana (Artemisa, FEDER, IDIFEDER/2018/048), Ministry of Science and Innovation (MCIN \& NextGenEU PCI2022-135018-2, MICIN \& FEDER PID2021-125273NB, RYC2019-028510-I, RYC2020-030254-I, RYC2021-031273-I, RYC2022-038164-I); Sweden: Carl Trygger Foundation (Carl Trygger Foundation CTS 22:2312), Swedish Research Council (Swedish Research Council 2023-04654, VR 2021-03651, VR 2022-03845, VR 2022-04683, VR 2023-03403, VR 2024-05451), Knut and Alice Wallenberg Foundation (KAW 2018.0458, KAW 2022.0358, KAW 2023.0366); Switzerland: Swiss National Science Foundation (SNSF - PCEFP2\_194658); United Kingdom: Leverhulme Trust (Leverhulme Trust RPG-2020-004), Royal Society (NIF-R1-231091); United States of America: U.S. Department of Energy (ECA DE-AC02-76SF00515), Neubauer Family Foundation.

%
%


%
%
%
%
%
%
%

%
\clearpage
\printbibliography
\clearpage
 
\begin{flushleft}
\hypersetup{urlcolor=black}
{\Large The ATLAS Collaboration}

\bigskip

\AtlasOrcid[0000-0002-6665-4934]{G.~Aad}$^\textrm{\scriptsize 104}$,
\AtlasOrcid[0000-0001-7616-1554]{E.~Aakvaag}$^\textrm{\scriptsize 17}$,
\AtlasOrcid[0000-0002-5888-2734]{B.~Abbott}$^\textrm{\scriptsize 123}$,
\AtlasOrcid[0000-0002-0287-5869]{S.~Abdelhameed}$^\textrm{\scriptsize 119a}$,
\AtlasOrcid[0000-0002-1002-1652]{K.~Abeling}$^\textrm{\scriptsize 55}$,
\AtlasOrcid[0000-0001-5763-2760]{N.J.~Abicht}$^\textrm{\scriptsize 49}$,
\AtlasOrcid[0000-0002-8496-9294]{S.H.~Abidi}$^\textrm{\scriptsize 30}$,
\AtlasOrcid[0009-0003-6578-220X]{M.~Aboelela}$^\textrm{\scriptsize 45}$,
\AtlasOrcid[0000-0002-9987-2292]{A.~Aboulhorma}$^\textrm{\scriptsize 36e}$,
\AtlasOrcid[0000-0001-5329-6640]{H.~Abramowicz}$^\textrm{\scriptsize 157}$
% ...
\bigskip
\\

% ...
$^{17}$Department for Physics and Technology, University of Bergen, Bergen; Norway.\\
% ...
$^{30}$Physics Department, Brookhaven National Laboratory, Upton NY; United States of America.\\
% ...
$^{36}$$^{(a)}$Facult\'e des Sciences Ain Chock, Universit\'e Hassan II de Casablanca;$^{(b)}$Facult\'{e} des Sciences, Universit\'{e} Ibn-Tofail, K\'{e}nitra;$^{(c)}$Facult\'e des Sciences Semlalia, Universit\'e Cadi Ayyad, LPHEA-Marrakech;$^{(d)}$LPMR, Facult\'e des Sciences, Universit\'e Mohamed Premier, Oujda;$^{(e)}$Facult\'e des sciences, Universit\'e Mohammed V, Rabat;$^{(f)}$Institute of Applied Physics, Mohammed VI Polytechnic University, Ben Guerir; Morocco.\\
% ...
$^{45}$Physics Department, Southern Methodist University, Dallas TX; United States of America.\\
% ...
$^{49}$Fakult\"{a}t Physik , Technische Universit{\"a}t Dortmund, Dortmund; Germany.\\
% ...
$^{55}$II. Physikalisches Institut, Georg-August-Universit\"{a}t G\"ottingen, G\"ottingen; Germany.\\
% ...
$^{104}$CPPM, Aix-Marseille Universit\'e, CNRS/IN2P3, Marseille; France.\\
% ...
$^{119}$$^{(a)}$New York University Abu Dhabi, Abu Dhabi;$^{(b)}$United Arab Emirates University, Al Ain; United Arab Emirates.\\
% ...
$^{123}$Homer L. Dodge Department of Physics and Astronomy, University of Oklahoma, Norman OK; United States of America.\\
% ...
$^{157}$Raymond and Beverly Sackler School of Physics and Astronomy, Tel Aviv University, Tel Aviv; Israel.\\
% ...
$^{a}$ Also at Affiliated with an institute formerly covered by a cooperation agreement with CERN.\\
$^{b}$ Also at An-Najah National University, Nablus; Palestine.\\
$^{c}$ Also at Borough of Manhattan Community College, City University of New York, New York NY; United States of America.\\
$^{d}$ Also at Center for Interdisciplinary Research and Innovation (CIRI-AUTH), Thessaloniki; Greece.\\
$^{e}$ Also at Centre of Physics of the Universities of Minho and Porto (CF-UM-UP); Portugal.\\
$^{f}$ Also at CERN, Geneva; Switzerland.\\
% ...

\end{flushleft}


%

%
%

%

%
%
%
%

\end{document}